\documentclass[twocolumn,superscriptaddress,amsmath,amssymb,aps,prx,nofootinbib, 10pt,longbibliography]{revtex4-2}
\usepackage{graphicx}
\usepackage[caption=false]{subfig}
\usepackage{float,wrapfig}
\usepackage{mathtools}
\usepackage{bbm}
\usepackage[normalem]{ulem}
\usepackage[dvipsnames]{xcolor}
\usepackage{bbold}
\usepackage{comment}
\usepackage{amsthm}
\usepackage{amssymb}
\usepackage{enumitem}
\usepackage{physics}
\usepackage{braket}
\usepackage{tikz}
\usepackage[english]{babel}
\usepackage{algorithm}
\usepackage[noend]{algpseudocode}
\usetikzlibrary{calc,fadings,decorations.pathmorphing,arrows.meta,calc,decorations.markings}
\tikzset{midarrow/.style={postaction={decorate,decoration={markings,mark=at position 0.5 with {\arrow{Stealth}}}}}}
\tikzset{endarrow/.style={postaction={decorate,decoration={markings,mark=at position 0.9 with {\arrow{Stealth}}}}}}

\usepackage{lattice-setup} 
\usepackage{xspace}
\usepackage{pgfplots}
\pgfplotsset{compat=newest}

\usepackage[colorlinks, linkcolor=blue, citecolor=blue]{hyperref}

\renewcommand{\phi}{\varphi}
\renewcommand{\epsilon}{\varepsilon}

\renewcommand{\vec}[1]{\boldsymbol{#1}}
\def\U{\mathrm{U}(1)}

\newcommand{\bb}[1]{\mathbb{#1}}

\newcommand{\xkcd}{$X{\cal C}^{(k)}_d$}
\newcommand{\psiTQD}{\ket{\psi_{{\cal D}^\omega(\mathbb{Z}_p^n)}}}

\newcommand{\hide}[1]{}

\theoremstyle{definition}

\newtheorem{theorem}{Theorem}[section]

\newtheorem{lemma}[theorem]{Lemma}
\newtheorem{corollary}[theorem]{Corollary}

\newtheorem{definition}[theorem]{Definition}

\newtheorem{remark}[theorem]{Remark}

\newtheorem{example}[theorem]{Example}
\newtheorem{question}[theorem]{Question}
\newtheorem{conjecture}[theorem]{Conjecture}

\begin{document}
\title{Clifford-hierarchy stabilizer formalism with applications to twisted quantum doubles}

\author{Christopher Fechisin}\affiliation{Joint Quantum Institute, NIST/University of Maryland, College Park, MD, 20742, USA}\affiliation{Joint Center for Quantum Information and Computer Science, NIST/University of Maryland, College Park, MD, 20742, USA}
\author{Kieran Cooney}\affiliation{Joint Quantum Institute, NIST/University of Maryland, College Park, MD, 20742, USA}
\author{Mathi Raja}\affiliation{Joint Quantum Institute, NIST/University of Maryland, College Park, MD, 20742, USA}\affiliation{Joint Center for Quantum Information and Computer Science, NIST/University of Maryland, College Park, MD, 20742, USA}
\author{Victor~V.~Albert}\affiliation{Joint Center for Quantum Information and Computer Science, NIST/University of Maryland, College Park, MD, 20742, USA}
\author{Dominic J. Williamson}\affiliation{School of Physics, The University of Sydney, NSW 2006, Australia}
\author{Kyle Kawagoe}\affiliation{Institute for Physical Sciences and Technology, University of Maryland, College Park, Maryland 20742, USA}
\author{Seth Musser}\affiliation{Condensed Matter Theory Center and Joint Quantum Institute, Department of Physics, University of Maryland, College Park, Maryland 20742, USA}

\begin{abstract}
    Recent work has shown that quantum double models can be used to circumvent the Bravyi-Koenig bound. In particular, this is achieved through code-switching protocols in which non-Pauli stabilizer phases appear as intermediate codes. We introduce the $X\mathcal C^{(k)}_d$ stabilizer formalism, which generalizes the Pauli stabilizer formalism by combining Pauli-$X$ operators with diagonal operators drawn from a fixed level $k$ of the Clifford hierarchy on prime-dimensional qudits. This yields a natural class of beyond-Pauli stabilizer models, including many twisted quantum double phases. We classify a large class of these $X\mathcal C^{(k)}_d$ stabilizer states, showing that they are uniform superpositions over affine-linear subspaces with diagonal phases. In particular, when the stabilizers of a state require operators from level $k$ of the Clifford hierarchy, we show that the diagonal gates required to produce the phases in the wavefunction sit one level higher. We further show that every 2D twisted quantum double ${\cal D}^\omega(\mathbb{Z}_p^n)$ with $\omega$ a product of type-III cocycles and $p$ an odd prime admits a realization using only Clifford ($k=2$) stabilizers, and we conjecture that phases where $\omega$ contains a type-I or type-II cocycle require level-$p$ Clifford-hierarchy stabilizers for $p$-dimensional qudits. Finally, we demonstrate that a pair of qudit toric codes may be deterministically prepared in a logical magic state by measuring $X\mathcal C^{(k)}_d$ stabilizers. Together, these results provide a concrete stabilizer framework for topological codes beyond Pauli stabilizer models and their logical gates.
\end{abstract}

\buildcell{\cellB}{blue}
\buildcell{\cellR}{red}
\buildcell{\cellG}{Green}

\sbox\czbox{\begin{tikzpicture}[x=2.5cm,y=2.5cm,line width=1.5pt,line cap=round]
  \useasboundingbox (0,0);
  \draw[pale=black] ( 90:0.28) to[bend left=20] (150:0.28)
                    (210:0.28) to[bend left=20] (270:0.28)
                    (330:0.28) to[bend left=20] ( 30:0.28);
  \draw[pale=brown] ( 30:0.28) to[bend left=20] ( 90:0.28)
                    (150:0.28) to[bend left=20] (210:0.28)
                    (270:0.28) to[bend left=20] (330:0.28);
\end{tikzpicture}}

\maketitle

\tableofcontents

\section{Introduction}

Stabilizer codes -- quantum error correcting codes wherein the codespace is specified as the mutual $+1$ eigenspace of a group of operators -- form a central framework in quantum error correction. They are highly valued for their combination of broad expressive power with a structure that facilitates a wide range of analysis. The earliest and most extensively studied examples are qubit Pauli stabilizer codes \cite{Gottesman_StabilizerCodesQuantum_1997,CalderbankEtAl_QuantumErrorCorrection_97}, specified by an abelian subgroup $\mathcal{S}\subseteq \mathcal{P}_n$ of the $n$-qubit Pauli group. The codespace is the mutual $+1$ eigenspace of all of the stabilizers $s\in\mathcal{S}$. Given a Pauli stabilizer code, one can efficiently compute a number of key properties, such as the code dimension, logical states, and some logical operators \cite{Gottesman_StabilizerCodesQuantum_1997}. It is also straightforward to extend the Pauli stabilizer formalism from qubits to prime-dimensional qudits, with many analogous results following naturally \cite{Knill_NonbinaryUnitaryError_96,Gottesman_StabilizerCodesQuantum_1997,AshikhminKnill_NonbinaryQuantumStabilizer_01,KetkarEtAl_NonbinaryStabilizerCodes_06, AharonovBen-Or_FaultTolerantQuantumComputation_08a,NadellaKlappenecker_StabilizerCodesFrobenius_12}.

The Pauli stabilizer formalism has also become a useful tool for condensed matter theorists, as it provides a unified language capable of describing both quantum error correcting codes and topological quantum matter. When a stabilizer group $\mathcal{S}$ admits a set of generators $\mathcal{G}\subseteq \mathcal{S}$ that are spatially local on a $d$-dimensional lattice, one can construct a \textit{stabilizer Hamiltonian} from these generators by taking the equal-weight negative sum $H=-\sum_{s\in \cal G}s$ whose groundstate subspace is precisely the codespace of $\mathcal{S}$ \cite{Kitaev_FaulttolerantQuantumComputation_03,DennisEtAl_TopologicalQuantumMemory_02}. When a stabilizer Hamiltonian realizes a topologically ordered phase, the corresponding code is called a topological stabilizer code and enjoys topological protection from local errors \cite{Kitaev_FaulttolerantQuantumComputation_03,DennisEtAl_TopologicalQuantumMemory_02,BombinMartin-Delgado_TopologicalQuantumDistillation_06a,KoenigEtAl_QuantumComputationTuraev_10,SucharaEtAl_ConstructionsNoiseThreshold_11,Bombin_IntroductionTopologicalQuantum_13}. Conversely, any frustration-free commuting Pauli Hamiltonian can be interpreted as a Pauli stabilizer code. 

The paradigmatic model at the intersection of quantum error correction and condensed matter theory is the celebrated Kitaev toric code \cite{Kitaev_FaulttolerantQuantumComputation_03}. In this model, one can identify the syndromes caused by Pauli errors on the codespace with anyonic excitations above the vacuum. Knowledge of topological order further enables the identification of logical errors with the transport of anyons around noncontractible cycles of the torus and motivates decoding via matching algorithms which implement pairwise anyon annihilation \cite{DennisEtAl_TopologicalQuantumMemory_02,Brown_ConservationLawsQuantum_22,HiggottGidney_SparseBlossomCorrecting_25}. Topological insights also facilitate the design of new codes. These include surface codes with boundaries \cite{BravyiKitaev_QuantumCodesLattice_98}, punctures \cite{RaussendorfHarrington_FaultTolerantQuantumComputation_07,BrownEtAl_PokingHolesCutting_17}, and twist defects \cite{Bombin_TopologicalOrderTwist_10,HastingsGeller_ReducedSpacetimeTime_15}, which utilize the physics of anyon condensation at gapped boundaries \cite{BeigiEtAl_QuantumDoubleModel_11,KitaevKong_ModelsGappedBoundaries_12,Burnell_AnyonCondensationIts_18} and topological defects \cite{BarkeshliEtAl_TwistDefectsProjective_13,BarkeshliEtAl_SymmetryFractionalizationDefects_19} to encode logical information.

The correspondence between quantum error correction and topological order leads to a natural question of interest to both communities: \textit{What topological phases of matter can be realized by stabilizer Hamiltonians?} A restricted version of this question was answered by Haah in Ref.~\cite{Haah_ClassificationTranslationInvariant_21}. In particular, Haah proved that every translationally invariant Pauli stabilizer Hamiltonian on prime $p$-dimensional qudits lies in the same phase as a decoupled stack of some number of $\mathbb{Z}_p$ toric codes \cite{Haah_ClassificationTranslationInvariant_21}. One can seek natural generalizations of this result by relaxing one or both of the following assumptions:
\renewcommand{\labelenumi}{(\roman{enumi})}
\begin{enumerate}
    \item prime-dimensional qudits,
    \item stabilizers drawn from the Pauli group.
\end{enumerate}
Either of these relaxations lead to models that realize more general topological orders. In particular, the topological orders realizable by Pauli stabilizer models on composite-dimensional qudits --  relaxing (i) while enforcing (ii) -- have been closely studied \cite{EllisonEtAl_PauliStabilizerModels_22}. Ellison et al.~demonstrate in Ref.~\cite{EllisonEtAl_PauliStabilizerModels_22} that models of this kind can realize every two-dimensional abelian topological order which admits a gapped boundary. Yang and Yu further clarify qualitative distinctions between the topological phases realizable by lattice Pauli stabilizer models and those described by continuum TQFTs \cite{YangYu_ClassificationPauliStabilizer_26}.

Over the past decade, there has been substantial progress in the design of fault-tolerant non-Clifford logical operations \cite{Bombin_2DQuantumComputation_18,LaubscherEtAl_UniversalQuantumComputation_19,Brown_FaulttolerantNonCliffordGate_20,ScrubyEtAl_NumericalImplementationJustInTime_22,HuangChen_GeneratingLogicalMagic_25,DavydovaEtAl_UniversalFaultTolerant_25,KobayashiEtAl_CliffordHierarchyStabilizer_25, WarmanSchafer-Nameki_TransversalCliffordHierarchyGates_25,ManjunathEtAl_UniversalQuantumComputation_26,zhu2026nonabelianqldpctqftformalism,williamson2026fastmagicstatepreparation,christos2026nonabelianquantumlowdensityparity}, with an intense renewed interest driven in part by insights from the study of topological quantum matter. A common theme in these constructions is the use of code-switching protocols~\cite{AndersonEtAl_FaultTolerantConversionSteane_14,KubicaBeverland_UniversalTransversalGates_15,Bombin_DimensionalJumpQuantum_16} involving intermediate codes associated with \textit{non-abelian} topological orders, phases of matter which cannot be realized by composite-qudit Pauli stabilizer codes. In particular, the intermediate codes are generically \textit{twisted quantum doubles}, generalizations of the surface code based on finite groups which can host non-abelian anyons \cite{Kitaev_FaulttolerantQuantumComputation_03}. This strategy circumvents two fundamental limitations of quantum computation with surface codes: 
the Eastin-Knill theorem \cite{EastinKnill_RestrictionsTransversalEncoded_09}, which states that no single quantum error correcting code can transversally implement a universal gate set, and the Bravyi-Koenig bound \cite{BravyiKoenig_ClassificationTopologicallyProtected_13}, which states that a 2D Pauli stabilizer code cannot implement topologically-protected gates outside of the Clifford group. 

The quantum double stabilizer models discussed in Refs.~\cite{DavydovaEtAl_UniversalFaultTolerant_25,KobayashiEtAl_CliffordHierarchyStabilizer_25, WarmanSchafer-Nameki_TransversalCliffordHierarchyGates_25,ManjunathEtAl_UniversalQuantumComputation_26} are able to circumvent the Bravyi-Koenig bound because they require non-Pauli gates such as $CZ$, $S$, and $T$, to be included in the stabilizers, which belong to higher levels of the Clifford hierarchy \cite{GottesmanChuang_DemonstratingViabilityUniversal_99}. Accordingly, these codes are sometimes called \textit{Clifford hierarchy stabilizer codes} \cite{KobayashiEtAl_CliffordHierarchyStabilizer_25}. Despite their clear utility, a general framework for constructing and characterizing Clifford hierarchy stabilizer codes is lacking. 

In this work, we construct a general framework for analyzing a large class of Clifford hierarchy stabilizer codes and explore the expressibility of this code family, especially as it pertains to non-abelian topological orders in general, and quantum double models in particular. We refer to the codes we investigate as \xkcd codes, pronounced ``xkcd,'' codes. These stabilizer models on prime-dimensional qudits are constructed from products of Pauli-$X$ operators and diagonal operators drawn from level $k$ of the Clifford hierarchy. This choice ensures that the operator set is a group, even though level $k$ of the Clifford hierarchy is not a group when $k>2$. Furthermore, nearly all of the Clifford stabilizer codes introduced in Refs.~\cite{DavydovaEtAl_UniversalFaultTolerant_25,KobayashiEtAl_CliffordHierarchyStabilizer_25, WarmanSchafer-Nameki_TransversalCliffordHierarchyGates_25,ManjunathEtAl_UniversalQuantumComputation_26} are of this form. We define these codes in Sec.~\ref{sec:XkCd_intro}.

We remark that \xkcd codes are a special case of monomial stabilizer codes~\cite{nest2011monomial}: codes where every stabilizer can be expressed as a matrix with only one nonzero entry per row and column in some basis. Accordingly, \xkcd codes inherit the properties shared by all monomial stabilizer codes, such as a canonical basis for the codespace formed by orbits of a permutation group~\cite{nest2011monomial}. However, \xkcd codes introduce additional structure not guaranteed in an arbitrary monomial stabilizer code---namely, stabilizers belonging to the Clifford hierarchy. This information is closely linked to practical considerations arising in quantum computation, which motivates us to study these codes.

It is natural to consider the class of states which are uniquely stabilized by \xkcd codes. Groundstates of topological codes in general and TQDs in particular may be studied in this framework by placing them on trivial topologies or including logical operators in the stabilizer group, such that the codespace becomes one-dimensional. In Sec.~\ref{sec:XkCd_class} we consider this problem and in Thm.~\ref{thm:unique_state} we prove that the most generic computational-basis expansion of states uniquely stabilized by \xkcd stabilizers is
\begin{equation}
|\psi\rangle = \frac{1}{p^{\ell/2}}\sum_{\mathbf{x}\in \mathbb{Z}_p^\ell} e^{2\pi i \varphi(\mathbf{x})}|\mathbf{x}, M\mathbf{x}+\mathbf{c}\rangle,
\end{equation}
where $M$ is a matrix, $\mathbf{c}$ is a vector, and $e^{2\pi i \varphi}$ is a diagonal operator at most at level $k+1$ of the Clifford hierarchy. In other words, if we want phases in our wavefunction ``at level $k+1$,'' then we need an \xkcd code with stabilizers at level $k$. For the case of qubits, $p=2$, this result is consistent with the wavefunctions known to be preparable by either Pauli stabilizers, $k=1$, or XS stabilizers, $k=2$ \cite{HostensEtAl_StabilizerStatesClifford_05a, NiEtAl_NonCommutingStabilizerFormalism_15}.

The questions of what phases arise in the twisted quantum double wavefunction and what operators stabilize it are intimately connected to the universal data that characterizes the anyons in the TQD, as suggested by the methods of Ref.~\cite{Haah_ClassificationTranslationInvariant_21}. There, products of on-site stabilizers forming a Wilson line were used to compute the topological spin of anyons in the topological order, linking the stabilizer formulation to universal data. Furthermore, the cocycle $\omega \in H^3(G, U(1))$ which defines the twisted quantum double ${\cal D}^\omega(G)$ figures prominently in both microscopic realizations and macroscopic descriptions of the phase. 

When $G$ is a finite abelian group, the physical consequences of the twist $\omega$ for the anyon theory are especially clear. Here every 3-cocycle of $G$ may be decomposed into a product of three qualitatively distinct types of contributions, known as type-I, -II, and -III cocycles~\cite{deWildPropitius_TopologicalInteractionsBroken_95}. Of these cocycles Type-III cocycles have the most dramatic effect on the anyon theory. They give rise to non-abelian anyons even though the underlying gauge group $G$ is abelian.

Given the connection between universal data and stabilizer algebra, one may ask a natural question, discussed in Sec.~\ref{sec:realizing_with}:
\begin{question}\label{q:main_intro}
    Given a twisted abelian quantum double topological order ${\cal D}^\omega(\mathbb{Z}_p^n)$ with $p$ prime, what is the minimal level $k$ such that there exists an \xkcd codestate realizing the topological order ${\cal D}^\omega(\mathbb{Z}_p^n)$?
\end{question}
\noindent This is challenging to answer in full generality and we are not able to fully address it here. Nonetheless we make a conjecture that we substantiate in Sec.~\ref{sec:realizing_with}:
\begin{conjecture}\label{conj:intro_level_k}
   Given a TQD topological order ${\cal D}^\omega(\mathbb{Z}_p^n)$, the minimum $k$ such that there exists an \xkcd stabilizer code whose codestate lies in the ${\cal D}^\omega(\mathbb{Z}_p^n)$ phase is given by:
   \begin{equation}
       k = \begin{cases}
           2,&\text{if }\omega\text{ is a product of type-III cocycles}\\
           p,&\text{otherwise}
       \end{cases},
   \end{equation}
   where we take $\omega$ to be nontrivial.
\end{conjecture}
\noindent In Sec.~\ref{sec:realizing_with}, we prove the first case by construction and provide supporting evidence for the second case. Our proof demonstrates that twisted quantum doubles $\mathcal{D}^\omega(\mathbb{Z}_p^n)$ for which $\omega$ is a product of type-III cocycles are realizable with Clifford stabilizers ($k=2$). Because these topological orders host non-abelian anyons, they cannot be realized with Pauli stabilizers ($k=1$), so $k=2$ is in fact a lower bound. We also highlight a surprising consequence of Conjecture~\ref{conj:intro_level_k}: some qudit TQD models with non-abelian anyons (namely, those for which $\omega$ is a product of type-III cocycles) can be realized with lower-level stabilizers than TQD models with only abelian anyons. We emphasize that this unexpected conclusion only holds for qudits of prime dimension $\geq 3$, as the two cases in Conjecture~\ref{conj:intro_level_k} coincide for $p=2$, i.e. for qubits. 

In Sec.~\ref{sec:gates}, we present an application of our formalism to magic state preparation for logical qudits. We construct a code-switching protocol which includes measurement of \xkcd stabilizers as a key step, generalizing a protocol discussed in Ref.~\cite{DavydovaEtAl_UniversalFaultTolerant_25} to odd prime qudits. Notably, while the qubit protocol probabilistically prepares a magic state, our protocol does so deterministically.

Finally, in Sec.~\ref{sec:discussion} we discuss future directions prompted by our introduction of \xkcd codes.

\section{Review of twisted quantum double models}\label{sec:TQD_review}

In this section, we will review the definition and some key features of twisted quantum double models~\cite{DijkgraafWitten_TopologicalGaugeTheories_90,HuEtAl_TwistedQuantumDouble_13}. While this section helps to motivate the stabilizer set from which \xkcd codes are constructed, readers primarily interested in the quantum error-correcting properties of \xkcd codes (Secs.~\ref{sec:XkCd_intro} and \ref{sec:XkCd_class}) may safely move on to Sec.~\ref{sec:XkCd_intro}. 

The operator set we consider here is naturally suited to the task of realizing twisted quantum double models. Twisted quantum doubles are 2D lattice models whose low-energy physics is described by twisted $G$-gauge theory for a finite group $G$, also known as Dijkgraaf-Witten theory~\cite{DijkgraafWitten_TopologicalGaugeTheories_90}. These models generalize the quantum double based on the group $G$ (which realizes ordinary $G$-gauge theory) by ``twisting'' the $X$-type stabilizers by $U(1)$ phases~\cite{HuEtAl_TwistedQuantumDouble_13}. We will specialize to the case $G=\mathbb{Z}_p^n$, which corresponds to a stack of $n$ toric codes on $p$-dimensional qudits with an additional twist. Given such a stack, the possible twists are classified by the cohomology classes $\omega\in H^3(\mathbb{Z}_p^n,U(1))$~\cite{DijkgraafWitten_TopologicalGaugeTheories_90}. 

The resulting topological order, denoted ${\cal D}^\omega(\mathbb{Z}_p^n)$, can be realized by a generalized stabilizer model with qudits valued in $\mathbb{Z}_p^n$ placed on the edges of an arbitrary triangulation of a 2$d$ manifold with branching structure\footnote{A branching structure of a 2D triangulation is a choice of orientation for each edge such that in each plaquette, at least one edge is directed clockwise, and at least one edge is directed counterclockwise. This guarantees that the vertices of a triangle can be ordered, with edges directed from lower-ordered indices to higher ones.}. 
We denote the vertices $v\in V$, edges $e\in E$, and faces $f\in F$. The canonical Hamiltonian takes the form~\cite{HuEtAl_TwistedQuantumDouble_13}
\begin{equation}\label{eq:can_TQD_ham}
    H_{{\cal D}^\omega(\mathbb{Z}_p^n)}=-\sum_{v\in V}\frac{1}{p^n}\sum_{g\in \mathbb{Z}_p^n}A_g^\omega(v)-\sum_{f\in F} B(f), 
\end{equation}
where the $A_g^\omega(v)$ and $B(f)$ are the ``star'' and ``plaquette'' terms, to be defined shortly. 

\begin{figure}
    \newcommand\scale{0.88}

\centering
\begin{minipage}{.2\textwidth}
\begin{tikzpicture}[scale=\scale]

\tikzset{
    vertex/.style={circle, fill, inner sep=2pt},
    edgedot/.style={circle,draw=black, fill=white, inner sep=2pt},
    edge/.style={line width=1pt}
}

\coordinate (v1) at (90:2);
\coordinate (v2) at (30:2);
\coordinate (v3) at (-30:2);
\coordinate (v4) at (-90:2);
\coordinate (v5) at (-150:2);
\coordinate (v6) at (150:2);

\coordinate (c) at (0,0);

\fill[fill=red!20] (v6)--(c)--(v1);

\tikzfading[name=fade along, top color=transparent!100, bottom color=transparent!0]

\foreach \x/\y/\zF/\zB/\a in {v1/v2/240/60/black,v2/v3/180/0/black,v3/v4/120/300/black,v4/v5/0/180/black,v5/v6/0/180/black,v6/v1/0/180/blue}{\draw[edge,endarrow,\a] (\y) -- (\x);
\draw[edge,path fading=fade along,fading angle=\zF] (\y) -- ($(\x)!1.2!(\y)$);
\draw[edge,path fading=fade along,fading angle=\zB] (\x) -- ($(\x)!-0.2!(\y)$);
}

\foreach \x/\y/\z in {v1/0/blue,v2/300/black,v3/240/black,v4/180/black,v5/120/black,v6/60/blue}{
\draw[edge,endarrow,\z] (c) -- (\x);
\draw[edge,path fading=fade along,fading angle=\y] (\x) -- ($(c)!1.13!(\x)$);}

\node[edgedot] at ($(v1)!0.5!(v2)$) {};
\node[edgedot] at ($(v2)!0.5!(v3)$) {};
\node[edgedot] at ($(v3)!0.5!(v4)$) {};
\node[edgedot] at ($(v4)!0.5!(v5)$) {};
\node[edgedot] at ($(v5)!0.5!(v6)$) {};
\node[edgedot,label={[label distance=12pt,text=blue,font=\large,centered]110:$\partial(f)$}] at ($(v6)!0.5!(v1)$) {};
\node[edgedot,label={[label distance=13pt,text=red,font=\large,centered]180:$f$}] at ($(v1)!0.5!(c)$) {};
\node[edgedot] at ($(v2)!0.5!(c)$) {};
\node[edgedot] at ($(v3)!0.5!(c)$) {};
\node[edgedot] at ($(v4)!0.5!(c)$) {};
\node[edgedot] at ($(v5)!0.5!(c)$) {};
\node[edgedot] at ($(v6)!0.5!(c)$) {};
\pgfresetboundingbox 
\path[use as bounding box] (-2.15,-2.3) rectangle + (4.25,4.7);
\end{tikzpicture}
\end{minipage}\quad\;
\begin{minipage}{.2\textwidth}
\begin{tikzpicture}[scale=\scale]
\tikzset{
    vertex/.style={circle, fill=Plum, inner sep=3pt},
    edgedot/.style={circle,draw=black, fill=white, inner sep=2pt},
    edge/.style={line width=1pt}
}

\coordinate (v1) at (90:2);
\coordinate (v2) at (30:2);
\coordinate (v3) at (-30:2);
\coordinate (v4) at (-90:2);
\coordinate (v5) at (-150:2);
\coordinate (v6) at (150:2);

\coordinate (c) at (0,0);

\tikzfading[name=fade along, top color=transparent!100, bottom color=transparent!0]

\foreach \x/\y/\zF/\zB in {v1/v2/240/60,v2/v3/180/0,v3/v4/120/300,v4/v5/0/180,v5/v6/0/180,v6/v1/0/180}{\draw[edge,endarrow,ForestGreen] (\y) -- (\x);
\draw[edge,path fading=fade along,fading angle=\zF] (\y) -- ($(\x)!1.2!(\y)$);
\draw[edge,path fading=fade along,fading angle=\zB] (\x) -- ($(\x)!-0.2!(\y)$);
}

\foreach \x/\y in {v1/0,v2/300,v3/240,v4/180,v5/120,v6/60}{
\draw[edge,endarrow,Dandelion!90!black] (c) -- (\x);
\draw[edge,path fading=fade along,fading angle=\y] (\x) -- ($(c)!1.13!(\x)$);}

\node[vertex,label={[label distance=8pt,text=Plum,font=\large,centered]60:$v$}] at (c) {};

\node[edgedot,label={[label distance=12pt,text=ForestGreen,font=\large,centered]70:$\mathrm{Lk}(v)$}] at ($(v1)!0.5!(v2)$) {};
\node[edgedot] at ($(v2)!0.5!(v3)$) {};
\node[edgedot] at ($(v3)!0.5!(v4)$) {};
\node[edgedot] at ($(v4)!0.5!(v5)$) {};
\node[edgedot] at ($(v5)!0.5!(v6)$) {};
\node[edgedot] at ($(v6)!0.5!(v1)$) {};
\node[edgedot] at ($(v1)!0.5!(c)$) {};
\node[edgedot] at ($(v2)!0.5!(c)$) {};
\node[edgedot] at ($(v3)!0.5!(c)$) {};
\node[edgedot,label={[label distance=15pt,text=Dandelion!90!black,font=\large,centered]0:$\delta(v)$}] at ($(v4)!0.5!(c)$) {};
\node[edgedot] at ($(v5)!0.5!(c)$) {};
\node[edgedot] at ($(v6)!0.5!(c)$) {};

\pgfresetboundingbox 
\path[use as bounding box] (-2.15,-2.3) rectangle + (4.25,4.7);

\end{tikzpicture}
\end{minipage}
    \caption{The geometry associated to the twisted quantum double stabilizers $B(f)$ (left) and $A^\omega_g(v)$ (right) defined in Eqs.~\ref{eq:TQD_Zs} and \ref{eq:TQD_Xs}, for an example triangulation. All colored edges are in the support of the corresponding stabilizer.
    }
    \label{fig:TQD_geometry}
\end{figure}

Borrowing language from lattice gauge theory, the plaquette stabilizers are projectors onto the ``zero-flux'' or ``flat'' subspace, which take the form
\begin{equation}\label{eq:TQD_Zs}
B(f)=\sum_{\{e_i\}\in\mathbb{Z}_p^n}\delta_{e_1+e_2-e_3,0}\ketbra{e_1,e_2,e_3}{e_1,e_2,e_3},
\end{equation}
for all faces $f\in F$, where $e_1,e_2,e_3\in \partial(f)$, the set of edges bordering $f$, and $e_3$ is taken without loss of generality to be the edge oriented against the other two. 
For example, 
\begin{equation}
B(f)\vcenter{\hbox{\newcommand\scale{0.8}
\begin{tikzpicture}[scale=\scale,]
\tikzset{
    vertex/.style={circle, fill, inner sep=2pt},
    edgedot/.style={circle,draw=black, fill=white, inner sep=2pt},
    edge/.style={line width=1pt}
}

\coordinate (v1) at (120:2);
\coordinate (v6) at (180:2);

\coordinate (c) at (0,0);

\tikzfading[name=fade along, top color=transparent!100, bottom color=transparent!0]


\foreach \x/\y/\zF/\zB/\a in {v6/v1/0/180/black}{\draw[edge,endarrow,\a] (\y) -- (\x);
}

\foreach \x/\y/\z in {v1/0/black,v6/60/black}{
\draw[edge,endarrow,\z] (c) -- (\x);}

\node[edgedot,label={[label distance=8pt,font=\normalsize,centered]150:$\ket{h}$}] at ($(v6)!0.5!(v1)$) {};
\node[edgedot,label={[label distance=8pt,text=black,font=\normalsize,centered]30:$\ket{g}$}] at ($(v1)!0.5!(c)$) {};
\node[edgedot,label={[label distance=8pt,text=black,font=\normalsize,centered]270:$\ket{k}$}] at ($(v6)!0.5!(c)$) {};
\node[label={[label distance=\scale*34,font=\huge,centered,rotate=270]230:$\circlearrowleft$}] at (c) {};



\pgfresetboundingbox 
\path[use as bounding box] (-2.25,-0.8) rectangle + (2.5,2.6);


\end{tikzpicture}}}=\delta_{g+h-k,0}\vcenter{\hbox{\newcommand\scale{0.8}
\begin{tikzpicture}[scale=\scale,]
\tikzset{
    vertex/.style={circle, fill, inner sep=2pt},
    edgedot/.style={circle,draw=black, fill=white, inner sep=2pt},
    edge/.style={line width=1pt}
}

\coordinate (v1) at (120:2);
\coordinate (v6) at (180:2);

\coordinate (c) at (0,0);

\tikzfading[name=fade along, top color=transparent!100, bottom color=transparent!0]


\foreach \x/\y/\zF/\zB/\a in {v6/v1/0/180/black}{\draw[edge,endarrow,\a] (\y) -- (\x);
}

\foreach \x/\y/\z in {v1/0/black,v6/60/black}{
\draw[edge,endarrow,\z] (c) -- (\x);}

\node[edgedot,label={[label distance=8pt,font=\normalsize,centered]150:$\ket{h}$}] at ($(v6)!0.5!(v1)$) {};
\node[edgedot,label={[label distance=8pt,text=black,font=\normalsize,centered]30:$\ket{g}$}] at ($(v1)!0.5!(c)$) {};
\node[edgedot,label={[label distance=8pt,text=black,font=\normalsize,centered]270:$\ket{k}$}] at ($(v6)!0.5!(c)$) {};
\node[label={[label distance=\scale*34,font=\huge,centered,rotate=270]230:$\circlearrowleft$}] at (c) {};



\pgfresetboundingbox 
\path[use as bounding box] (-2.25,-0.8) rectangle + (2.5,2.6);


\end{tikzpicture}}}.
\end{equation}

The vertex stabilizers, labeled by $g\in\mathbb{Z}_p^n$ and $\omega\in H^3(\mathbb{Z}_p^n,U(1))$, act on all edges bordering the faces which meet at the vertex $v$:
\begin{equation}\label{eq:TQD_Xs}
\begin{split}
    A_g^\omega(v)=\sum_{\{e_i\}\in\mathbb{Z}_p^n}f^\omega_g(e)&\left(\bigotimes_{e_j\in\delta(v)}\ketbra{e_j\pm g}{e_j}\right)\\
    &\qquad\;\times\left(\bigotimes_{e_k\in\mathrm{Lk}(v)}\ketbra{e_k}{e_k}\right),
\end{split}
\end{equation}
where $f^\omega_g(e)\in U(1)$ and the sets\footnote{The notation for $\partial$, $\delta$, and $\mathrm{Lk}$ is standard and originates in the study of homology. In particular, $\partial$ is the boundary operator which maps $d$-dimensional objects to $(d-1)$-dimensional objects and $\delta$ is the coboundary operator which acts in the reverse direction. Furthermore, $\mathrm{Lk}(v)$ denotes the ``link'' of $v$, which for a 2D complex corresponds to the edges and vertices on the perimeter of the faces which meet at $v$.} $\delta(v)$ and $\mathrm{Lk}(v)$ are defined for an example triangulation in Fig.~\ref{fig:TQD_geometry}. The sign appearing in $|e_j\pm g\rangle$ depends on the orientation of the edge $e_j$; the plus sign corresponds to $e_j$ directed out of $v$, and the minus sign to $e_j$ directed into $v$. Schematically, the function $f^\omega_g:\mathbb{Z}_p^x\to U(1)$ is a product over 3-cocycles evaluated on different group elements, where each factor is associated with a face adjacent to $v$. For the triangulation presented in Fig.~\ref{fig:TQD_geometry}, we have
\begin{equation}
    f_g^\omega(e)=\prod_{f\in\delta(v)}\omega|_f,\quad \omega|_f=\omega(g,a,b),
\end{equation}
where $a$ and $b$ are defined for each plaquette as
\begin{equation}
    \vcenter{\hbox{\newcommand\scale{0.8}
\begin{tikzpicture}[scale=\scale,]
\tikzset{
    vertex/.style={circle, fill, inner sep=2pt},
    edgedot/.style={circle,draw=black, fill=white, inner sep=2pt},
    edge/.style={line width=1pt}
}

\coordinate (v1) at (120:2);
\coordinate (v6) at (180:2);

\coordinate (c) at (0,0);

\tikzfading[name=fade along, top color=transparent!100, bottom color=transparent!0]

\fill[fill=red!20] (v6)--(c)--(v1);
\node[label={[label distance=9pt,text=red,font=\large,centered]205:$f$}] at ($(v1)!0.5!(c)$) {};

\foreach \x/\y/\zF/\zB/\a in {v6/v1/0/180/black}{\draw[edge,endarrow,\a] (\y) -- (\x);
}

\foreach \x/\y/\z in {v1/0/black,v6/60/black}{
\draw[edge,endarrow,\z] (c) -- (\x);}

\node[vertex,fill=Plum,label={[label distance=8pt,text=Plum,font=\large,centered]330:$v$}] at (c) {};
\node[vertex] at (v1) {};
\node[vertex] at (v6) {};

\node[edgedot,label={[label distance=10pt,centered]170:$\ket{b}$}] at ($(v6)!0.5!(v1)$) {};
\node[edgedot,label={[label distance=10pt,text=black,centered]10:$\ket{a}$}] at ($(v1)!0.5!(c)$) {};
\node[edgedot,label={[label distance=6pt,text=black,centered]270:$\ket{c}$}] at ($(v6)!0.5!(c)$) {};

\pgfresetboundingbox 
\path[use as bounding box] (-2.25,-0.8) rectangle + (2.5,2.6);


\end{tikzpicture}}}.
\end{equation}
The algorithm for computing $\omega|_f$ in a given triangulation is detailed in Fig.~\ref{fig:D_omega}.

\begin{figure*}
    \centering
    $\vcenter{\hbox{\newcommand\scale{0.8}
\begin{tikzpicture}[scale=\scale,]
\tikzset{
    vertex/.style={circle, fill, inner sep=2pt},
    edgedot/.style={circle,draw=black, fill=white, inner sep=2pt},
    edge/.style={line width=1pt}
}

\coordinate (v1) at (120:2);
\coordinate (v6) at (180:2);

\coordinate (c) at (0,0);

\tikzfading[name=fade along, top color=transparent!100, bottom color=transparent!0]

\fill[fill=red!20] (v6)--(c)--(v1);
\node[label={[label distance=9pt,text=red,font=\large,centered]205:$f$}] at ($(v1)!0.5!(c)$) {};

\foreach \x/\y/\zF/\zB/\a in {v6/v1/0/180/black}{\draw[edge,endarrow,\a] (\y) -- (\x);
}

\foreach \x/\y/\z in {v1/0/black,v6/60/black}{
\draw[edge,endarrow,\z] (c) -- (\x);}

\node[vertex,fill=Plum,label={[label distance=8pt,text=Plum,font=\large,centered]330:$v$}] at (c) {};
\node[vertex] at (v1) {};
\node[vertex] at (v6) {};

\node[edgedot,label={[label distance=10pt,centered]170:$\ket{b}$}] at ($(v6)!0.5!(v1)$) {};
\node[edgedot,label={[label distance=10pt,text=black,centered]10:$\ket{a}$}] at ($(v1)!0.5!(c)$) {};
\node[edgedot,label={[label distance=6pt,text=black,centered]270:$\ket{c}$}] at ($(v6)!0.5!(c)$) {};

\pgfresetboundingbox 
\path[use as bounding box] (-2.25,-0.8) rectangle + (2.5,2.6);


\end{tikzpicture}}}\quad\Rightarrow\quad\vcenter{\hbox{\newcommand\scale{0.8}
\begin{tikzpicture}[scale=\scale,]
\tikzset{
    vertex/.style={circle, fill, inner sep=2pt},
    edgedot/.style={circle,draw=black, fill=white, inner sep=2pt},
    edge/.style={line width=1pt}
}

\coordinate (v1) at (120:2);
\coordinate (v6) at (180:2);

\coordinate (c) at (0,0);

\tikzfading[name=fade along, top color=transparent!100, bottom color=transparent!0]

\fill[fill=red!20] (v6)--(c)--(v1);
\node[label={[label distance=9pt,text=red,font=\large,centered]205:$f$}] at ($(v1)!0.5!(c)$) {};

\foreach \x/\y/\zF/\zB/\a in {v6/v1/0/180/black}{\draw[edge,endarrow,\a] (\y) -- (\x);
}

\foreach \x/\y/\z in {v1/0/black,v6/60/black}{
\draw[edge,endarrow,\z] (c) -- (\x);}

\node[vertex,fill=Plum,label={[label distance=8pt,text=Plum,font=\large,centered]330:$v_1$}] at (c) {};
\node[vertex,label={[label distance=8pt,font=\large,centered]90:$v_2$}] at (v1) {};
\node[vertex,label={[label distance=8pt,font=\large,centered]240:$v_3$}] at (v6) {};

\node[edgedot,label={[label distance=10pt,centered]170:$\ket{b}$}] at ($(v6)!0.5!(v1)$) {};
\node[edgedot,label={[label distance=10pt,text=black,centered]10:$\ket{a}$}] at ($(v1)!0.5!(c)$) {};
\node[edgedot,label={[label distance=6pt,text=black,centered]270:$\ket{c}$}] at ($(v6)!0.5!(c)$) {};

\pgfresetboundingbox 
\path[use as bounding box] (-2.25,-0.8) rectangle + (2.5,2.6);


\end{tikzpicture}}}\quad\Rightarrow\quad\vcenter{\hbox{\newcommand\scale{0.8}
\begin{tikzpicture}[scale=\scale,]
\tikzset{
    vertex/.style={circle, fill, inner sep=2pt},
    edgedot/.style={circle,draw=black, fill=white, inner sep=2pt},
    edge/.style={line width=1pt}
}

\coordinate (v1) at (162:2);
\coordinate (v6) at (198:2);
\coordinate (newlabel) at (140:2.5);

\coordinate (c) at (0,0);
\coordinate (newc) at (0,2);

\tikzfading[name=fade along, top color=transparent!100, bottom color=transparent!0]

\fill[fill=red!20] (v6)--(c)--(v1);

\foreach \x/\y/\a in {v6/v1/black,v6/newc/black,v1/newc/black,c/newc/black}{\draw[edge,endarrow,\a] (\y) -- (\x);
}

\foreach \x/\y/\z in {v1/0/dashed,v6/60/black}{
\draw[edge,endarrow,\z] (c) -- (\x);}

\node[vertex,fill=Plum,label={[label distance=8pt,text=Plum,font=\large,centered]330:$v_1$}] at (c) {};
\node[vertex,label={[label distance=8pt,font=\large,centered]150:$v_2$}] at (v1) {};
\node[vertex,label={[label distance=8pt,font=\large,centered]240:$v_3$}] at (v6) {};

\node[vertex,label={[label distance=8pt,font=\large,centered]45:$v_1'$}] at (newc) {};


\node[edgedot,label={[label distance=6pt,centered]180:$\ket{b}$}] at ($(v6)!0.5!(v1)$) {};
\node[edgedot,label={[label distance=10pt,text=black,centered]20:$\ket{a}$}] at ($(v1)!0.5!(c)$) {};
\node[edgedot,label={[label distance=8pt,text=black,centered]280:$\ket{c}$}] at ($(v6)!0.5!(c)$) {};

\node[edgedot] at ($(v6)!0.5!(newc)$) {};
\node[edgedot] at ($(v1)!0.5!(newc)$) {};
\node[edgedot] at ($(newc)!0.5!(c)$) {};


\pgfresetboundingbox 
\path[use as bounding box] (-2.25,-0.8) rectangle + (2.5,2.6);


\end{tikzpicture}}}\quad\Rightarrow\quad\vcenter{\hbox{\newcommand\scale{0.8}
\begin{tikzpicture}[scale=\scale,]
\tikzset{
    vertex/.style={circle, fill, inner sep=2pt},
    edgedot/.style={circle,draw=black, fill=white, inner sep=2pt},
    edge/.style={line width=1pt}
}

\coordinate (v1) at (162:2);
\coordinate (v6) at (198:2);
\coordinate (newlabel) at (140:2.5);

\coordinate (c) at (0,0);
\coordinate (newc) at (0,2);

\tikzfading[name=fade along, top color=transparent!100, bottom color=transparent!0]

\fill[fill=red!20] (v6)--(c)--(v1);

\foreach \x/\y/\a in {v6/v1/black,v6/newc/black,v1/newc/black,c/newc/black}{\draw[edge,endarrow,\a] (\y) -- (\x);
}

\foreach \x/\y/\z in {v1/0/dashed,v6/60/black}{
\draw[edge,endarrow,\z] (c) -- (\x);}

\node[vertex,fill=Plum,label={[label distance=8pt,text=Plum,font=\large,centered]330:$v_1$}] at (c) {};
\node[vertex,label={[label distance=8pt,font=\large,centered]150:$v_2$}] at (v1) {};
\node[vertex,label={[label distance=8pt,font=\large,centered]240:$v_3$}] at (v6) {};

\node[vertex,label={[label distance=8pt,font=\large,centered]45:$v_1'$}] at (newc) {};

\node[label={[label distance=6pt,centered]140:$\ket{gc}$}] at (newlabel) {};

\node[edgedot,label={[label distance=6pt,centered]180:$\ket{b}$}] at ($(v6)!0.5!(v1)$) {};
\node[edgedot,label={[label distance=10pt,text=black,centered]20:$\ket{a}$}] at ($(v1)!0.5!(c)$) {};
\node[edgedot,label={[label distance=8pt,text=black,centered]280:$\ket{c}$}] at ($(v6)!0.5!(c)$) {};

\node[edgedot] at ($(v6)!0.5!(newc)$) {};
\node[edgedot,label={[label distance=10pt,text=black,centered]100:$\ket{ga}$}] at ($(v1)!0.5!(newc)$) {};
\node[edgedot,label={[label distance=8pt,text=black,centered]0:$\ket{g}$}] at ($(newc)!0.5!(c)$) {};

\draw[] ($(newc)!0.5!(v6)$) -- (newlabel);

\pgfresetboundingbox 
\path[use as bounding box] (-2.25,-0.8) rectangle + (2.5,2.6);


\end{tikzpicture}}}\quad\Rightarrow\quad \omega|_f = \omega(g,a,b)$
    \caption{A depiction of the procedure used to evaluate the 3-cocycle on a triangle $f$. In order to evaluate the phase $\omega|_f$ associated to the face $f$ adjacent to the vertex $v$ at which $A_g^\omega(v)$ is being applied, we perform the following algorithm~\cite{HuEtAl_TwistedQuantumDouble_13}. \textit{Step 1.} Label the vertices of the face $f$ with labels $v_i$, such that the branching structure gives the ordering $v_1<v_2<v_3$. \textit{Step 2.} Introduce a new vertex $v_i'$ corresponding to the vertex $v_i$ at which $A_g^\omega(v)$ is being applied. (This is the purple vertex in the figure.) With four vertices, we now have a tetrahedron. The new edges should be oriented so that $v_i'$ appears immediately before $v_i$ in the ordering. \textit{Step 3.} Place the state $\ket{g}$ on the edge directed from $v_i'$ to $v_i$.
    Then assign group elements to the other new edges such that the new plaquettes satisfy the flatness condition $B(f)$. \textit{Step 4.} We now have four vertices with an ordering. The 3-cocycle $\omega|_f$ is evaluated at the edges connecting adjacent sites in the ordering. For instance, if $v=v_2$, the ordering is $v_1<v_2'<v_2<v_3$, and $\omega|_f=\omega(v_1\to v_2',v_2'\to v_2,v_2\to v_3)^{\pm 1}$. Finally, the exponent $\pm 1$ is determined by the orientation of the tetrahedron.}
    \label{fig:D_omega}
\end{figure*}

We here recall the algebraic relations between the terms appearing in Eq.~\ref{eq:can_TQD_ham}, as first discussed in Ref.~\cite{HuEtAl_TwistedQuantumDouble_13}. Each operator $A_g^\omega(v)$ commutes with every $B(f)$ and the projectors $B(f)$ commute among themselves. Furthermore, $A_g^\omega(v)$ and $A_h^\omega(v')$ for $v\neq v'$ commute on the flat subspace, where $B(f)=1$ for every face $f$. If we additionally take $G$ to be abelian, as in the $\mathcal{D}^\omega(\mathbb{Z}_p^n)$ case we primarily discuss, then $A_g^\omega(v)$ and $A_h^\omega(v)$ also commute on the flat subspace.

Notice that the off-diagonal stabilizers defined in Eq.~\ref{eq:TQD_Xs} can be decomposed into a nontrivial multi-site phase $\Lambda_g^\omega$ and an on-site action of $X_g$ on the sites in $\delta(v)$. In other words, we can write 

\begin{equation}
    A^\omega_g(v)=\left(\bigotimes_{e_i\in\delta(v)}X_{\pm g}^{(e_i)}\right)\Lambda_g^\omega,
\end{equation}
where $\Lambda_g^\omega$ is a multi-site diagonal unitary acting on the edges in $\delta(v)\cup \mathrm{Lk}(v)$. Examining the functional form of 3-cocycles over $\mathbb{Z}_p^n$, one can show that $\Lambda^\omega_g$ generically belongs to the Clifford hierarchy. This discussion will be made more precise in Sec.~\ref{sec:realizing_with}.

\section{Introduction to \texorpdfstring{\xkcd}{xkcd} stabilizer models and states}\label{sec:XkCd_intro}

In this section we introduce the $X{\cal{C}}^{(k)}_d$ stabilizer formalism.
The stabilizers we consider are built from products of qudit Pauli $X$ operators and diagonal operators drawn from a fixed level of the Clifford hierarchy. This construction is a natural generalization of the XS~\cite{NiEtAl_NonCommutingStabilizerFormalism_15} and XP~\cite{WebsterEtAl_XPStabiliserFormalism_22} stabilizer formalisms\footnote{XP codes, defined for qubits, are only \xkcd codes when the precision $N$ is a power of 2.}, which allow products of Pauli $X$ operators and on-site diagonal operators. Our construction expands on these works by considering prime-dimensional qudits in addition to qubits and allowing multi-site diagonal gates. (We consider prime-dimensional qudits because the diagonal elements of the Clifford hierarchy are well-understood in this case~\cite{CuiEtAl_DiagonalGatesClifford_17}.) At the same time, our stabilizers are more structured than those of the monomial stabilizer formalism~\cite{nest2011monomial}, which are products of a permutation of computational basis states and an arbitrary diagonal unitary. We begin with a discussion of the diagonal Clifford hierarchy, define the stabilizer models we consider, and conclude with some examples.

For qudits, the Clifford hierarchy \cite{GottesmanChuang_DemonstratingViabilityUniversal_99} is a familiar measure of how ``non-Clifford'' a gate is. If ${\cal P}_n$ is the $n$-qudit Pauli group then the first level of the Clifford hierarchy for $n$ qudits is defined as ${\cal C}^{(1)} = {\cal P}_n$ and the $k$th level for $n$ qudits is defined recursively as 
\begin{equation}
{\cal C}^{(k)} = \{U \in {\rm U}(d^n) \mid U{\cal P}_n U^\dagger \subseteq {\cal C}^{(k-1)}\},
\end{equation}
where $d$ is the qudit dimension. The usual Clifford gates that map Pauli operators to Pauli operators can be found at the second level of the Clifford hierarchy.

A characterization of ${\cal C}^{(k)}$ is not known in full generality, though progress has been made on simpler problems. For \textit{single} qudits a full description was recently given in Ref.~\cite{deSilvaLautsch_CliffordHierarchyOne_25} if the qudit dimension is prime.
Unfortunately, ${\cal C}^{(k)}$ is a group if and only if $k=1$ or $2$ (see App.~\ref{app:useful_results}). To avoid this complication, Ref.~\cite{CuiEtAl_DiagonalGatesClifford_17} focused on the subset of operators in ${\cal C}^{(k)}$ that are \textit{diagonal} in the computational basis, denoted ${\cal C}^{(k)}_d$ with the $d$ subscript signifying diagonal. This subset forms a group under matrix multiplication for any $k$. More specifically it was shown in Ref.~\cite{CuiEtAl_DiagonalGatesClifford_17} that in the $n$-qudit case
\begin{equation}
{\cal C}^{(k),n}_d = \left\langle U_{m,\mathbf{a}} \mid (p-1)(m-1)+{\sum_i} a_i\leq k\right\rangle,
\end{equation}
where
\begin{equation}\label{eq:U-m-a}
U_{m,\mathbf{a}} = \sum_{\mathbf{x}\in \bb{Z}_p^n}\exp\left(\frac{2\pi i}{p^m}x_1^{a_1}\cdots x_n^{a_n}\right)\ketbra{\mathbf{x}}{\mathbf{x}},
\end{equation}
with $m \in \bb{Z}_{>0}$ and $0\leq a_i\leq p-1$ for all $1\leq i\leq n$, and where the prime $p$ is the qudit dimension. 

In this work we use slightly different notation to that used in Ref.~\cite{CuiEtAl_DiagonalGatesClifford_17}:
\begin{definition}\label{def:cliff_hier_def}
For ease of notation when discussing diagonal gates, we let
\begin{equation}\label{def:Dmf}
D_{m}[f] = \sum_{\mathbf{x}\in \mathbb{Z}_p^n} \exp\left(\frac{2\pi i}{p^m}f(\mathbf{x})\right)\ketbra{\mathbf{x}}{\mathbf{x}},
\end{equation}
where $f\colon \mathbb{Z}_p^n\rightarrow \mathbb{Z}_{p^m}$.
\end{definition}
\noindent From Ref.~\cite{CuiEtAl_DiagonalGatesClifford_17}, we have that
\begin{equation}\label{Dmf_level}
   D_m[f]\in\mathcal{C}_d^{(k)},\quad k=(m-1)(p-1)+\mathrm{deg}(f),
\end{equation}
where $\mathrm{deg}(f)$ is the polynomial degree of $f$, which we define below. 

The phase function $f$ takes in a computational basis state $\mathbf{x}$ and returns a value in $\mathbb{Z}_{p^m}$, which specifies a $p^m$-th root of unity. In order to place $D_m[f]$ in the Clifford hierarchy using Eq.~\ref{Dmf_level}, it is necessary to unambiguously define the degree of an arbitrary function $f:\mathbb{Z}_p^n\to \mathbb{Z}_{p^m}$. 
This is made possible by the fact that every function $f\colon \mathbb{Z}_{p}^n \rightarrow \mathbb{Z}_{p^m}$ can be uniquely expressed in a basis of monomials $x^{k_1}\cdots x^{k_n}$ of total degree less than or equal to $ n(p-1)$~\cite{CuiEtAl_DiagonalGatesClifford_17}:
\begin{equation}\label{eq:poly_decomp}
f(\mathbf{x}) = \sum_{\mathbf{k}\in \mathbb{Z}_p^n}a_{\mathbf{k}}x_1^{k_1}\cdots x_n^{k_n}.
\end{equation}
Each monomial encodes an independent $||\mathbf{k}||_0$-body gate, where the $\ell_0$ norm gives the number of nonzero entries in $\mathbf{k}$. For instance, if we take $p=2$, then $D_1[x_i]=Z_i$ and $D_1[x_ix_jx_k]=CCZ_{i,j,k}$.
A polynomial, being a sum over monomials as in Eq.~\ref{eq:poly_decomp}, encodes the composition of multiple independent gates.

Because the function $f\colon \mathbb{Z}_{p}^n \rightarrow \mathbb{Z}_{p^m}$ admits the unique polynomial decomposition given in Eq.~\ref{eq:poly_decomp}, there exists a well-defined notion of the degree of $f$. For a function $f$ with monomial coefficients $a_{\mathbf{k}}$,
we define
\begin{equation}
\mathrm{deg}(f) = \mathrm{max}\left\{\sum_{i=1}^{n}k_i \mid \mathbf{k}\in \mathbb{Z}_p^n \ \mathrm{s.t.} \ a_{\mathbf{k}}\neq 0\right\}.
\end{equation}
This coincides with the way one would normally define the degree of a multivariate polynomial over $\mathbb{R}^n$, for instance.

To facilitate the comparison of gates with different ``precision'' $m$ we often discuss these gates in a unified way.

\begin{definition}\label{defn:Df}
Let $f\colon \mathbb{Z}_p^n \rightarrow \mathbb{R}/\mathbb{Z}$ and define
\begin{equation}
D[f] = \sum_{\mathbf{x}\in \mathbb{Z}_p^n}\exp\left(2\pi i f(\mathbf{x})\right)|\mathbf{x}\rangle\langle \mathbf{x}|.
\end{equation}
\end{definition}
\noindent The drawback of this definition is that without information about the degree of the function it may not be immediately obvious how to compute the level of the Clifford hierarchy. However, in App.~\ref{app:useful_results} we see that the finite difference operator always lowers the level of the Clifford hierarchy, and thus we can measure the level of the Clifford hierarchy by how many applications of the finite difference operator it takes to trivialize $f$.

\subsection{The \texorpdfstring{$n$}{n}-qudit Clifford hierarchy}

In this section, we briefly discuss the key definitions of the $n$-qudit Clifford hierarchy. Level $k$ of the $n$-qudit Clifford hierarchy, denoted ${\cal C}^{(k),n}$, is the subset of ${\cal C}^{(k)}$ consisting of operators which have support on $n$ qudits or fewer. By definition, we see that ${\cal C}^{(k),n}\subset {\cal C}^{(k),n+1}$. To further specify the set of level-$k$ Clifford operators with support on \textit{exactly} $n$ qudits, we define ${\cal S}^{k,\Delta n} = {\cal C}^{(k),n}\setminus {\cal C}^{(k),n-1}$. Similarly, the Clifford hierarchy operators which act on $n$ or fewer qudits and are \textit{precisely} level-$k$ are denoted by ${\cal S}^{\Delta k,n} = {\cal C}^{(k),n}\setminus {\cal C}^{(k-1),n}$.

In this work, we focus on the diagonal subgroups of the Clifford hierarchy. We define ${\cal C}^{(k),n}_d\subset {\cal C}^{(k)}_d$ to be the subgroup of level-$k$ diagonal Clifford operators which have support on $n$ qudits or fewer. The set ${\cal S}^{k,\Delta n}_d$ is defined to be ${\cal C}^{(k),n}_d\setminus {\cal C}^{(k),n-1}_d$, i.e.~the set of diagonal Clifford-hierarchy operators at level $k$ or lower which have support on exactly $n$ qudits.

We note that the set ${\cal C}^{(k),n}_d$ forms a group under matrix multiplication. This is because composing elements of ${\cal C}^{(k),n}_d$ can make them act on \textit{fewer} qudits, but not more. However,  ${\cal S}^{k,\Delta n}_d$ is not a group for $n\geq 1$ since it does not contain the identity.

\subsection{Stabilizer setup}

We first define the group of potential stabilizers that we consider. We define $G^{(k),n}$ to be the set generated by products of strings of qudit Pauli-$X$ operators and unitaries drawn from the group $\mathcal{C}^{(k),n}_d$, for some fixed $k$. Our stabilizers are therefore comprised of on-site $X$ operators and multi-qudit diagonal gates. In App.~\ref{app:on_site_clifford} we discuss the case where only on-site stabilizers are allowed. It is not hard to see that every element $g\in G^{(k),n}$ can be written in the form
\begin{equation}\label{eq:main_stabilizer}
g = X(\mathbf{v})D\left[f\right], \ f\colon \mathbb{Z}_p^n\rightarrow \mathbb{R}/\mathbb{Z}, \mathbf{v}\in \mathbb{Z}_p^n,
\end{equation}
where $X(\mathbf{v})=\bigotimes _{i=1}^n X_i^{v_i}$ and $D[f]$ is a diagonal $n$-qudit Clifford operator at level $k$. Indeed, commuting any $X(\mathbf{v})$ past a diagonal operator $D[f]$ results in a constant shift in the argument of $f$, which does not change the level at which $D[f]$ sits in the Clifford hierarchy.
This allows an arbitrary product of generators to be brought into the canonical form given in Eq.~\ref{eq:main_stabilizer} and endows $G^{(k),n}$ with the structure of a group. Furthermore, one may immediately verify that $G^{(k),n}\subset {\cal C}^{(k),n}$.

Next, we define stabilizer models constructed from $G^{(k),n}$.
Let $H\subseteq G^{(k),n}$ be a subgroup. If there exists some $n$-qudit state such that $h|\psi\rangle = |\psi\rangle$ for every $h\in H$, then we call $H$ an $n$-qudit \xkcd stabilizer group. The space of all states stabilized by $H$ is denoted by ${\cal L}_{H}$ and referred to as the $n$-qudit \xkcd stabilizer code associated with $H$. If there is only one state in ${\cal L}_{H}$ then we call this an $n$-qudit \xkcd stabilizer state.

We motivated the \xkcd stabilizer formalism as a way to classify stabilizer Hamiltonians with ${\cal D}^\omega(\mathbb{Z}_p^n)$ topological order. On a torus, the number of ground states is given by the number of anyons in ${\cal D}^\omega(\mathbb{Z}_p^n)$, but if we instead consider the Hamiltonian on a sphere there is a single ground state wavefunction \cite{BarkeshliEtAl_SymmetryFractionalizationDefects_19}. 
First, we restrict our attention to the sphere and consider the case of a unique stabilizer state for simplicity of presentation. This motivates the following definition of the sets of possible stabilizer states:
\begin{definition}\label{defn:stab_sets}
We denote by $\Psi^{(k),n}_G$ the set of states that can be realized as the unique codeword of an \xkcd stabilizer code on $n$ qudits.
\end{definition}

We now give an example of states that require stabilizers at level $k$.

\subsection{Example states that require level \texorpdfstring{$k$}{k}}

A natural class of states to study are those that \textit{require} stabilizers at level $k+1$ of the Clifford hierarchy. These are precisely the states that belong to 
$\Psi^{(k+1),n}_{G}\setminus \Psi^{(k),n}_{G}$. We can use a simple procedure to construct a state in this set.

\begin{example}\label{ex:k_wavefct}
We claim that
\begin{equation}
|\psi\rangle = \frac{1}{p^{n/2}} \sum_{\mathbf{x}\in \mathbb{Z}_p^n} e^{2\pi i \phi(\mathbf{x})}|\mathbf{x}\rangle,
\end{equation}
where $\phi \colon \mathbb{Z}_{p}^n\rightarrow \mathbb{R}/\mathbb{Z}$ is such that $D[\phi]\in {\cal C}^{(k+1),n}_d$, lies in $\Psi^{(k),n}_G$.

This is easiest to see if we begin with the stabilizer group $H_0 = \langle X(\mathbf{e}_i)\rangle_{i=1,\ldots, n}$, where $\mathbf{e}_i$ is the unit vector in the $i$th direction. Clearly all the generators commute and have order $p$ so $H_0\simeq \mathbb{Z}_p^n$. It is easy to see that they uniquely stabilize the state
\begin{equation}
|\psi_0\rangle = \frac{1}{p^{n/2}}\sum_{\mathbf{x}\in \mathbb{Z}_p^n}|\mathbf{x}\rangle.
\end{equation}
We now define the group $H = \langle D[\phi]X(\mathbf{e}_i)D[\phi]^\dagger\rangle$. Clearly this group is isomorphic to $H_0$. The group $H$ uniquely stabilizes the state
\begin{equation}
|\psi\rangle = D[\phi]|\psi_0\rangle = \frac{1}{p^{n/2}}\sum_{\mathbf{x}\in \mathbb{Z}_p^n}e^{2\pi i \phi(\mathbf{x})}|\mathbf{x}\rangle.
\end{equation}

We note that $D[\phi]X(\mathbf{e}_i)D[\phi]^\dagger = X(\mathbf{e}_i)D[\Delta_{\mathbf{e}_i}\phi(\mathbf{x})]$, where $\Delta_{\mathbf{e}_i}\phi(\mathbf{x}) = \phi(\mathbf{x}+\mathbf{e}_i)-\phi(\mathbf{x})$. Then by Corr.~\ref{cor:diff_lowers} we know that $D[\Delta_{\mathbf{e}_i}\phi(\mathbf{x})]\in {\cal C}^{(k),n}_d$ for every $\mathbf{e}_i$. Thus this stabilizer lies in $G^{(k),n}$ if $\phi$ is a function such that $D[\phi]\in {\cal C}^{(k+1),n}_d$.
\end{example}

This example suggests that if the phase in the wavefunction is level $k+2$, then the stabilizers cannot be at level $k$. We prove this for a specific choice below.

\begin{lemma}\label{lem:state_req_stabilizers}
The state 
\begin{equation}
|\psi\rangle = \frac{1}{p^{n/2}} \sum_{\mathbf{x}\in \mathbb{Z}_p^n} e^{2\pi i x_0^a/p^m}|\mathbf{x}\rangle
\end{equation}
lies in $\Psi^{(k+1),n}_G\setminus \Psi^{(k),n}_G$ for $m\in\mathbb{Z}_{>0}$ and $a\in\{1,\ldots,p-1\}$ chosen so that $(p-1)(m-1)+a = k+2$.
\end{lemma}
\begin{proof} 
We prove this in App.~\ref{ap:state_req_stabilizers}.
\end{proof}

From this example we can guess that states which \textit{require} stabilizers at level $k$ are precisely those that have ``phases'' in their wavefunction at level $k+1$. In the next section we prove that this is indeed the case.

\section{Classification of \texorpdfstring{\xkcd}{xkcd} stabilizer states}\label{sec:XkCd_class}
 
The objective of this section is to understand the structural form of \xkcd stabilizer states. This effort culminates in the proof of Theorem \ref{thm:unique_state} at the end of Sec.~\ref{sec:class_theory}, which states that a given wavefunction is an \xkcd stabilizer state if and only if it can be brought into the form presented in Eq.~\ref{eq:state_form}. This result is valuable for state classification, as it can be used to determine if a given wavefunction is an \xkcd stabilizer state, and if so, the minimal value of $k$ such that it is. Furthermore, we present in Sec.~\ref{sec:class_alg} an algorithm which takes in an \xkcd stabilizer subgroup with unique stabilized state and returns the stabilizer state in the canonical form of Eq.~\ref{eq:state_form}.

\subsection{Theory}\label{sec:class_theory}
The main result of this section, Theorem \ref{thm:unique_state}, generalizes the well-known result~\cite{DehaeneDeMoor_CliffordGroupStabilizer_03} (see also Ref.~\cite{VanDenNes_ClassicalSimulationQuantum_10}) that every Pauli stabilizer state $\ket{\psi}$ may be expressed as
\begin{equation}
    \ket{\psi}\propto \sum_{\mathbf{v}\in A}i^{l(\mathbf{v})}(-1)^{q(\mathbf{v})}\ket{\mathbf{v}},
\end{equation}
where $A=\{ M\mathbf{x}+\mathbf{c}\mid\mathbf{x}\in \mathbb{Z}_2^m\}\subseteq{\mathbb{Z}_2^N}$ is an \textit{affine linear subspace} and $l(\mathbf{v})$ and $q(\mathbf{v})$ are linear and quadratic functions on $\mathbb{Z}_2^n$, respectively. As we will see, the result we present in Eq.~\ref{eq:state_form} takes quite an analogous form. Throughout this section, we will prove intermediate results which are necessary in order to prove Theorem \ref{thm:unique_state}.

We begin by stating a commonly used result on the structure of stabilizer states for any stabilizer group.
\begin{lemma}\label{lem:group_comp_wavefct}
Suppose that $|\psi\rangle \in \Psi^{(k),n}_{G}$ with a stabilizer subgroup $H\subseteq G^{(k),n}$. Then 
\begin{equation}\label{eq:orbit_psi}
|\psi\rangle \propto \sum_{h\in H}h|\mathbf{x}\rangle,
\end{equation}
for any $\mathbf{x}\in \mathbb{Z}_p^n$ such that 
\begin{equation}\label{eq:x_nonzero}
    \sum_{h\in H}h|\mathbf{x}\rangle \neq 0
\end{equation}
\end{lemma}
\begin{proof}
It can be seen directly that $h'|\psi\rangle = |\psi\rangle$ for any $h'\in H$ by relabeling group elements in the sum.
\end{proof}
\noindent This lemma is quite broad and does not yet tell us very much about the structure of \xkcd stabilizer states. However, we are not dealing with an arbitrary stabilizer group, but subgroups $H$ of the \xkcd operator group $G^{(k),n}$. These groups are highly structured, and we will use this structure to impose further constraints on \xkcd stabilizer states.

In the discussion to follow, the \textit{diagonal subgroup} of $H$ -- denoted $D_H$ and comprised of operators diagonal in the computational basis -- will play an important role. We will now precisely define this subgroup and prove that it is normal in $H$.
The diagonal subgroup $D_H\subseteq H\subseteq G^{(k),n}$ is defined as
\begin{equation}
    D_H := H\cap {\cal C}^{(k),n}_d.
\end{equation}
Since an arbitrary \xkcd stabilizer is of the form $X(\mathbf{v})D[f]$, $D_H$ contains all the stabilizers where $\mathbf{v}=\mathbf{0}$. Furthermore, because $D_H$ is comprised of all diagonal matrices in $H$, it is clear that $D_H$ is an abelian subgroup of $H$. Let us now show that $D_H$ is additionally a normal subgroup of $H$, i.e., $hD_H h^{-1}=D_H$ for any $h\in H$. An arbitrary element of $H$ is of the form $h=X(\mathbf{v})D[g]$ and an arbitrary element of $D_H$ is of the form $d=D[f]$, so we have
\begin{equation}\label{eq:diag_norm}
    \begin{split}
        h d h^{-1}\ket{\mathbf{x}}&=(X(\mathbf{v})D[g])D[f](D[-g]X(\mathbf{-v}))\ket{\mathbf{x}},\\
        &=D[f(\mathbf{x}-\mathbf{v})]\ket{\mathbf{x}},
    \end{split}
\end{equation}
for any $\mathbf{x}\in\mathbb{Z}_p^n$. Since $d,h\in H$, we know that $h d h^{-1}\in H$ due to closure of $H$ under multiplication. Furthermore, we see from Eq.~\ref{eq:diag_norm} that $h d h^{-1}$ is diagonal, so we conclude that $h d h^{-1}\in D_H$ for any $h\in H$ and $d\in D_H$. This is precisely the statement that $D_H$ is normal in $H$.

We can further partition $H$ into cosets $h D_H\in H/D_H$, such that every element $h\in H$ belongs to exactly one coset. Let us now consider one of these cosets. If we let $h=X(\mathbf{v})D[g]$, then 
\begin{equation}
    hD_H=\{X(\mathbf{v})D[g+f]\mid D[f]\in D_H\}.
\end{equation}
It will be useful to choose a representative from each coset, such that every element of $H$ may be expressed as a product of a coset representative and a diagonal operator. For every coset $h D_H\in H/D_H$, choose a distinguished element of the coset and call it $r$. The set of all such $r$ is denoted $R_{H/D_H}\subset H$ and called the set of coset representatives, or, more technically, a transversal of $D_H$ in $H$. By a slight abuse of notation, we will write $r\in {H/D_H}$ when we mean $r\in R_{H/D_H}$, and similarly for other quotient groups. Finally, we may uniquely express any element $h\in H$ as $h=r d$ for some coset representative $r\in {H/D_H}$ and some $d\in D_H$. 
\begin{remark}\label{rem:orbit}
Not only can every element $h\in H$ be uniquely written as $h=rd$, but each $r$ may be uniquely identified with some $\mathbf{v}\in \mathbb{Z}_p^n$. In other words, two elements of $H$, call them $X(\mathbf{v})D[f]$ and $X(\mathbf{w})D[g]$, belong to the same coset if and only if $\mathbf{v}=\mathbf{w}$. It is clear that they may only be in the same coset if $\mathbf{v}=\mathbf{w}$, as multiplication by diagonal matrices cannot change the argument of $X(\mathbf{v})$. To see that they \textit{must} be in the same coset, notice that if $h=X(\mathbf{v})D[f]$ and $h'=X(\mathbf{v})D[g]$ are both in $H$, then 
\begin{equation}
    h^{-1}h'=(X(\mathbf{v})D[f])^\dagger (X(\mathbf{v})D[g])=D[g-f]\in D_H,
\end{equation}
so 
\begin{equation}
    (X(\mathbf{v})D[f])(D[g-f])=h'\in h D_H.
\end{equation}
\end{remark}

With the coset machinery in place, we are now able to prove a strong constraint on the $\mathbf{x}\in \mathbb{Z}_p^n$ which satisfy 
$\sum_{h\in H}h|\mathbf{x}\rangle \neq 0$ for a given $H\subseteq G^{(k),n}$, as in Lemma~\ref{lem:group_comp_wavefct}. The following lemma will have significant consequences for the structure of \xkcd stabilizer states:
\begin{lemma}\label{lem:wavefct_construction}
Suppose we have an \xkcd stabilizer subgroup 
$H\subseteq G^{(k),n}$ with unique stabilized state. Then a fixed $\mathbf{x}\in\mathbb{Z}_p^n$ will satisfy $\sum_{h\in H}h|\mathbf{x}\rangle \neq 0$ if and only if
\begin{equation}
    f(\mathbf{x}) = 0\!\!\!\!\pmod{1}\;\forall f\;\text{such that}\;D[f]\in D_H.
\end{equation}

\end{lemma}
\begin{proof}
We will first prove the direction assuming that $\mathbf{x}$ is chosen to satisfy $\sum_{h\in H}h|\mathbf{x}\rangle \neq 0$. As we saw above, for every $h\in H$ there is a unique $r\in H/D_H$ and $d\in D_H$ such that $h=rd$. This implies that we may separate the sum $\sum_{h\in H}h|\mathbf{x}\rangle$ as follows:
\begin{align}
\sum_{h\in H}h|\mathbf{x}\rangle=\sum_{r\in H/D_H}\sum_{d\in D_H}rd|\mathbf{x}\rangle.
\end{align}
Now, choose any diagonal operator $D[f]\in D_H$. Because $D[f]\in \mathcal{C}^{(k)}_d$, it may be written as $D_m[f p^m]$ for some integer $m$. Notice that the subgroup $\braket{D_m[f p^m]}\subseteq D_H$ generated by $D_m[f p^m]$ is normal in $D_H$, so we may further quotient $D_H$ by $\braket{D_m[f p^m]}$. We may then write
\begin{align}
\sum_{h\in H}h|\mathbf{x}\rangle= \sum_{r\in H/D_H}\sum_{d\in D_H/\braket{D[f]}}rd \left(\sum_{k=0}^{p^m-1}D_m[k p^m f(\mathbf{x})]\right)|\mathbf{x}\rangle,
\label{eq:lemma_III_5_wide}
\end{align}
The term in parentheses is equal to zero if $p^m f (\mathbf{x})\neq 0\pmod{p^m}$, otherwise it is equal to $p^m$. Since we have taken the left hand side to be nonzero by assumption, then it must be the case that $p^m f (\mathbf{x}) = 0\pmod{p^m}$, which implies that $f(\mathbf{x}) = 0\pmod{1}$. Further since $D[f]\in D_H$ was arbitrary, we conclude that it must be the case that if $\sum_{h\in H}h|\mathbf{x}\rangle \neq 0$, then $f(\mathbf{x}) = 0\pmod{1}$ for \textit{every} $D[f]\in D_H$. 

Now we prove the reverse direction. Suppose that we fix $\mathbf{x}$ such that $f(\mathbf{x})=0\pmod{1}$ for all $D[f]\in D_H$. Then, as above, $d|\mathbf{x}\rangle = |\mathbf{x}\rangle$ for all $d\in D_H$. In particular this means that
\begin{equation}
\sum_{h\in H}h|\mathbf{x}\rangle
=|D_H|\sum_{r\in H/D_H}r|\mathbf{x}\rangle.
\end{equation}
Next, we take the inner product of both sides with $\ket{\mathbf{x}}$, obtaining
\begin{equation}\label{eq:sum_nonzero}
\begin{split}
\bra{\mathbf{x}}\left(\sum_{h\in H}h|\mathbf{x}\rangle\right)
&=|D_H|\sum_{r\in H/D_H}\bra{\mathbf{x}}r|\mathbf{x}\rangle=|D_H|,\\
\end{split}
\end{equation}
where the second equality is due to the fact presented in Rem.~\ref{rem:orbit} that there is a unique correspondence between $r$ and $\mathbf{v}\in \mathbb{Z}_p^n$ such that $r = X(\mathbf{v})D[h]$ for some $D[h]$. In particular there is only one $r$ such that $\langle\mathbf{x}|r|\mathbf{x}\rangle\neq 0$, and this is $r=I$. Eq.~\ref{eq:sum_nonzero} tells us that   $\sum_{h\in H}h|\mathbf{x}\rangle$ has a nonzero component and is therefore not the zero vector.
\end{proof}

From Lemma~\ref{lem:wavefct_construction}, it immediately follows that \xkcd stabilizer states can be constructed as orbits under the coset representatives $r\in H/D_H$ alone, rather than the full $H$:
\begin{corollary}
Suppose that $|\psi\rangle\in \Psi^{(k),n}_G$ with stabilizer subgroup $H\subseteq G^{(k),n}$ and fix $\mathbf{x}\in \mathbb{Z}_p^n$ such that $\sum_{h\in H}h|\mathbf{x}\rangle \neq 0$. Then by Lemma~\ref{lem:wavefct_construction}, we have
\begin{equation}\label{eq:coset_wvfnct}
    |\psi\rangle = \sqrt{\frac{|D_H|}{|H|}}\sum_{r\in {H/D_H}}r|\mathbf{x}\rangle.
\end{equation}
\end{corollary}
\begin{proof}
    Using our coset construction, we may rewrite Eq.~\ref{eq:orbit_psi} as
    \begin{equation}
        |\psi\rangle \propto \sum_{r\in {H/D_H}}\sum_{d\in {D_H}}rd|\mathbf{x}\rangle
    \end{equation}
    for an \xkcd stabilizer subgroup $H\subseteq G^{(k),n}$. Then by Lem.~\ref{lem:wavefct_construction}, $\ket{\mathbf{x}}$ is a +1 eigenstate of each $d\in D_H$, so we have
    \begin{equation}
        |\psi\rangle \propto \sum_{r\in {H/D_H}}r|\mathbf{x}\rangle.
    \end{equation}
    Finally, we fix normalization $\ket{\psi}\mapsto \mathcal{N}^{-1/2}\ket{\psi}$ by observing that 
    \begin{equation}
    \begin{split}
        \mathcal{N}&=\sum_{r,s\in {H/D_H}}\bra{\mathbf{x}}s^{-1}r|\mathbf{x}\rangle,\\
        &=\sum_{r,s\in {H/D_H}}\delta_{r,s}
=|H|/|D_H|,        
    \end{split}
    \end{equation}
    where the fact that $\bra{\mathbf{x}}s^{-1}r|\mathbf{x}\rangle=\delta_{r,s}$ is true for $r,s\in R_{H/D_H}$ and follows directly from Rem.~\ref{rem:orbit}.
\end{proof}

The simplified form of the wavefunction presented in Eq.~\ref{eq:coset_wvfnct} has an immediate implication for the computational basis states on which the wavefunction can have support. 
We define the \textit{support} $V_\psi$ of $\ket{\psi}\in \Psi^{(k),n}$ to be the subset of $\mathbb{Z}_p^n$ where the wavefunction is nonzero, i.e.
\begin{equation}
V_\psi = \{\mathbf{x}\in \mathbb{Z}_p^n \mid \langle \mathbf{x}|\psi\rangle \neq 0\}\subseteq \mathbb{Z}_p^n.
\end{equation}
Because $I\in H/D_H$, Eq.~\ref{eq:coset_wvfnct} tells us that every $\mathbf{x}$ such that $\sum_{h\in H}h|\mathbf{x}\rangle \neq 0$ is in $V_\psi$. It is not obvious, however, that no other computational basis states may appear in $V_\psi$, but this will indeed be the case. As we will prove shortly, it turns out that $V_\psi$ is precisely the set of computational basis states satisfying $\sum_{h\in H}h|\mathbf{x}\rangle \neq 0$. 

\begin{lemma}\label{lem:Vpsi_descrip}
Suppose that $|\psi\rangle\in \Psi^{(k),n}_G$ with stabilizer subgroup $H\subseteq G^{(k),n}$. 
An element $\mathbf{x}\in V_\psi$ if and only if $\sum_{h\in H}h|\mathbf{x}\rangle \neq 0$. In particular, 
this means that
\begin{equation}
V_\psi = \{\mathbf{x}\in \mathbb{Z}_p^n\mid f(\mathbf{x}) = 0\!\!\!\! \pmod{1}\;\forall D[f]\in D_H\},
\end{equation}
and
\begin{equation}
|\psi\rangle = \sqrt{\frac{|D_H|}{|H|}}\sum_{r\in H/D_H}r|\mathbf{x}\rangle
\end{equation}
for any choice of $\mathbf{x}\in V_\psi$.
\end{lemma}
\begin{proof}
We prove this in App.~\ref{ap:Vpsi_descrip}.
\end{proof}

This lemma tells us that \xkcd codes are naturally endowed with a ``flat'' subspace $V_\psi$ which is the mutual $+1$ eigenspace of all the diagonal stabilizers. This structure generalizes the flatness constraint from lattice gauge theory, which requires all plaquette holonomies to be trivial. Just as local Gauss law operators map flat states to one another in a lattice gauge theory, here the off-diagonal stabilizers preserve $V_\psi$, mapping ``flat'' configurations to ``flat'' configurations

Thus far, we have been leveraging properties of the stabilizer subgroup $H\subseteq G^{(k),n}$ to prove statements about \xkcd stabilizer states $\ket{\psi}\in\Psi^{(k),n}$. However, the functional form of the wavefunction presented in Eq.~\ref{eq:coset_wvfnct} also imposes a consistency condition on $H$ (if $H$ is to admit a nonvanishing mutual +1 eigenspace):
all of the elements $X(\mathbf{v})D[f]\in H$ must be such that $\mathbf{v}$ is the difference of two elements of $V_\psi$.

\begin{lemma}\label{lem:Vpsi_stabsub}
Suppose that $|\psi\rangle\in \Psi^{(k),n}_G$ with stabilizer subgroup $H\subseteq G^{(k),n}$. Let $\mathbf{x}_0 \in V_\psi$. Then $\mathbf{v}\in V_\psi$ if and only if $X(\mathbf{v}-\mathbf{x}_0)D[f]\in H$ for some $D[f]$. In particular this means that if $\mathbf{x}_0\in V_\psi$ and $h\in H$ then there exists a unique $\mathbf{v}\in V_\psi$ and $D[f]$ such that $h = X(\mathbf{v}-\mathbf{x}_0)D[f]$.
\end{lemma}
\begin{proof}
Let $h = X(\mathbf{v}-\mathbf{x}_0)D[f]$ be an arbitrary element in $H$. Then we know from Rem.~\ref{rem:orbit} that there is some unique $r\in H/D_H$ such that $r = X(\mathbf{v}-\mathbf{x}_0)D[g]$ for some $g$. But then from Lem.~\ref{lem:Vpsi_descrip} we see that
\begin{align}
\langle \mathbf{v}|\psi\rangle &= \sqrt{\frac{|D_H|}{|H|}}\sum_{r\in H/D_H}\langle \mathbf{v}|r|\mathbf{x}_0\rangle \nonumber \\
&= \sqrt{\frac{|D_H|}{|H|}}D[g(\mathbf{x}_0)]\neq 0.
\end{align}
So then $\mathbf{v}\in V_\psi$.

On the other hand suppose that $\mathbf{v}\in V_\psi$. Then we have that 
\begin{equation}
0 \neq \langle \mathbf{v}|\psi\rangle = \sqrt{\frac{|D_H|}{|H|}}\sum_{r\in H/D_H}\langle \mathbf{v}|r|\mathbf{x}_0\rangle.
\end{equation}
From Rem.~\ref{rem:orbit} we know that for each $r\in H/D_H$ we have $r = X(\mathbf{v}')D[f]$ where each $r$ is labeled by a unique $\mathbf{v}'$. So then by the above we conclude that there must be some $\mathbf{v}' = \mathbf{v} - \mathbf{x}_0$ such that $X(\mathbf{v}')D[f] \in H$.
\end{proof}

The result above also has the immediate consequence that the support is an affine linear subspace.

\begin{corollary}\label{corr:aff_lin_sub}
Suppose that $|\psi\rangle\in\Psi^{(k),n}_G$. Then $V_\psi\subseteq \mathbb{Z}_p^n$ is an affine linear subspace of $\mathbb{Z}_p^n$. In particular if $V_\psi$ has dimension $\ell$ then there exists an $(n-\ell) \times \ell$ matrix $M$, with entries in $\mathbb{Z}_p$, and a vector $\mathbf{c}\in \mathbb{Z}_p^{n-\ell}$ such that
\begin{equation}\label{eq:V_psi}
V_\psi = \{(\mathbf{x}, M\mathbf{x}+\mathbf{c}) \mid \mathbf{x}\in \mathbb{Z}_p^\ell \}
\end{equation}
up to a permutation. Note that the notation $(\mathbf{a},\mathbf{b})$ indicates the vertical concatenation of column vectors, i.e. $(\mathbf{a},\mathbf{b}):=\begin{pmatrix}
    \mathbf{a}\\\mathbf{b}
\end{pmatrix}$.
\end{corollary}
\begin{proof}
Let $\mathbf{x}_0\in V_\psi$ be a fixed point. Next, choose arbitrary $\mathbf{v},\mathbf{v}'\in V_\psi$. From Lem.~\ref{lem:Vpsi_stabsub} we see this means that there exists some elements of the form $X(\mathbf{v}-\mathbf{x}_0)D[f], X(\mathbf{v}'-\mathbf{x}_0)D[g]\in H$. This implies that $X(\mathbf{v}+\mathbf{v}'-2\mathbf{x}_0)D[h]\in H$ since $H$ is a group. By applying Lem.~\ref{lem:Vpsi_stabsub}, we see that $\mathbf{v}+\mathbf{v}'-\mathbf{x}_0\in V_\psi$. This means that
\begin{equation}
    (\mathbf{v}-\mathbf{x}_0)+(\mathbf{v}'-\mathbf{x}_0)=(\mathbf{v}+\mathbf{v}'-\mathbf{x}_0)-\mathbf{x}_0\in V_\psi-\mathbf{x}_0.
\end{equation}
In other words, $V_\psi - \mathbf{x}_0$ is closed under addition and by repeated addition under scalar multiplication.
This allows us to conclude that $V_\psi - \mathbf{x}_0$ is a linear subspace, which means that $V_\psi$ is an affine linear subspace.

We can now use this fact to write $V_\psi$ in a simplified form. If $V_\psi$ has dimension $\ell$, then $W = V_\psi -\mathbf x_0$ is a linear subspace of $\mathbb Z_p^n$ of dimension $\ell$ and we choose a basis, $w_1, \dots, w_\ell$, for $W$. As the basis vectors are linearly independent, we can always choose $\ell$ indices that determine an invertible $\ell \times \ell$ matrix. We can then permute the indices so these $\ell$ coordinates come first.
Using this coordinate choice, we can freely choose the first $\ell$ components of an element of $W$ as $\mathbf x \in \mathbb Z_p^\ell$, and the remaining $n-\ell$ components are forced to be linearly dependent on $\mathbf x$, or specifically a linear function of $\mathbf x$ since $W$ is $\ell$-dimensional.
Specifically, there exists an $(n-\ell) \times \ell$ matrix $M$ over $\mathbb Z_p$ such that $W = \{(\mathbf x, M\mathbf x) \mid \mathbf x \in \mathbb Z_p^\ell \}$. Since $V_\psi = \mathbf x_0 + W$, we can write $\mathbf x_0 = (\mathbf x_0^{(1)}, \mathbf x_0^{(2)})$ in the same block decomposition to see that $V_\psi = \{(\mathbf x_0^{(1)}+\mathbf x, \mathbf x_0^{(2)} + M\mathbf x) \mid \mathbf x \in \mathbb Z_p^\ell \}$. Here $\mathbf{x}_0^{(1)}$ has dimension $\ell$ and $\mathbf{x}_0^{(2)}$ has dimension $n-\ell$.
Without loss of generality, we can relabel $\mathbf x$ by shifting it by $\mathbf{x}_0^{(1)}$ and define $\mathbf c =  \mathbf x_0^{(2)} - M\mathbf{x}_0^{(1)} \in \mathbb Z_p^{n-\ell}$, which gives us our general form $V_\psi = \{(\mathbf{x}, M\mathbf{x}+\mathbf{c}) \mid \mathbf{x}\in \mathbb{Z}_p^\ell \}$.
\end{proof}

We are now prepared to prove the central theorem of this section. As discussed earlier, this theorem places restrictions on the stabilizer states of \xkcd codes, with a natural connection between the degree of the polynomial appearing as a phase in the wavefunction and the level of the Clifford hierarchy in which the stabilizers are contained. It generalizes the classic result of Ref.~\cite{DehaeneDeMoor_CliffordGroupStabilizer_03} for Pauli stabilizer states.

\begin{theorem}\label{thm:unique_state}
A state $|\psi\rangle$ is contained in $\Psi^{(k),n}_G$ if and only if it takes the form 
\begin{equation}\label{eq:state_form_alg}
|\psi\rangle = \frac{1}{p^{\ell/2}}\sum_{\mathbf{x}\in \mathbb{Z}_p^\ell} D[\phi]\otimes \mathbbm{1}|\mathbf{x},M\mathbf{x}+\mathbf{c}\rangle
\end{equation}
after a suitable permutation of qudits, where $M$ is an $(n-\ell)\times \ell$ matrix with entries in $\mathbb{Z}_p$, $\mathbf{c}\in \mathbb{Z}_p^{n-\ell}$, and $\phi$ is such that $D[\phi] \in {\cal C}^{(k+1),\ell}_d$.
\end{theorem}
\begin{proof}
Showing that such a state is contained in $\Psi^{(k),n}_G$ is straightforward. Consider the stabilizer subgroup
\begin{equation}
\begin{split}
    H = \langle X(\mathbf{e}_i)D[\Delta_{\mathbf{e}_i}\phi]\otimes X(M\mathbf{e}_i)\rangle_{i=1,\ldots, \ell}\\\times \langle \zeta_p^{c_i}Z(M_{i1})\otimes\ldots \otimes Z(M_{i\ell})\otimes Z(-\mathbf{e}_i)\rangle_{i=1,\ldots, n-\ell},
\end{split}
\end{equation}
where $\zeta_p:=e^{2\pi i/p}$ and $\{\mathbf{e}_i\}_{i=1}^{\ell}$ denotes the standard basis of $\mathbb{Z}_p^\ell$; that is, $\mathbf{e}_i$ has a single nonzero entry, equal to $1$, in its $i$th component.
Since we assumed that $D[\phi]\in {\cal C}^{(k+1),\ell}_d$ then by Corr.~\ref{cor:diff_lowers} we know that $D[\Delta_{\mathbf{v}_i}\phi] \in {\cal C}^{(k),\ell}_d$ for every $\mathbf{v}_i$. So then $H\subseteq G^{(k),n}$. Moreover, it is straightforward to check that this is the unique state stabilized by this group.

Now we consider the other direction. We know from Cor.~\ref{corr:aff_lin_sub} that up to a permutation $V_\psi = \{(\mathbf{x}, M\mathbf{x}+\mathbf{c}) \mid \mathbf{x}\in \mathbb{Z}_p^\ell\}$. Without loss of generality we take this ordering of the qudits. Because $(\mathbf{0},\mathbf{c}),(\mathbf{x},M\mathbf{x}+\mathbf{c})\in V_\psi$, then for every $\mathbf{x}\in \mathbb{Z}_p^\ell$ there will be some $g_{\vec x} \in H$ such that $g_{\vec x}=X(\mathbf{x})\otimes X(M\mathbf{x})D[f_{\vec x}]$ for some $D[f]\in {\cal C}^{(k),n}_d$, as required by Lem.~\ref{lem:Vpsi_stabsub}. Now from Lem.~\ref{lem:group_comp_wavefct} we know that we construct our wavefunction by repeated action of these $g$ on $|\mathbf{y}\rangle$ for any choice of $\mathbf{y}\in V_\psi$. Simplifying via Lem.~\ref{lem:wavefct_construction}, we have
\begin{equation}
    \ket \psi = \frac{1 }{p^{\ell/2}} \sum_{\vec x \in \mathbb Z_p^\ell} X(\vec x) \otimes X(M\vec x) D[f_{\vec x}] \ket {\vec y}.
\end{equation}
Without loss of generality, we can pick $\vec y = (\vec 0, \vec c)$. Thus, the $D[f_{\mathbf{x}}]$ operator simply gives us a phase of the form $\exp\left(2\pi if_{\vec x}((\vec 0, \vec c))\right)$ for each group element indexed by $\vec x \in \mathbb Z_p^\ell $. Simplifying by acting with $X(\vec x) \otimes X (M\vec x)$ gives us

\begin{equation}
    \ket \psi = \frac{1 }{p^{\ell/2}} \sum_{\vec x \in \mathbb Z_p^\ell} \exp\left(2 \pi i f_{\vec x}((\vec 0, \vec c))\right) \ket{\vec x, M\vec x+\vec c}.
\end{equation}
Define $\varphi(\vec x) = f_{\vec x}(\vec 0, \vec c)$, so we can write our state as
\begin{equation}
|\psi\rangle = \frac{1}{p^{\ell/2}}\sum_{\mathbf{x}\in \mathbb{Z}_p^\ell} D[\phi]\otimes \mathbbm{1}|\mathbf{x},M\mathbf{x}+\mathbf{c}\rangle.
\end{equation}
From here, we just need to show that $D[\varphi] \in \mathcal C_d^{(k+1), \ell}$. Fixing $\vec v \in \mathbb Z_p^\ell$, for each $\vec x \in \mathbb Z_p^\ell$, the operators $g_{\vec{v}} g_{\vec{x}}$ and $g_{\vec x+ \vec v}$ only differ by an element of $D_H$ which acts trivially on $\ket{\vec 0, \vec c}$. Thus, 
\begin{equation}
    g_{\vec{v}}g_{\vec{x}} \ket{\vec 0, \vec c} = g_{\vec x+ \vec v} \ket{\vec 0, \vec c},
\end{equation}
so taking finite differences yields
\begin{equation}
    \Delta_{\vec{v}} \varphi(\vec x) = f_{\vec v}(\vec x, M\vec x + \vec c),
\end{equation}
up to a constant. Since $D[f] \in \mathcal C_d^{(k),n}$ and $(\vec x, M\vec x + \vec c)$ is affine-linear, Thm.~\ref{thm:lin_preserves_level} implies $D[\Delta_{\vec{v}} \varphi] \in \mathcal C_d^{(k),\ell}$ for all $\vec v \in \mathbb Z_p^\ell$. Therefore, by Cor.~\ref{cor:diff_lowers}, $D[\varphi] \in \mathcal C_d^{(k+1), \ell}$, and we are done.
\end{proof}

In App.~\ref{app:on_site_clifford} we formulate and justify a conjecture for a similar result involving only on-site stabilizers.

\subsection{Algorithm}\label{sec:class_alg}

In this section, we formalize the algorithm for computing the canonical wavefunction which was implicit in the proofs of Cor.~\ref{corr:aff_lin_sub} and Thm.~\ref{thm:unique_state}. Suppose we are given an \xkcd stabilizer subgroup $H\subseteq G^{(k),n}$ and told that it uniquely stabilizes a state $\ket{\psi}$. This subgroup is fully encoded by the following set of pairs
\begin{equation}\label{eq:H_data}
    \{(\mathbf{v}_i,f_i)\mid X(\mathbf{v}_i)D[f_i]\in H\}.
\end{equation}
Given such a subgroup, we are promised by Thm.~\ref{thm:unique_state} that there exists a positive integer $\ell\leq n$, an $(n-\ell)\times \ell$ matrix $M$, an $(n-\ell)$-dimensional vector $\mathbf{c}$, and a function $\phi:\mathbb{Z}_p^\ell\to \mathbb{R}/\mathbb{Z}$ such that
\begin{equation}\label{eq:state_form}
|\psi\rangle = \frac{1}{p^{\ell/2}}\sum_{\mathbf{x}\in \mathbb{Z}_p^\ell} D[\phi]\otimes \mathbbm{1}|\mathbf{x},M\mathbf{x}+\mathbf{c}\rangle,
\end{equation}
up to a permutation of the site ordering encoded in a permutation matrix $Q$. An algorithm for computing the canonical wavefunction should take the pairs in Eq.~\ref{eq:H_data} as input and return $\ell$, $M$, $\mathbf{c}$, and $Q$.

The diagonal subgroup $D_H\subset H$ corresponds to the pairs
\begin{equation}
    \{(\mathbf{v}_i,f_i)\mid X(\mathbf{v}_i)D[f_i]\in H, \mathbf{v}_i=\mathbf{0}\}.
\end{equation}
From Lem.~\ref{lem:Vpsi_descrip}, we may construct $V_\psi$ by computing $f(\mathbf{y})$ for all $D[f]\in D_H$ and $\mathbf{y}\in\mathbb{Z}_p^n$ and identifying the intersection of zero sets. Furthermore, we are guaranteed by Cor.~\ref{corr:aff_lin_sub} that $V_\psi$ is of the form given in Eq.~\ref{eq:V_psi} after suitable permutation of the site indices. Suppose $V_\psi$ has dimension $\ell$, to be determined shortly. We would like to find a permutation matrix $Q$ such that
\begin{equation}
    Q^\top V_\psi = \left\{\begin{pmatrix}
        \mathbf{x} \\M\mathbf{x}+\mathbf{c}
    \end{pmatrix} \mid \mathbf{x}\in \mathbb{Z}_p^\ell \right\}.
\end{equation}

Because $V_\psi$ is an affine linear subspace, $L_\psi := V_\psi-\mathbf{y}$ is an $\ell$-dimensional linear subspace for any choice of $\mathbf{y}\in V_\psi$. 

First, construct a basis $\{\mathbf{b}_i\}$ for $L_\psi$ via Gaussian elimination on the matrix whose rows are given by $\mathbf{y}_i\in L_\psi$. Once this matrix is brought into reduced row-echelon form, the number of nonzero rows provides the dimension $\ell$ of $L_\psi$ and the nonzero rows form a basis $\{\mathbf{b}_i\}$ for $L_\psi$. Next, form the $\ell\times n$ matrix $B$ whose rows are the $\mathbf{b}_i$, i.e.
\begin{equation}
    B=\begin{pmatrix}
        \quad \mathbf{b}_1^\top\quad \\
        \vdots\\
         \mathbf{b}_\ell^\top
    \end{pmatrix}.
\end{equation}
By construction, we know that $L_\psi=\{B^\top\mathbf{x}\mid \mathbf{x}\in\mathbb{Z}_p^\ell\}$. $B$ is guaranteed to have $\ell$ pivot columns because it has rank $\ell$. We define $Q$ to be an $n\times n$ permutation matrix which brings the pivot columns to the left, preserving their ordering. The matrix $BQ$ then takes the form $(\mathbbm{1}|W)$, where $W$ is an $\ell \times (n-\ell)$ matrix. This tells us that
\begin{equation}
\begin{split}
    Q^\top L_\psi &= \{(BQ)^\top\mathbf{x}\mid \mathbf{x}\in\mathbb{Z}_p^\ell\},\\
    &=\left\{\begin{pmatrix}
        \mathbf{x}\\
        W^\top \mathbf{x}
    \end{pmatrix}\mid \mathbf{x}\in\mathbb{Z}_p^\ell\right\}.
\end{split}
\end{equation}

Recall that $L_\psi := V_\psi-\mathbf{y}_0$, so that
\begin{equation}
\begin{split}
    Q^\top V_\psi &= Q^\top L_\psi+Q^\top \mathbf{y}_0,\\
    &= \left\{\begin{pmatrix}
        \mathbf{x}+(Q^\top \mathbf{y}_0)^{(1)}\\
        W^\top \mathbf{x}+(Q^\top \mathbf{y}_0)^{(2)}
    \end{pmatrix}\mid \mathbf{x}\in\mathbb{Z}_p^\ell\right\},\\
    &= \left\{\begin{pmatrix}
        \mathbf{x}\\
        W^\top \mathbf{x}-W^\top(Q^\top \mathbf{y}_0)^{(1)}+(Q^\top \mathbf{y}_0)^{(2)}
    \end{pmatrix}\mid \mathbf{x}\in\mathbb{Z}_p^\ell\right\},\\
    &= \left\{\begin{pmatrix}
        \mathbf{x}\\
        M \mathbf{x}+\mathbf{c}
    \end{pmatrix}\mid \mathbf{x}\in\mathbb{Z}_p^\ell\right\},
\end{split}
\end{equation}
where we have defined 
\begin{equation}
    M=W^\top,\quad \mathbf{c}=-W^\top(Q^\top \mathbf{y}_0)^{(1)}+(Q^\top \mathbf{y}_0)^{(2)}.
\end{equation}
At this point, we have determined $\ell$, $M$, and $\mathbf{c}$, with only $\phi$ left to determine.

The following discussion closely mirrors the proof of Thm.~\ref{thm:unique_state}. Because $(\mathbf{0},\mathbf{c})$ and $(\mathbf{x},M\mathbf{x}+\mathbf{c})$ are both contained in $Q^\top V_\psi$, then we know by Lem.~\ref{lem:Vpsi_stabsub} that for every $\mathbf{x}\in \mathbb{Z}_p^\ell$ there will be some $g_{\vec x} \in H$ such that $g_{\vec x}=X(\mathbf{x})\otimes X(M\mathbf{x})D[f_{\mathbf x}]$ for some $D[f_{\mathbf x}]\in {\cal C}^{(k),n}_d$ after permutation by $Q$. We construct our wavefunction by repeated action of these $g$ (the coset representatives) on $|\mathbf{y}\rangle$ for any choice of $\mathbf{y}\in V_\psi$. Without loss of generality, choose $\mathbf{y}=(\mathbf{0},\mathbf{c})$, so that
\begin{equation}
\begin{split}
    \ket \psi &= \frac{1 }{p^{\ell/2}} \sum_{\vec x \in \mathbb Z_p^\ell} X(\vec x) \otimes X(M\vec x) D[f_{\vec x}] \ket {\vec y}\\
    & = \frac{1 }{p^{\ell/2}} \sum_{\vec x \in \mathbb Z_p^\ell} \exp\left(2 \pi i f_{\vec x}((\vec 0, \vec c))\right) \ket{\vec x, M\vec x+\vec c}.
\end{split}
\end{equation}
Define $\varphi(\vec x) = f_{\vec x}(\vec 0, \vec c)$, so we can write our state as
\begin{equation}
|\psi\rangle = \frac{1}{p^{\ell/2}}\sum_{\mathbf{x}\in \mathbb{Z}_p^\ell} D[\phi(\mathbf{x})]|\mathbf{x},M\mathbf{x}+\mathbf{c}\rangle.
\end{equation}
To determine $\phi(\mathbf{x})$, we choose a representative $\mathbf{v}_i$ such that $Q^\top \mathbf{v}_i$ begins with $\mathbf{x}$. We take $\phi(\mathbf{x})$ to be $f_i(Q(\vec 0, \vec c))$ and repeat for all $\mathbf x\in \mathbb{Z}_p^\ell$, thereby constructing $\phi$. We emphasize that $Q(\vec 0, \vec c)\in V_\psi$, so $f_i(Q(\vec 0, \vec c))$ is invariant of the coset representative, as these differ by multiplication by $D[g]\in D_H$ and $g(\mathbf{y})=0\pmod{1}$ for any $\mathbf{y}\in V_\psi$.

We now present the procedure we have constructed in compact form as an explicit algorithm:
\algrenewcommand\algorithmicrequire{\textbf{Input:}}
\algrenewcommand\algorithmicensure{\textbf{Output:}}

\begin{algorithm}[H]
\caption{Procedure to compute the canonical form of an \xkcd stabilizer state}
\begin{algorithmic}[1]
\label{alg:canonical-form}
\Require pairs specifying $H\subseteq G^{(k),n}$:
$\{(\mathbf{v}_i,f_i)\mid X(\mathbf{v}_i)D[f_i]\in H\}$
\Ensure $\ell$, $Q$, $M$, $\mathbf{c}$, $\phi$

\State Extract the diagonal subgroup
\[
D_H=\{(\mathbf{0},f_i)\mid (\mathbf{v}_i,f_i)\in H,\ \mathbf{v}_i=\mathbf{0}\}.
\]

\State Compute
\[
V_\psi=\{\mathbf{y}\in\mathbb Z_p^n\mid
f(\mathbf{y})=0\pmod 1,\ \forall D[f]\in D_H\}.
\]

\State Choose any $\mathbf{y}_0\in V_\psi$

\State Compute the linear subspace
\[
L_\psi=V_\psi-\mathbf{y}_0.
\]

\State Construct a basis $\{\mathbf{b}_1,\ldots,\mathbf{b}_\ell\}$ of $L_\psi$ via Gaussian elimination

\State Form the basis matrix
\[
B=\begin{pmatrix}
\quad \mathbf{b}_1^\top \quad \\
\vdots\\
\mathbf{b}_\ell^\top 
\end{pmatrix}.
\]

\State Identify the pivot columns of $B$

\State Construct the permutation matrix $Q$ that moves the pivot columns to the first $\ell$ positions

\State Compute
\[
(\mathbbm 1\,|\,W)\gets BQ.
\]

\State Set
\[
M\gets W^\top.
\]

\State Compute
\[
\mathbf c\gets
(Q^\top\mathbf y_0)^{(2)}
-
M(Q^\top\mathbf y_0)^{(1)}.
\]

\ForAll{$\mathbf{x}\in\mathbb Z_p^\ell$}
    \State Choose a coset representative
    $X(\mathbf{x})\otimes X(M\mathbf{x})D[f_{\mathbf{x}}]$
    in the permuted stabilizer group 
    \[\{(Q^\top \mathbf{v}_i,f_i\circ Q)\mid X(\mathbf{v}_i)D[f_i]\in H\}\]
    \State $\phi(\mathbf{x})\gets
    f_{\mathbf{x}}(\mathbf{0},\mathbf c)$
\EndFor

\State \Return $\ell$, $Q$, $M$, $\mathbf c$, $\phi$

\end{algorithmic}
\end{algorithm}

\section{Realizing \texorpdfstring{${\cal D}^\omega(\mathbb{Z}_p^n)$}{Dω(Zpn)} topological order with \texorpdfstring{\xkcd}{xkcd} codes}\label{sec:realizing_with}

In this section, we return to the twisted quantum double models discussed in Sec.~\ref{sec:TQD_review} using the stabilizer formalism developed in Secs.~\ref{sec:XkCd_intro} and \ref{sec:XkCd_class}. We are ultimately interested in the following question:
\begin{question}\label{q:main}
    Given a twisted abelian quantum double topological order ${\cal D}^\omega(\mathbb{Z}_p^n)$ with $p$ prime, what is the minimal level $k$ such that there exists an \xkcd codestate realizing the topological order ${\cal D}^\omega(\mathbb{Z}_p^n)$?
\end{question}
\noindent Although we do not answer this question in full, we give a physically motivated conjecture which is contextualized by the following formalism.

As we demonstrate below, the minimal $k$ strongly depends on the structure of $\omega$. We therefore begin this section by reviewing the functional form of 3-cocycles over $\mathbb{Z}_p^n$. A 3-cocycle on $\mathbb{Z}_p^n$ with coefficients in $U(1)$ is a function 
\begin{equation}
    \omega:\mathbb{Z}_p^n\times \mathbb{Z}_p^n\times \mathbb{Z}_p^n\to U(1)
\end{equation} which additionally satisfies the 3-cocycle condition:
\begin{equation}
    \frac{
\omega(g_1,g_2,g_3)\,
\omega(g_0,g_1g_2,g_3)\,
\omega(g_0,g_1,g_2)
}{
\omega(g_0g_1,g_2,g_3)\,
\omega(g_0,g_1,g_2g_3)
}
=1.
\end{equation}
The group $H^3(G,U(1))$ is comprised of equivalence classes of such functions under multiplication by coboundaries.
Any $\omega\in H^3(\mathbb{Z}_p^n, U(1))$ can be written (up to coboundary) as a product over type-I, type-II, and type-III 3-cocycles: special functions on $\mathbb{Z}_p^n$ of the form $\omega(a,b,c)=\omega(a_i,b_j,c_k)$ (i.e. functions which depend on only a single component of each argument) \cite{deWildPropitius_TopologicalInteractionsBroken_95}. In particular, for every $\omega\in H^3(\mathbb{Z}_p^n, U(1))$ there exist $n_i,n_{ij},n_{ijk}\in \mathbb{Z}_p$ such that
\begin{equation}
\begin{split}
    \omega(g,h,k)\cong\prod_{i=1}^n\left[\omega_\mathrm{I}^{(i)}(g,h,k)\right]^{n_i}\prod_{i<j}\left[\omega_\mathrm{II}^{(ij)}(g,h,k)\right]^{n_{ij}}\\\times \prod_{i<j<k}\left[\omega_\mathrm{III}^{(ijk)}(g,h,k)\right]^{n_{ijk}},
\end{split}
\end{equation}
where ``$\cong$'' denotes equivalence up to coboundary and
\begin{align}\label{eq:type_coycles}
&\omega_\mathrm{I}^{(i)}(a,b,c)=e^{\frac{2\pi i}{p}a_i C(b_i,c_i)},\quad  \nonumber\\
&\omega_\mathrm{II}^{(ij)}(a,b,c)=e^{\frac{2\pi i}{p}a_i C(b_j,c_j)}, \quad  \nonumber\\&\omega_\mathrm{III}^{(ijk)}(a,b,c)=e^{\frac{2\pi i}{p}a_ib_jc_k}.
\end{align}
The \textit{carry function} $C(x,y)$ is equal to one if $x+y\geq p$ and zero otherwise. Because $a\in\mathbb{Z}_p^n$, the notation $a_i$ denotes the $i$th $\mathbb{Z}_p$-valued entry, and similarly for $b$ and $c$.

The ``type'' nomenclature indicates the number of independent factors of $\mathbb{Z}_p^n$ on which the cocycle acts. Physically, we can imagine a stack of $n$ $\mathbb{Z}_p$-toric codes, with type-I cocycles encoding twists within a single layer, type-II encoding twists between two layers, and so on. Importantly for our purposes, ${\cal D}^\omega(\mathbb{Z}_p^n)$ is a non-abelian topological order if and only if $\omega$ contains a type-III cocycle in its decomposition \cite{deWildPropitius_TopologicalInteractionsBroken_95}.

We are now prepared to introduce the central object of study in this subsection:
\begin{definition}
    We define the \textit{canonical TQD wavefunction} $\psiTQD$ to be the groundstate of the canonical TQD Hamiltonian $H_{{\cal D}^\omega(\mathbb{Z}_p^n)}$ defined in Eq.~\ref{eq:can_TQD_ham}. To ensure that this groundstate is unique, we assume that the Hamiltonian is defined on a triangulation of a space with the topology of the 2-sphere.
\end{definition}

As suggested in Sec.~\ref{sec:TQD_review}, there is an \xkcd code naturally associated to the canonical TQD Hamiltonian whose unique codeword is the canonical TQD wavefunction:
\begin{lemma}\label{lem:TQD_code}
    The groundstate subspace of the canonical TQD Hamiltonian $H_{{\cal D}^\omega(\mathbb{Z}_p^n)}$ coincides with the codespace of an \xkcd code. This code can be defined via the following stabilizer generators:
    \begin{equation}\label{eq:TQD_xkcd_stabs}
    \begin{gathered}
        Z^{(e_{1,l})}Z^{(e_{2,l})}\left(Z^{(e_{3,l})}\right)^\dagger,\; \forall f=(e_1,e_2,e_3)\in F,\;l\in\mathbb{Z}_n,\\ A^\omega_g(v),\;\forall g\in \mathbb{Z}_p^n,\;v\in V
    \end{gathered}
    \end{equation}
    where $Z^{(e_{i,l})}$ assigns a phase proportional to the computational basis label located on the edge $e_i$ in the $l$th layer. The $Z$-type operators generate a flatness constraint within each layer, which ensures flatness on the composite variable which combines all layers. The $A^\omega_g$ stabilizers defined in Eq.~\ref{eq:TQD_xkcd_stabs} are only present in the Clifford hierarchy at and above the following levels:
	    \begin{equation}\label{eq:k_bound}
	       k = \begin{cases}
	           2,&\text{if }\omega\text{ is a product of type-III cocycles}\\
	           p,&\text{otherwise}
	       \end{cases},
	   \end{equation}
    where we have assumed that $\omega$ is nontrivial
\end{lemma}
\begin{proof}
    First, we prove that the code defined in Eq.~\ref{eq:TQD_xkcd_stabs} stabilizes the groundstate of $H_{{\cal D}^\omega(\mathbb{Z}_p^n)}$. The $A_g^\omega(v)$ are simply the star terms appearing in $H_{{\cal D}^\omega(\mathbb{Z}_p^n)}$, so there is nothing to prove here. The nontrivial point is that the terms $Z^{(e_{1,l})}Z^{(e_{2,l})}\left(Z^{(e_{3,l})}\right)^\dagger$ evaluate to $+1$ on the flux-free subspace specified by $B(f)$. On the flux-free subspace, $\ket{e_3}=\ket{e_1e_2}$, so $Z^{(e_{1,l})}Z^{(e_{2,l})}\left(Z^{(e_{3,l})}\right)^\dagger=\mathbbm{1}$ on this subspace. This relation implies that the groundstate of $H_{{\cal D}^\omega(\mathbb{Z}_p^n)}$ belongs to the codespace. Furthermore, the code defined in Eq.~\ref{eq:TQD_xkcd_stabs} ensures that all terms in $H_{{\cal D}^\omega(\mathbb{Z}_p^n)}$ are satisfied, so its codespace must be contained in the groundstate space of $H_{{\cal D}^\omega(\mathbb{Z}_p^n)}$. We therefore conclude that the \xkcd code defined in Eq.~\ref{eq:TQD_xkcd_stabs} uniquely stabilizes the canonical TQD wavefunction $\psiTQD$.

    We now prove the statement given in Eq.~\ref{eq:k_bound} regarding the Clifford hierarchy level of $A^\omega_g$. First, note that $A^\omega_g$ admits the following decomposition:
    \begin{equation}
    \begin{gathered}A^\omega_g(v)=\left(\bigotimes_{e_i\in\delta(v)}X_{\pm g}^{(e_i)}\right)\Lambda_g^\omega,\\
    \Lambda^\omega_g=\prod_{f\in\delta(v)}\sum_{\{e_i\}\in \mathbb{Z}_p^n}\omega|_f\ketbra{e_1,e_2,e_3}{e_1,e_2,e_3},
    \end{gathered}
    \end{equation}
where $\{e_i\}\in\partial(f)$ and $\omega|_f$ is defined for an example face in Fig.~\ref{fig:D_omega}.    

We now want to place the diagonal operator $\Lambda_g^\omega$ in the Clifford hierarchy. If $\omega$ is a type-III cocycle, then $\Lambda^\omega_g$ is a diagonal operator whose phase function is quadratic in the qudit variables $\{e_i\}$ with precision $2\pi/p$ (i.e., $m=1$). We emphasize that this operator is only quadratic in the edge variables, despite $\omega_\mathrm{III}$ having a cubic function in its exponent, because one of the arguments is always set to $g$, which is fixed. For example, when evaluating a type-III cocycle on the face depicted in Fig.~\ref{fig:D_omega}, we have $\omega_\mathrm{III}(g,a,b)=e^{\frac{2\pi i}{p}g_i (a_j b_k)}$, which is quadratic in the edge variables $\{a,b,c\}$. From the definition of the diagonal multi-qudit Clifford hierarchy (Def.~\ref{def:cliff_hier_def}), we conclude that $\Lambda^\omega_g$ is first available at level 2 when $\omega$ only contains type-III cocycles in its decomposition.

In order to classify the operator $\Lambda^\omega_g$ when $\omega$ contains type-I and type-II cocycles in its decomposition, we need to understand the degree of the multivariate polynomial in the exponent of the type-I and -II cocycles with one argument fixed. This amounts to calculating the degree of
\begin{equation}
\begin{split}
    f:\;& \mathbb{Z}_p^3\rightarrow \mathbb{Z}_p\\
    &(x,y,z)\mapsto xC(y,z)\pmod{p} 
\end{split}
\end{equation}
 with one of $x$, $y$, and $z$ fixed. The most straightforward way to do this is to note that 
 \begin{equation}
     C(y+1,z)-C(y,z) = \delta_{y+z,p-1}-\delta_{y,p-1} \pmod{p}
 \end{equation}
 by the definition of $C(y,z)$. But $\delta_{x,p-1} = 1-(x+1)^{p-1}\pmod{p}$ by Fermat's little theorem. So then we note that since $\Delta_{(1,0)}C(y,z)=C(y+1,z)-C(y,z)$ is a degree-$(p-1)$ polynomial, then $C(y,z)$ must be a degree-$p$ polynomial. It follows directly that $f(x,y,z)=xC(y,z)$ is a degree-$p$ polynomial if one of its arguments is held fixed to a nonzero value.

We thus see that $\Lambda^\omega_g$ is a diagonal operator whose phase function is degree-$p$ in the qudit variables $\{e_i\}$ with precision $2\pi/p$ (i.e., $m=1$) if $\omega$ contains a type-I or -II cocycle in its decomposition. We therefore conclude that $\Lambda^\omega_g$ is first available at level $p$ in this case.
\end{proof}

From the equivalence established in Lemma~\ref{lem:TQD_code}, we know that $\psiTQD$ is stabilized by the \xkcd code defined in Eq.~\ref{eq:TQD_xkcd_stabs}. However, it is not yet obvious that there is no clever way to construct a different \xkcd code, with lower values of $k$ than those stated in Eq.~\ref{eq:k_bound}, which still stabilizes $\psiTQD$. Indeed, if we were to prove this for $\psiTQD$ this would only set an upper bound on the level necessary to realize the TQD topological order ${\cal D}^\omega(\mathbb{Z}_p^n)$, but would not necessarily preclude the existence of another state in the same phase which is stabilized at a lower level. However, we conjecture that no such state exists:
\begin{conjecture}\label{conj:level-k-TQD}
   Given a TQD topological order ${\cal D}^\omega(\mathbb{Z}_p^n)$, the minimum $k$ such that there exists an \xkcd stabilizer code whose codestate lies in the ${\cal D}^\omega(\mathbb{Z}_p^n)$ phase is given by:
   \begin{equation}
       k = \begin{cases}
           2,&\text{if }\omega\text{ is a product of type-III cocycles}\\
           p,&\text{otherwise}
       \end{cases},
   \end{equation}
where we take $\omega$ to be nontrivial. Below we are able to prove the type-III case, and make a conjecture about type-I and II.
\end{conjecture}
\noindent 

Suppose $\omega$ is a type-III cocycle. In this case we know that ${\cal D}^\omega(\mathbb{Z}_p^n)$ must contain non-abelian anyons \cite{deWildPropitius_TopologicalInteractionsBroken_95}. By Ref.~\cite{PotterVasseur_SymmetryConstraintsManybody_16} we know that it is not possible to realize a topological order with non-abelian anyons in a Pauli stabilizer model. Thus it is not possible to realize ${\cal D}^\omega(\mathbb{Z}_p^n)$ using an \xkcd code with $k=1$, so $k=2$ is the smallest possible. By Lem.~\ref{lem:TQD_code} we see that it is possible to realize this topological order with $k=2$, so the lower bound is saturated. We therefore conclude that in this case $k=2$ is the minimum $k$ such that there exists an \xkcd stabilizer code whose codestate lies in the ${\cal D}^\omega(\mathbb{Z}_p^n)$ phase when $\omega$ is a type-III cocycle.

If $\omega$ contains a type-I or -II cocycle, it is much more difficult to establish a lower bound on $k$. Instead we attempt to substantiate our conjecture. We begin with a solid fact.
If $p$ is an odd prime and $\omega$ has any type-I or -II cocycles then by Lem.~\ref{lem:types} of App.~\ref{app:quant_double_rest} it can be proven there exists at least one anyon of ${\cal D}^\omega(\mathbb{Z}_p^n)$ with topological spin that is a primitive $p^2$ root of unity. Proceeding in a conjectural fashion we take inspiration from Ref.~\cite{Haah_ClassificationTranslationInvariant_21}. There, a Wilson line for an anyon was defined in terms of products of on-site stabilizers. The topological spin was then read off from products of these stabilizers. Now assume that the same procedure generalizes to TQDs. Construct the Wilson line for the anyon with topological spin a primitive $p^2$ root of unity. Because this topological spin must arise from the products of stabilizers, this means that there must be a primitive $p^2$ root of unity appearing in our \xkcd stabilizers. But the smallest level of the Clifford hierarchy for which a primitive $p^2$ root of unity can appear is an on-site stabilizer at level $p$. This suggests that it may not be possible to prepare ${\cal D}^\omega(\mathbb{Z}_p^n)$ using \xkcd stabilizers with $k<p$ if $\omega$ contains a type-I or -II cocycle. 

\section{Preparing a logical magic state in a pair of qudit toric code blocks}\label{sec:gates}

\begin{figure*}
  \centering
  $\hfill\vcenter{\hbox{\begin{tikzpicture}[x=2.5cm,y=2.5cm]
  \path[use as bounding box,clip] (-1.1,-1.01) rectangle (3.6,1.01);

  \lattice{\cellB}{(0,-0.5773503)}
  \lattice{\cellR}{(0.5,-0.2886751)}
  \lattice{\cellG}{(0,0)}

  \begin{scope}[line width=1.5pt,line cap=round]
    \draw        ( 90:0.28) to[bend left=20] (150:0.28)
                 (210:0.28) to[bend left=20] (270:0.28)
                 (330:0.28) to[bend left=20] ( 30:0.28);
    \draw[brown] ( 30:0.28) to[bend left=20] ( 90:0.28)
                 (150:0.28) to[bend left=20] (210:0.28)
                 (270:0.28) to[bend left=20] (330:0.28);
  \end{scope}

  \begin{scope}[Green,line width=1.5pt,line cap=round]
    \draw[dirA] ( 0  , 0   ) -- ( 1  , 0   );
    \draw[dirA] (-1  , 0   ) -- ( 0  , 0   );
    \draw[dirB] ( 0  , 0   ) -- ( 0.5, \hh );
    \draw[dirB] (-0.5,-\hh ) -- ( 0  , 0   );
    \draw[dirC] ( 0  , 0   ) -- (-0.5, \hh );
    \draw[dirC] ( 0.5,-\hh ) -- ( 0  , 0   );
    \draw[dirA] (1.5,-\hh) -- (2.5,-\hh);
    \draw[dirB] (1.5,-\hh) -- (2  , 0  );
    \draw[dirC] (2.5,-\hh) -- (2  , 0  );
    \draw[dirA] (2.5,\hh) -- (3.5,\hh);
    \draw[dirB] (2.5,\hh) -- (3  , 0  );
    \draw[dirC] (3.5,\hh) -- (3  , 0  );
  \end{scope}

  \foreach \a in {120,180,240}{\node[op,fill=Green] at (\a:0.5) {\fixw{$X^\dagger$}};}
  \foreach \a in {300,0,60}  {\node[op,fill=Green] at (\a:0.5) {$X$};}
  \node[op,fill=Green] at (1.75,-0.5*\hh) {\fixw{$Z^\dagger$}};
  \node[op,fill=Green] at (2.25,-0.5*\hh) {\fixw{$Z^\dagger$}};
  \node[op,fill=Green] at (2   ,-\hh)     {$Z$};
  \node[op,fill=Green] at (2.75, 0.5*\hh) {$Z$};
  \node[op,fill=Green] at (3.25, 0.5*\hh) {$Z$};
  \node[op,fill=Green] at (3   , \hh)     {\fixw{$Z^\dagger$}};

  \node[fill=white,draw] (czlab) at (-0.50,0.25) {$CZ$};
  \draw (czlab) -- (130:0.28);
  \node[brown,fill=white,draw] (czdlab) at (0.50,0.25) {$CZ^\dagger$};
  \draw[brown] (czdlab) -- (50:0.28);
  \node[Green,fill=white,draw] at (0.0,0.8)  {\Large$\mathcal{A}_{v,g}$};
  \node[Green,fill=white,draw] at (2.7,-0.3) {\Large${B}_{f,g}$};
  \node[fill=white] at (-0.96,0.9) {\large(a)};
\end{tikzpicture}}}\hfill\vcenter{\hbox{\begin{tikzpicture}[x=2.5cm,y=2.5cm]
  \path[use as bounding box,clip] (-1.0,-1.01) rectangle (1.0,1.01);

  \lattice{\cellB}{(0,-0.5773503)}
  \lattice{\cellR}{(0.5,-0.2886751)}
  \lattice{\cellG}{(0,0)}

  \begin{scope}[line width=1.5pt,line cap=round]
    \foreach \x in {-1,0,1}{%
    \foreach \y in {-1,0,1}{%
      \begin{scope}[shift={(\x+0.5*\y,\hh*\y)}]
        \draw        ( 90:0.28) to[bend left=20] (150:0.28)
                     (210:0.28) to[bend left=20] (270:0.28)
                     (330:0.28) to[bend left=20] ( 30:0.28);
        \draw[brown] ( 30:0.28) to[bend left=20] ( 90:0.28)
                     (150:0.28) to[bend left=20] (210:0.28)
                     (270:0.28) to[bend left=20] (330:0.28);
      \end{scope}}}
  \end{scope}

  \node[Green,fill=white,draw] at (-0.5,0.3*\hh) {\Large$\hat C_g$};
  \node[fill=white] at (-0.9,0.9) {\large(b)};
\end{tikzpicture}}}\hfill$
  \caption{(a) One layer of stabilizer generators of the TQD $D^{\omega_\mathrm{III}}(\mathbb{Z}_p^3)$. The stabilizer generators in the other layers are defined analogously. (b) The product of all TQD vertex generators in the green layer is $\color{Green}\hat C_g$, a diagonal operator which is only supported on the red and blue layers. Given a pair of $\mathbb{Z}_p$ toric codes in the red and blue layers, $\color{Green}\hat C_g$ acts as a nontrivial logical operator on this code.}
  \label{fig:color-stabs}
\end{figure*}

In this section, we demonstrate how to prepare a logical magic state encoded into a pair of $\mathbb{Z}_p$ toric codes by introducing ancillae, measuring stabilizers of an intermediate \xkcd code, and subsequently disentangling from the ancillae. This procedure was discussed for the qubit toric code in Ref.~\cite{DavydovaEtAl_UniversalFaultTolerant_25}, and the following construction is the generalization from $\mathbb{Z}_2$ to $\mathbb{Z}_p$. This procedure may also be understood as gauging and ungauging an automorphism of the initial topological order~\cite{DavydovaEtAl_UniversalFaultTolerant_25,Williamson_2026Fault}, as we discuss below.

\subsection{Model and stabilizers}

To define the model, we begin with three copies of a triangulation of the torus, equipped with a branching structure, and place a $\mathbb{Z}_p$ qudit on each edge in each copy. We label the copies with the colors $\{ {\color{red}\text{red}},{\color{blue}\text{blue}},{\color{Green}\text{green}}\}$. In the figures appearing throughout this section, we depict the lattices of different colors offset from one another for ease of visualization and to make the underlying mathematical structure more apparent. 

In the protocol below, we switch through a series of stabilizer codes via measurement of new stabilizers. We begin and end in the groundstate space of the untwisted quantum double $D(\mathbb{Z}_p\times\mathbb{Z}_p)$, i.e., two copies of the qudit toric code, passing through the TQD $D^{\omega_\mathrm{III}}(\mathbb{Z}_p^3)$. We now introduce the conventions for these stabilizers to be used throughout this section. 

We denote the vertex stabilizer generators for the untwisted quantum double $D(\mathbb{Z}_p)$ in the layer $c$ by
\begin{equation}
    A_{v,c}=\prod_{e\in\delta(v)}X_{e,c}^{(\dagger)},
\end{equation}
where the dagger is applied for edges oriented into the vertex $v$. The corresponding TQD stabilizer generator\footnote{This model differs slightly from the canonical TQD Hamiltonian in its placement of interlayer diagonal operators but realizes the same phase. We use it in this section because it directly generalizes the model discussed in Ref.~\cite{DavydovaEtAl_UniversalFaultTolerant_25}.} is denoted $\mathcal{A}_{v,c}$ and may be obtained from $A_{v,c}$ by decorating $CZ$ and $CZ^\dagger$ between nearby sites of the other two colors, as depicted in Fig.~\ref{fig:color-stabs}. The plaquette stabilizer generator at the face $f$ is denoted $B_{f,c}$ and is also depicted in Fig.~\ref{fig:color-stabs}. Note that the plaquette stabilizer generator is the same operator in the twisted or untwisted quantum double. 

In the notation we have established, the stabilizer group for three copies of the $\mathbb{Z}_p$ toric code is given by
\begin{equation}\label{eq:Zp3-stab}
    \mathcal{S}_{D(\mathbb{Z}_p^3)}=\langle {\color{red}A_{v,r}},{\color{red}B_{f,r}},{\color{blue}A_{v,b}},{\color{blue}B_{f,b}},{\color{Green}A_{v,g}},{\color{Green}B_{f,g}}\rangle,
\end{equation}
where $v$ and $f$ run over all vertices and faces of the triangulation, respectively. In contrast, the stabilizer group for the TQD code is given by
\begin{equation}\label{eq:TQD-stab}
    \mathcal{S}_{D^{\omega_\mathrm{III}}(\mathbb{Z}_p^3)}=\langle {\color{red}\mathcal{A}_{v,r}},{\color{red}B_{f,r}},{\color{blue}\mathcal{A}_{v,b}},{\color{blue}B_{f,b}},{\color{Green}\mathcal{A}_{v,g}},{\color{Green}B_{f,g}}\rangle.
\end{equation}
In the protocol described below, we switch between these codes by performing measurements that modify the vertex stabilizers.

The global product of the TQD vertex stabilizer generators of a single color plays a central role:
\begin{equation}\label{eq:charge-def}
    \hat C_{c}:=\prod_v\mathcal{A}_{v,c}.
\end{equation}
As an example of direct relevance to the following section, consider the green version of this operator, denoted $\color{Green}\hat C_g$.
The product
${\color{Green}\hat C_g}$ reduces to a diagonal operator supported only on the red and blue layers, as the off-diagonal operators in the green layer cancel pairwise. Further, we find that $\color{Green}\hat C_g$ preserves the codespace of ${{\color{red}D(\mathbb{Z}_p)_r}\times\color{blue}D(\mathbb{Z}_p)_b}$ because it maps stabilizer generators to products of stabilizer generators, as shown in App.~\ref{ap:magic-state}.

Further, ${\color{Green}\hat C_g}$ enacts a nontrivial logical transformation on the codespace of ${{\color{red}D(\mathbb{Z}_p)_r}\times\color{blue}D(\mathbb{Z}_p)_b}$ given by
\begin{equation}\label{eq:Cg-log-action}
    {\color{Green}\hat C_g}=\overline{CZ}_{{\color{red}rV},{\color{blue}bH}}^\dagger\overline{CZ}_{{\color{red}rH},{\color{blue}bV}},
\end{equation}
where $H$ and $V$ label the two non-contractible cycles of the torus and $CZ\ket{m,n}=\mathbbm{1}\otimes Z^m\ket{m,n}=\zeta^{mn}\ket{m,n}$ is the qudit controlled-$Z$ (with $\zeta:=e^{2\pi i/p}$). To derive this result, we verify that it holds for a particular representative flat configuration of the loop gas wavefunction, then show that the result is independent of the representative chosen. 

We show in Fig.~\ref{fig:color-logs} that $\color{Green}\hat C_g$ yields a phase $\zeta^{-mn}$ when acting on a representative flat configuration in the ${\color{red}\ket{\overline{m}}_{rV}}{\color{blue}\ket{\overline{n}}_{bH}}$ homology class. Accordingly, $\color{Green}\hat C_g$ acts on this logical subsystem as $\overline{CZ}_{{\color{red}rV},{\color{blue}bH}}^\dagger$. A similar calculation (presented in App.~\ref{ap:magic-state}) shows that $\color{Green}\hat C_g$ acts independently on the other logical subsystem as $\overline{CZ}_{{\color{red}rH},{\color{blue}bV}}$, producing the overall logical action presented in Eq.~\ref{eq:Cg-log-action}.

To see that this result is independent of the configuration chosen, note that the loop gas may be obtained by acting on a representative flat configuration with the projector into the gauge-invariant subspace, given by
\begin{equation}
    \Pi_A:=\prod_v\left(\frac{1}{p}\sum_{n\in\mathbb{Z}_p}\left({\color{red}
    A_{v,r}}\right)^{n}\right)\left(\frac{1}{p}\sum_{n\in\mathbb{Z}_p}\left({\color{blue}
    A_{v,b}}\right)^{n}\right).
\end{equation}
As discussed above, conjugation by $\color{Green}\hat C_g$ attaches products of plaquette stabilizers $B_{f,r/b}$ to vertex stabilizers $A_{v,b/r}$. Because all stabilizers of $D(\mathbb{Z}_p)$ commute with one another, we find that $\Pi_A$ commutes with $\color{Green}\hat C_g$ on the flat subspace. In turn, we may conclude that $\color{Green}\hat C_g$ must return the same phase when acting on any state in a fixed homology class, so that our result is independent of the representative chosen.

\subsection{Measurement protocol}

In this section, we describe the progression through a series of stabilizer codes via measurement of new local stabilizers, effectively measuring a non-local logical operator. When initialized in an appropriate logical state, this protocol deterministically produces a logical magic state. 

The initial stabilizer group is given by
\begin{equation}
    \mathcal{S}_1
    =\langle {\color{red}A_{v,r}},{\color{red}B_{f,r}},{\color{blue}A_{v,b}},{\color{blue}B_{f,b}},{\color{Green}Z_{e,g}}\rangle.
\end{equation}
This code has a $p^4$-dimensional codespace consisting of logical qudits along the two non-contractible cycles of the torus in the red and blue layers. In the green layer, the state $\ket{0}$ is placed on each edge. Because all qudits in the green layer are in the $\ket{0}$ state, the TQD stabilizers $\color{red} \mathcal{A}_{v,r}$ and $\color{blue} \mathcal{A}_{v,b}$ are also satisfied, so we may equivalently write
\begin{equation}
    \mathcal{S}_1
    =\langle {\color{red}\mathcal{A}_{v,r}},{\color{red}B_{f,r}},{\color{blue}\mathcal{A}_{v,b}},{\color{blue}B_{f,b}},{\color{Green}Z_{e,g}},{\color{Green}B_{f,g}},{\color{Green}\overline{Z}_{H,g}},{\color{Green}\overline{Z}_{V,g}}\rangle.
\end{equation}
Following Ref.~\cite{DavydovaEtAl_UniversalFaultTolerant_25}, we have written redundant stabilizer generators on the green layer for easier bookkeeping when we begin measuring new stabilizers. 

Comparing with Eq.~\ref{eq:TQD-stab}, we see that $\mathcal{S}_1$ contains all of the TQD stabilizer generators except the $\color{Green}\mathcal{A}_{v,g}$. To bring the code into the TQD phase, we measure these stabilizers. Although $\color{Green}\mathcal{A}_{v,g}$ is not Hermitian, we may nonetheless define the set of projectors $\{P_n:n\in\mathbb{Z}_p\}$, where $P_n$ projects onto the eigenspace of $\color{Green}\mathcal{A}_{v,g}$ corresponding to the eigenvalue $\zeta^n$, and perform a projective measurement. Suppose this measurement returns $-n_v\in\mathbb{Z}_p$. In this case, the post-measurement state is stabilized by $\zeta^{n_v}{\color{Green}\mathcal A_{v,g}}$. (We use this sign convention so that the new stabilizer generators carry the prefactor $\zeta^{n_v}$.) We define measurement of other non-Hermitian operators analogously.

After measuring all of the $\color{Green}\mathcal{A}_{v,g}$, the stabilizer group becomes 
\begin{equation}
    \mathcal{S}_{2}
    =\langle {\color{red}\mathcal{A}_{v,r}},{\color{red}B_{f,r}},{\color{blue}\mathcal{A}_{v,b}},{\color{blue}B_{f,b}},\zeta^{n_v}{\color{Green}\mathcal{A}_{v,g}},{\color{Green}B_{f,g}},{\color{Green}\overline{Z}_{H,g}},{\color{Green}\overline{Z}_{V,g}}\rangle.
\end{equation}
To construct this group, we have added stabilizers corresponding to the new measurement outcomes and removed stabilizers incompatible with the new stabilizers (namely, the single-site $\color{Green}Z_{e,g}$). Notice that the following operator belongs to $\mathcal{S}_2$:
\begin{equation}
    \prod_{v}\left(\zeta^{n_v}{\color{Green}\mathcal{A}_{v,g}}\right)=\zeta^{\sum_v n_v}{\color{Green}\hat C_g},
\end{equation}
where the total charge operator ${\color{Green}\hat C_g}$ was defined for an arbitrary color in Eq.~\ref{eq:charge-def}. As we show below, the non-trivial logical action of our protocol arises from the introduction of $\color{Green}\hat C_g$ into the stabilizer group.

To return to the $D(\mathbb{Z}_p^2)$ phase, we measure each green qudit in the $Z$ basis, obtaining measurement outcomes $\{{-n_e}\}$. The resultant stabilizer group is
\begin{equation}
\begin{split}
    \mathcal{S}_3=\langle {\color{red}\mathcal{A}_{v,r}},{\color{red}B_{f,r}},{\color{blue}\mathcal{A}_{v,b}},&{\color{blue}B_{f,b}},\zeta^{n_e}{\color{Green}Z_{e,g}},{\color{Green}B_{f,g}},\\ &\zeta^{\sum_v n_v}{\color{Green}\hat C_g},{\color{Green}\overline{Z}_{H,g}},{\color{Green}\overline{Z}_{V,g}}\rangle.
\end{split}
\end{equation}
Although each $\color{Green}\mathcal{A}_{v,g}$ fails to commute with some $\color{Green}Z_{e,g}$, the global product ${\color{Green}\hat C_g}$ commutes with all $\color{Green}Z_{e,g}$ and remains in the stabilizer group.

The code defined by $\mathcal{S}_3$ cannot yet be factorized into three decoupled layers. The $\color{Green}Z_{e,g}$ measurements leave the green layer in a product state which is in general distinct from $\bigotimes_e\color{Green}\ket{0}_{e,g}$, so $\mathcal{A}_{v,r/b}$ is not necessarily equal to ${A}_{v,r/b}$ on the codespace, in contrast to what we found for $\mathcal{S}_1$. However, because ${\color{Green}B_{f,g}}$, $\color{Green}\overline{Z}_{H,g}$, and $\color{Green}\overline{Z}_{V,g}$ belong to $\mathcal{S}_3$, the configuration in the green layer must be flat and homologically trivial. We may therefore apply a product of vertex operators $\color{Green}\mathcal{A}_{v,g}$ which ``corrects'' the green layer to the all-$\ket{0}$ state, leaving $\mathcal{S}_3$ invariant aside from mapping each $\zeta^{n_e}\color{Green}Z_{e,g}$ to $\color{Green}Z_{e,g}$.

\begin{figure}
  \centering
  $\vcenter{\hbox{

\begin{tikzpicture}[x=2.5cm,y=2.5cm]
  \path[use as bounding box,clip] (-1.5,-1) rectangle (1.5,1.4);

  \lattice{\cellB}{(0,-0.5773503)}
  \lattice{\cellR}{(0.5,-0.2886751)}
  \lattice{\cellG}{(0,0)}

  \draw[dash pattern={on 8pt off 5pt}, blue, line width=1pt] (-3,-\hh/6) -- (3,-\hh/6);
  \draw[dash pattern={on 8pt off 5pt}, red , line width=1pt] (-1,-7*\hh/3) -- (1,5*\hh/3);

  \foreach \y in {-1,0,1}{%
    \pgfmathtruncatemacro{\xmin}{ceil(-1.8-0.5*\y)}%
    \pgfmathtruncatemacro{\xmax}{floor(1.8-0.5*\y)}%
    \foreach \x in {\xmin,...,\xmax}{%
      \node[stamp] at (\x+0.5*\y,\hh*\y) {\usebox\czbox};}}

  \draw[line width=1.5pt,line cap=round]       (-30:0.28) to[bend left=20] ( 30:0.28);
  \draw[line width=1.5pt,line cap=round]       (210:0.28) to[bend left=20] (270:0.28);
  \draw[line width=1.5pt,line cap=round,brown] (270:0.28) to[bend left=20] (330:0.28);

  \begin{scope}[blue,line width=1.5pt,line cap=round]
    \foreach \xs in {0,1,2}{
      \draw[dirA] (-1.5+\xs,\hh/3) -- (-1.0+\xs,-2*\hh/3);
      \node[opl,fill=blue] at (-1.25+\xs,-\hh/6) {\scriptsize$-n$};
      \draw[dirA] (-1.0+\xs,-2*\hh/3) -- (-0.5+\xs,\hh/3);
      \node[op,fill=blue] at (-0.75+\xs,-\hh/6) {\scriptsize\fixw{${n}$}};}
  \end{scope}

  \begin{scope}[red,line width=1.5pt,line cap=round]
    \foreach \xs in {0,1,2}{
      \draw[dirA] (-1+\xs/2,-4*\hh/3+\hh*\xs) -- (\xs/2,-4*\hh/3+\hh*\xs);
      \node[op,fill=red] at (-0.5+\xs/2,-4*\hh/3+\hh*\xs) {\scriptsize\fixw{$m$}};
      \draw[dirA] (-0.5+\xs/2,-\hh/3+\hh*\xs) -- (\xs/2,-4*\hh/3+\hh*\xs);
      \node[op,fill=red] at (-0.25+\xs/2,-4*\hh/3+\hh*\xs+\hh/2) {\scriptsize\fixw{$m$}};}
  \end{scope}

\end{tikzpicture}}}$
  \caption{We take a representative state in the ${\color{red}\ket{\overline{m}}_{rV}}{\color{blue}\ket{\overline{n}}_{bH}}$ homology class, where all qudits not explicitly labeled are taken to be $\ket{0}$. We draw the three local operators which contribute a nontrivial phase to $\color{Green}\hat{C}_g$ in full saturation. These phases multiply to $\zeta^{-mn}$, producing an effective logical action $\overline{CZ}_{{\color{red}rV},{\color{blue}bH}}^\dagger$.}
  \label{fig:color-logs}
\end{figure}

After applying this correction, we arrive at the final stabilizer group
\begin{equation}
    \mathcal{S}_4=\langle {\color{red}A_{v,r}},{\color{red}B_{f,r}},{\color{blue}A_{v,b}},{\color{blue}B_{f,b}},{\color{Green}Z_{e,g}},\zeta^{\sum_v n_v}{\color{Green}\hat C_g}\rangle,
\end{equation}
which consists of the original stabilizer generators present in $\mathcal{S}_1$, along with the new generator $\zeta^{\sum_vn_v}{\color{Green}\hat C_g}$. In effect, we have measured the operator ${\color{Green}\hat C_g}=\overline{CZ}_{{\color{red}rV},{\color{blue}bH}}^\dagger\overline{CZ}_{{\color{red}rH},{\color{blue}bV}}$ and obtained the outcome $-\sum_v n_v$. 

Now, we explain how to leverage this measurement sequence to prepare magic states in the qudit toric code. Rather than an arbitrary initial logical state, suppose we begin with the logical state ${\color{red}\ket{+_p}_{rV}}{\color{blue}\ket{+_p}_{bH}}{\color{red}\ket{+_p}_{rH}}{\color{blue}\ket{+_p}_{bV}}$, where
\begin{equation}
    \ket{+_p}:=\frac{1}{\sqrt{p}}\sum_{n\in \mathbb{Z}_p}\ket{n}
\end{equation}
is the uniform superposition of computational basis states.
At the conclusion of the protocol, we will have projected this state into the eigenspace of $\color{Green}\hat C_g$ with eigenvalue $\zeta^{-\sum_v n_v}$. 

Let $m:=-\sum_v n_v$, so that the measured eigenvalue of $\color{Green}\hat C_g$ becomes $\zeta^m$. Projecting the specified logical state into this eigenspace, we obtain the output logical state
\begin{equation}
    \ket{\overline{\mathrm{out}}}\propto\sum_{\{s_i\}\in\mathbb{F}_p}{\color{red}\ket{s_1}_{rV}}{\color{blue}\ket{s_2}_{bH}}{\color{red}\ket{s_3}_{rH}}{\color{blue}\ket{s_4}_{bV}}\delta_{-s_1s_2+s_3s_4,m}.
\end{equation}
Notice that we may write
\begin{equation}
\ket{\overline{\mathrm{out}}}\propto\sum_{n}\sqrt{N_nN_{m+n}}\ket{\overline{CZ}_n}_{{\color{red}rV},{\color{blue}bH}}\ket{\overline{CZ}_{m+n}}_{{\color{red}rH},{\color{blue}bV}},
\end{equation}
where $\ket{CZ_m}$ is the uniform superposition in the $\zeta^m$ eigenspace of $CZ$:
\begin{equation}
    \ket{CZ_m}:=\frac{1}{\sqrt{N_m}}\sum_{\substack{j,k\in\mathbb{F}_p:\\jk=m}}\ket{j,k},
\end{equation}
and $N_m$ is the size of the computational basis support of $\ket{CZ_m}$:
\begin{equation}
    N_m=\begin{cases}
        2p-1,&m=0\\
        p-1,&m\neq 0
    \end{cases}.
\end{equation}
Recall that a prime qudit Pauli stabilizer state can be expressed as a sum over an affine subspace of $\mathbb{F}_p^n$~\cite{HostensEtAl_StabilizerStatesClifford_05a}, which must contain $p^k$ elements, for some $k$. Because $N_m$ is not a power of $p$ for any $m$ (when $p$ is taken to be an odd prime), we may conclude that all of the $\ket{CZ_m}$ are magic states. Note that by the same counting argument, we may conclude that $\ket{\overline{\text{out}}}$ is a logical magic state.

We may also construct a logical magic state on the ${\color{red}rV},{\color{blue}bH}$ logical subsystem, for instance, by measuring the ${\color{red}rH},{\color{blue}bV}$ subsystem in the logical $Z$ basis. Regardless of the outcome of this measurement, the result is a magic state on ${\color{red}rV},{\color{blue}bH}$. We emphasize that this result is distinct from the qubit case, where an analogous protocol yields a logical magic state only for certain outcomes of the final $Z$-basis measurement~\cite{DavydovaEtAl_UniversalFaultTolerant_25}.

\section{Discussion and Outlook}\label{sec:discussion}

We have constructed a formalism for classifying and characterizing \xkcd stabilizer models on prime-dimensional qudits, which are assembled from Pauli-$X$ operators and diagonal operators drawn from a fixed level $k$ of the Clifford hierarchy. After introducing the general structure of an \xkcd stabilizer code, we built Theorem~\ref{thm:unique_state}, which strongly constrains the form of unique codewords of \xkcd codes. We then turned to abelian twisted quantum double models, which can naturally be interpreted as \xkcd stabilizer codes. We discussed bounds on $k$ in these models, which culminated in Conjecture~\ref{conj:level-k-TQD}. Finally, we constructed a protocol for deterministic magic state preparation in prime qudit toric codes based on measuring \xkcd stabilizers.

In defining \xkcd codes, we allowed ourselves tensor products of Pauli-$X$ operators and many-body diagonal Clifford-hierarchy gates. 
However, it may be of interest in practical applications to consider \xkcd codes where the diagonal Clifford-hierarchy gates are restricted to be tensor products of single-site Clifford-hierarchy gates. We call such a code an \textit{on-site \xkcd code}. We discuss an analogous construction for on-site \xkcd codes and its relation to our primary construction in App.~\ref{app:on_site_clifford}. 

While there is some constraint on the support of the diagonal Clifford-hierarchy gates used in our construction from fixing the level $k$ (if $D_m[f]$ is an $n$-body gate, then $\mathrm{deg}(f)=n$ and this gate is at least level-$n$), we did not impose any notion of geometric locality on our stabilizers. This implies that our results are readily applicable to codes in higher dimensions and quantum low-density parity-check codes, which relax the notion of geometric locality altogether. In particular, we expect that the 3-dimensional twisted (abelian) quantum triple~\cite{WanEtAl_TwistedGaugeTheory_15,BullivantDelcamp_TubeAlgebrasExcitations_19,HuxfordEtAl_TwistedLatticeGauge_24} and twisted fracton models~\cite{SongEtAl_TwistedFractonModels_19} would be amenable to study in the \xkcd formalism, as they are formally quite similar to twisted quantum double models. 

We note several other interesting open directions. We have restricted ourselves to prime-dimensional qudits, where the Clifford-hierarchy is better understood, but we expect that there could be a suitable extension to composite-dimensional qudits. It would also be of practical value to determine the computational complexity of various problems based on \xkcd codes, such as computing the dimension of the codespace or expectation values of local operators in a stabilizer state. Ref.~\cite{nest2011monomial} showed that estimating local expectation values is NP-hard for general monomial stabilizer states, while identifying a subclass that can be simulated efficiently. Whether \xkcd stabilizer states fall into this subclass, and whether the additional Clifford-hierarchy structure makes further properties efficiently computable, remains open.

Finally we discuss the application of our results to the types of logical gates realizable by \xkcd codes. We first conjecture a generalized Bravyi-Koenig bound motivated by our consideration of Clifford-hierarchy stabilizer codes.
\begin{conjecture}
Locality preserving unitary gates on level-$k$ \xkcd codes in $d$-dimensions are restricted to lie in level $d+k-1$ of the Clifford hierarchy.
\end{conjecture}

An important open problem for our protocol and those of Refs.~\cite{KobayashiEtAl_CliffordHierarchyStabilizer_25, WarmanSchafer-Nameki_TransversalCliffordHierarchyGates_25,ManjunathEtAl_UniversalQuantumComputation_26} is to establish fault tolerance including code switching steps. Recently introduced efficient classical simulation methods for Clifford stabilizer codes are well-suited to this task~\cite{Surti_2026,takada2026exactefficientsimulationnoisy,delafuente2026efficientlyestimatingfailurerates,delafuente2026highthresholddecodingnonpaulicodes}. Related to this, it is an interesting problem to extend the static \xkcd codes to define spacetime fault-complexes~\cite{Gottesman_OpportunitiesChallengesFaultTolerant_22,BombinEtAl_LogicalBlocksFaultTolerant_23,GidneyBacon_LessBaconMore_23,DelfossePaetznick_SpacetimeCodesClifford_23,HillmannEtAl_SingleshotMeasurementbasedQuantum_25,delafuente2026highthresholddecodingnonpaulicodes}. 
Finally, it is interesting to explore whether \xkcd codes with abelian anyons can be used to implement higher-level Clifford gates or not. More generally, what are the necessary requirements for an \xkcd code to provide protected higher-level Clifford gates?

We leave these questions to future work.

\acknowledgments
D.J.W. thanks Amelie Skelton and Edmund Heng for useful discussions. C.F., K.C., and M.R. thank the Joint Quantum Institute at the University of Maryland for support through a JQI fellowship. C.F. was supported in part by the NSF STAQ program (awards No.~1818914 and No.~2325080). D.J.W. is supported by the Australian Research Council Discovery Early Career Research Award (DE220100625). K.K. was supported by the MathQuantum RTG which is funded by NSF DMS 2231533. S.M. was supported by the Laboratory for Physical Sciences through the Condensed Matter Theory Center. 

This manuscript was prepared with the assistance of OpenAI’s GPT-5.5, GPT-5.6, and GPT-6 as well as Anthropic’s Claude Opus 5, Opus 5.5, and Fable 5. These tools were used to check derivations, provide editorial feedback, and assist in developing proofs in accordance with the author’s instructions. All content, claims, and conclusions have been reviewed and verified by the authors to ensure accuracy and originality. Certain products, commercial and otherwise, are mentioned in this publication. These mentions are for informational purposes only, and do not imply recommendation or endorsement by NIST.

\bibliographystyle{apsrev4-2_custom}
\bibliography{clifford_codes.bib}

@article{AharonovBen-Or_FaultTolerantQuantumComputation_08a,
  title = {Fault-{{Tolerant Quantum Computation}} with {{Constant Error Rate}}},
  author = {Aharonov, Dorit and {Ben-Or}, Michael},
  year = 2008,
  month = jan,
  journal = {SIAM Journal on Computing},
  volume = {38},
  number = {4},
  pages = {1207--1282},
  issn = {0097-5397, 1095-7111},
  doi = {10.1137/S0097539799359385},
  urldate = {2026-03-04},
  langid = {english}
}

@article{AndersonEtAl_FaultTolerantConversionSteane_14,
  title = {Fault-{{Tolerant Conversion}} between the {{Steane}} and {{Reed-Muller Quantum Codes}}},
  author = {Anderson, Jonas T. and {Duclos-Cianci}, Guillaume and Poulin, David},
  year = 2014,
  month = aug,
  journal = {Physical Review Letters},
  volume = {113},
  number = {8},
  pages = {080501},
  publisher = {American Physical Society},
  doi = {10.1103/PhysRevLett.113.080501},
  urldate = {2026-03-07}
}

@article{AshikhminKnill_NonbinaryQuantumStabilizer_01,
  title = {Nonbinary Quantum Stabilizer Codes},
  author = {Ashikhmin, A. and Knill, E.},
  year = 2001,
  month = nov,
  journal = {IEEE Transactions on Information Theory},
  volume = {47},
  number = {7},
  pages = {3065--3072},
  issn = {1557-9654},
  doi = {10.1109/18.959288},
  urldate = {2026-03-04}
}

@article{BarkeshliEtAl_SymmetryFractionalizationDefects_19,
  title = {Symmetry Fractionalization, Defects, and Gauging of Topological Phases},
  author = {Barkeshli, Maissam and Bonderson, Parsa and Cheng, Meng and Wang, Zhenghan},
  year = 2019,
  month = sep,
  journal = {Physical Review B},
  volume = {100},
  number = {11},
  pages = {115147},
  issn = {2469-9950, 2469-9969},
  doi = {10.1103/PhysRevB.100.115147},
  urldate = {2023-05-11}
}

@article{BarkeshliEtAl_TwistDefectsProjective_13,
  title = {Twist Defects and Projective Non-{{Abelian}} Braiding Statistics},
  author = {Barkeshli, Maissam and Jian, Chao-Ming and Qi, Xiao-Liang},
  year = 2013,
  month = jan,
  journal = {Physical Review B},
  volume = {87},
  number = {4},
  pages = {045130},
  publisher = {American Physical Society},
  doi = {10.1103/PhysRevB.87.045130},
  urldate = {2026-03-07}
}

@article{BeigiEtAl_QuantumDoubleModel_11,
  title = {The {{Quantum Double Model}} with {{Boundary}}: {{Condensations}} and {{Symmetries}}},
  shorttitle = {The {{Quantum Double Model}} with {{Boundary}}},
  author = {Beigi, Salman and Shor, Peter W. and Whalen, Daniel},
  year = 2011,
  month = sep,
  journal = {Communications in Mathematical Physics},
  volume = {306},
  number = {3},
  pages = {663--694},
  issn = {1432-0916},
  doi = {10.1007/s00220-011-1294-x},
  urldate = {2026-03-04},
  langid = {english}
}

@misc{Bombin_2DQuantumComputation_18,
  title = {{{2D}} Quantum Computation with {{3D}} Topological Codes},
  author = {Bomb\'in, Hector},
  year = 2018,
  publisher = {arXiv},
  doi = {10.48550/ARXIV.1810.09571},
  urldate = {2026-06-23},
  copyright = {arXiv.org perpetual, non-exclusive license}
}

@article{Bombin_DimensionalJumpQuantum_16,
  title = {Dimensional Jump in Quantum Error Correction},
  author = {Bomb{\'i}n, H{\'e}ctor},
  year = 2016,
  month = apr,
  journal = {New Journal of Physics},
  volume = {18},
  number = {4},
  pages = {043038},
  publisher = {IOP Publishing},
  issn = {1367-2630},
  doi = {10.1088/1367-2630/18/4/043038},
  urldate = {2026-03-07},
  langid = {english}
}

@article{Bombin_IntroductionTopologicalQuantum_13,
  title = {An {{Introduction}} to {{Topological Quantum Codes}}},
  author = {Bomb\'in, H.},
  year = 2013,
  publisher = {arXiv},
  doi = {10.48550/ARXIV.1311.0277},
  urldate = {2026-03-06},
  copyright = {arXiv.org perpetual, non-exclusive license}
}

@article{Bombin_TopologicalOrderTwist_10,
  title = {Topological {{Order}} with a {{Twist}}: {{Ising Anyons}} from an {{Abelian Model}}},
  shorttitle = {Topological {{Order}} with a {{Twist}}},
  author = {Bomb\'in, H.},
  year = 2010,
  month = jul,
  journal = {Physical Review Letters},
  volume = {105},
  number = {3},
  pages = {030403},
  issn = {0031-9007, 1079-7114},
  doi = {10.1103/PhysRevLett.105.030403},
  urldate = {2026-03-05},
  copyright = {http://link.aps.org/licenses/aps-default-license},
  langid = {english}
}

@article{BombinEtAl_LogicalBlocksFaultTolerant_23,
  title = {Logical {{Blocks}} for {{Fault-Tolerant Topological Quantum Computation}}},
  author = {Bomb{\'i}n, H{\'e}ctor and Dawson, Chris and Mishmash, Ryan V. and Nickerson, Naomi and Pastawski, Fernando and Roberts, Sam},
  year = 2023,
  month = apr,
  journal = {PRX Quantum},
  volume = {4},
  number = {2},
  pages = {020303},
  issn = {2691-3399},
  doi = {10.1103/PRXQuantum.4.020303},
  urldate = {2026-06-30},
  langid = {english}
}

@article{BombinMartin-Delgado_TopologicalQuantumDistillation_06a,
  title = {Topological {{Quantum Distillation}}},
  author = {Bomb\'in, H. and {Martin-Delgado}, M. A.},
  year = 2006,
  month = oct,
  journal = {Physical Review Letters},
  volume = {97},
  number = {18},
  pages = {180501},
  publisher = {American Physical Society},
  doi = {10.1103/PhysRevLett.97.180501},
  urldate = {2026-03-07}
}

@misc{BravyiKitaev_QuantumCodesLattice_98,
  title = {Quantum Codes on a Lattice with Boundary},
  author = {Bravyi, S. B. and Kitaev, A. Yu},
  year = 1998,
  month = nov,
  number = {arXiv:quant-ph/9811052},
  eprint = {quant-ph/9811052},
  publisher = {arXiv},
  doi = {10.48550/arXiv.quant-ph/9811052},
  urldate = {2026-03-05},
  archiveprefix = {arXiv}
}

@article{BravyiKoenig_ClassificationTopologicallyProtected_13,
  title = {Classification of Topologically Protected Gates for Local Stabilizer Codes},
  author = {Bravyi, Sergey and Koenig, Robert},
  year = 2013,
  month = apr,
  journal = {Physical Review Letters},
  volume = {110},
  number = {17},
  eprint = {1206.1609},
  primaryclass = {quant-ph},
  pages = {170503},
  issn = {0031-9007, 1079-7114},
  doi = {10.1103/PhysRevLett.110.170503},
  urldate = {2026-03-05},
  archiveprefix = {arXiv}
}

@article{Brown_ConservationLawsQuantum_22,
  title = {Conservation {{Laws}} and {{Quantum Error Correction}}: {{Toward}} a {{Generalized Matching Decoder}}},
  shorttitle = {Conservation {{Laws}} and {{Quantum Error Correction}}},
  author = {Brown, Benjamin J.},
  year = 2022,
  month = dec,
  journal = {IEEE BITS the Information Theory Magazine},
  volume = {2},
  number = {3},
  pages = {5--19},
  issn = {2692-4110},
  doi = {10.1109/MBITS.2023.3246025},
  urldate = {2026-03-04}
}

@article{Brown_FaulttolerantNonCliffordGate_20,
  title = {A Fault-Tolerant Non-{{Clifford}} Gate for the Surface Code in Two Dimensions},
  author = {Brown, Benjamin J.},
  year = 2020,
  month = may,
  journal = {Science Advances},
  volume = {6},
  number = {21},
  pages = {eaay4929},
  issn = {2375-2548},
  doi = {10.1126/sciadv.aay4929},
  urldate = {2026-06-23},
  langid = {english}
}

@article{BrownEtAl_PokingHolesCutting_17,
  title = {Poking {{Holes}} and {{Cutting Corners}} to {{Achieve Clifford Gates}} with the {{Surface Code}}},
  author = {Brown, Benjamin J. and Laubscher, Katharina and Kesselring, Markus S. and Wootton, James R.},
  year = 2017,
  month = may,
  journal = {Physical Review X},
  volume = {7},
  number = {2},
  pages = {021029},
  issn = {2160-3308},
  doi = {10.1103/PhysRevX.7.021029},
  urldate = {2026-03-03},
  copyright = {https://creativecommons.org/licenses/by/4.0/},
  langid = {english}
}

@article{BullivantDelcamp_TubeAlgebrasExcitations_19,
  title = {Tube Algebras, Excitations Statistics and Compactification in Gauge Models of Topological Phases},
  author = {Bullivant, Alex and Delcamp, Clement},
  year = 2019,
  month = oct,
  journal = {Journal of High Energy Physics},
  volume = {2019},
  number = {10},
  eprint = {1905.08673},
  primaryclass = {cond-mat},
  pages = {216},
  issn = {1029-8479},
  doi = {10.1007/JHEP10(2019)216},
  urldate = {2025-07-08},
  archiveprefix = {arXiv}
}

@article{Burnell_AnyonCondensationIts_18,
  title = {Anyon {{Condensation}} and {{Its Applications}}},
  author = {Burnell, F. J.},
  year = 2018,
  month = mar,
  journal = {Annual Review of Condensed Matter Physics},
  volume = {9},
  number = {1},
  pages = {307--327},
  issn = {1947-5454, 1947-5462},
  doi = {10.1146/annurev-conmatphys-033117-054154},
  urldate = {2023-11-01}
}

@article{CosteEtAl_FiniteGroupModular_00,
  title = {Finite Group Modular Data},
  author = {Coste, Antoine and Gannon, Terry and Ruelle, Philippe},
  year = 2000,
  month = aug,
  journal = {Nuclear Physics B},
  volume = {581},
  number = {3},
  pages = {679--717},
  issn = {05503213},
  doi = {10.1016/S0550-3213(00)00285-6},
  urldate = {2026-03-17},
  copyright = {https://www.elsevier.com/tdm/userlicense/1.0/},
  langid = {english}
}

@article{CuiEtAl_DiagonalGatesClifford_17,
  title = {Diagonal Gates in the {{Clifford}} Hierarchy},
  author = {Cui, Shawn X. and Gottesman, Daniel and Krishna, Anirudh},
  year = 2017,
  month = jan,
  journal = {Physical Review A},
  volume = {95},
  number = {1},
  pages = {012329},
  issn = {2469-9926, 2469-9934},
  doi = {10.1103/PhysRevA.95.012329},
  urldate = {2026-03-03},
  copyright = {http://link.aps.org/licenses/aps-default-license},
  langid = {english}
}

@article{DavydovaEtAl_UniversalFaultTolerant_25,
  title={Universal Fault-Tolerant Quantum Computation in 2D without Getting Tied in Knots},
  author={Davydova, Margarita and Bauer, Andreas and Magdalena de la Fuente, Julio C and Webster, Mark and Williamson, Dominic J and Brown, Benjamin J},
  journal={Physical Review X},
  volume={16},
  number={3},
  pages={031001},
  year={2026},
  publisher={APS}
}

@article{DehaeneDeMoor_CliffordGroupStabilizer_03,
  title = {Clifford Group, Stabilizer States, and Linear and Quadratic Operations over {{GF}}(2)},
  author = {Dehaene, Jeroen and De Moor, Bart},
  year = 2003,
  month = oct,
  journal = {Physical Review A},
  volume = {68},
  number = {4},
  pages = {042318},
  issn = {1050-2947, 1094-1622},
  doi = {10.1103/PhysRevA.68.042318},
  urldate = {2026-06-23},
  copyright = {http://link.aps.org/licenses/aps-default-license},
  langid = {english}
}

@misc{DelfossePaetznick_SpacetimeCodesClifford_23,
  title = {Spacetime Codes of {{Clifford}} Circuits},
  author = {Delfosse, Nicolas and Paetznick, Adam},
  year = 2023,
  month = may,
  number = {arXiv:2304.05943},
  eprint = {2304.05943},
  primaryclass = {quant-ph},
  publisher = {arXiv},
  doi = {10.48550/arXiv.2304.05943},
  urldate = {2026-07-01},
  archiveprefix = {arXiv}
}

@article{DennisEtAl_TopologicalQuantumMemory_02,
  title = {Topological Quantum Memory},
  author = {Dennis, Eric and Kitaev, Alexei and Landahl, Andrew and Preskill, John},
  year = 2002,
  month = sep,
  journal = {Journal of Mathematical Physics},
  volume = {43},
  number = {9},
  pages = {4452--4505},
  issn = {0022-2488, 1089-7658},
  doi = {10.1063/1.1499754},
  urldate = {2026-03-03},
  langid = {english}
}

@article{deSilvaLautsch_CliffordHierarchyOne_25,
  title={The Clifford hierarchy for one qubit or qudit},
  author={de Silva, Nadish and Lautsch, Oscar},
  journal={Proceedings of the Royal Society A},
  volume={481},
  number={2324},
  pages={20250035},
  year={2025},
  publisher={The Royal Society}
}

@misc{deWildPropitius_TopologicalInteractionsBroken_95,
  title = {Topological Interactions in Broken Gauge Theories},
  author = {{de Wild Propitius}, Mark},
  year = 1995,
  eprint = {hep-th/9511195},
  archiveprefix = {arXiv}
}

@article{DijkgraafWitten_TopologicalGaugeTheories_90,
  title = {Topological Gauge Theories and Group Cohomology},
  author = {Dijkgraaf, Robbert and Witten, Edward},
  year = 1990,
  month = apr,
  journal = {Communications in Mathematical Physics},
  volume = {129},
  number = {2},
  pages = {393--429},
  issn = {0010-3616, 1432-0916},
  doi = {10.1007/BF02096988},
  urldate = {2026-06-23},
  copyright = {http://www.springer.com/tdm},
  langid = {english}
}

@article{DrinfeldEtAl_BraidedFusionCategories_10,
  title={On braided fusion categories I},
  author={Drinfeld, Vladimir and Gelaki, Shlomo and Nikshych, Dmitri and Ostrik, Victor},
  journal={Selecta Mathematica},
  volume={16},
  number={1},
  pages={1--119},
  year={2010},
  publisher={Springer}
}

@article{EastinKnill_RestrictionsTransversalEncoded_09,
  title = {Restrictions on {{Transversal Encoded Quantum Gate Sets}}},
  author = {Eastin, Bryan and Knill, Emanuel},
  year = 2009,
  month = mar,
  journal = {Physical Review Letters},
  volume = {102},
  number = {11},
  pages = {110502},
  publisher = {American Physical Society},
  doi = {10.1103/PhysRevLett.102.110502},
  urldate = {2026-03-05}
}

@article{EllisonEtAl_PauliStabilizerModels_22,
  title = {Pauli {{Stabilizer Models}} of {{Twisted Quantum Doubles}}},
  author = {Ellison, Tyler D. and Chen, Yu-An and Dua, Arpit and Shirley, Wilbur and Tantivasadakarn, Nathanan and Williamson, Dominic J.},
  year = 2022,
  month = mar,
  journal = {PRX Quantum},
  volume = {3},
  number = {1},
  pages = {010353},
  issn = {2691-3399},
  doi = {10.1103/PRXQuantum.3.010353},
  urldate = {2025-08-04}
}

@misc{GidneyBacon_LessBaconMore_23,
  title = {Less {{Bacon More Threshold}}},
  author = {Gidney, Craig and Bacon, Dave},
  year = 2023,
  month = may,
  number = {arXiv:2305.12046},
  eprint = {2305.12046},
  primaryclass = {quant-ph},
  publisher = {arXiv},
  doi = {10.48550/arXiv.2305.12046},
  urldate = {2026-06-30},
  archiveprefix = {arXiv}
}

@misc{Gottesman_OpportunitiesChallengesFaultTolerant_22,
  title = {Opportunities and {{Challenges}} in {{Fault-Tolerant Quantum Computation}}},
  author = {Gottesman, Daniel},
  year = 2022,
  month = oct,
  number = {arXiv:2210.15844},
  eprint = {2210.15844},
  primaryclass = {quant-ph},
  publisher = {arXiv},
  doi = {10.48550/arXiv.2210.15844},
  urldate = {2026-06-30},
  archiveprefix = {arXiv}
}

@book{Gottesman_StabilizerCodesQuantum_1997,
  title = {Stabilizer Codes and Quantum Error Correction},
  author = {Gottesman, Daniel},
  year = 1997,
  publisher = {California Institute of Technology}
}

@article{GottesmanChuang_DemonstratingViabilityUniversal_99,
  title = {Demonstrating the Viability of Universal Quantum Computation Using Teleportation and Single-Qubit Operations},
  author = {Gottesman, Daniel and Chuang, Isaac L.},
  year = 1999,
  month = nov,
  journal = {Nature},
  volume = {402},
  number = {6760},
  pages = {390--393},
  issn = {0028-0836, 1476-4687},
  doi = {10.1038/46503},
  urldate = {2026-03-07},
  copyright = {http://www.springer.com/tdm},
  langid = {english}
}

@article{Haah_ClassificationTranslationInvariant_21,
  title = {Classification of Translation Invariant Topological {{Pauli}} Stabilizer Codes for Prime Dimensional Qudits on Two-Dimensional Lattices},
  author = {Haah, Jeongwan},
  year = 2021,
  month = jan,
  journal = {Journal of Mathematical Physics},
  volume = {62},
  number = {1},
  pages = {012201},
  issn = {0022-2488, 1089-7658},
  doi = {10.1063/5.0021068},
  urldate = {2026-03-03},
  langid = {english}
}

@article{HastingsGeller_ReducedSpacetimeTime_15,
  title = {Reduced Space-Time and Time Costs {{Using}} Dislocation Codes and Arbitrary Ancillas},
  author = {Hastings, Matthew B. and Geller, A.},
  year = 2015,
  month = sep,
  journal = {Quantum Information and Computation},
  volume = {15},
  number = {11\&12},
  pages = {962--986},
  issn = {15337146, 15337146},
  doi = {10.26421/QIC15.11-12-6},
  urldate = {2026-03-05}
}

@article{HiggottGidney_SparseBlossomCorrecting_25,
  title = {Sparse {{Blossom}}: Correcting a Million Errors per Core Second with Minimum-Weight Matching},
  shorttitle = {Sparse {{Blossom}}},
  author = {Higgott, Oscar and Gidney, Craig},
  year = 2025,
  month = jan,
  journal = {Quantum},
  volume = {9},
  pages = {1600},
  publisher = {Verein zur F\"orderung des Open Access Publizierens in den Quantenwissenschaften},
  doi = {10.22331/q-2025-01-20-1600},
  urldate = {2026-03-04},
  langid = {british}
}

@article{HillmannEtAl_SingleshotMeasurementbasedQuantum_25,
  title = {Single-Shot and Measurement-Based Quantum Error Correction via Fault Complexes},
  author = {Hillmann, Timo and Dauphinais, Guillaume and Tzitrin, Ilan and Vasmer, Michael},
  year = 2025,
  month = oct,
  journal = {Physical Review A},
  volume = {112},
  number = {4},
  pages = {L040401},
  publisher = {American Physical Society},
  doi = {10.1103/cjb4-l57n},
  urldate = {2026-07-01}
}

@article{HostensEtAl_StabilizerStatesClifford_05a,
  title = {Stabilizer States and {{Clifford}} Operations for Systems of Arbitrary Dimensions and Modular Arithmetic},
  author = {Hostens, Erik and Dehaene, Jeroen and De Moor, Bart},
  year = 2005,
  month = apr,
  journal = {Physical Review A},
  volume = {71},
  number = {4},
  pages = {042315},
  publisher = {American Physical Society},
  doi = {10.1103/PhysRevA.71.042315},
  urldate = {2025-11-14}
}

@misc{HuangChen_GeneratingLogicalMagic_25,
  title = {Generating Logical Magic States with the Aid of Non-{{Abelian}} Topological Order},
  author = {Huang, Sheng-Jie and Chen, Yanzhu},
  year = 2025,
  month = jul,
  number = {arXiv:2502.00998},
  eprint = {2502.00998},
  primaryclass = {quant-ph},
  publisher = {arXiv},
  doi = {10.48550/arXiv.2502.00998},
  urldate = {2026-06-25},
  archiveprefix = {arXiv}
}

@article{HuEtAl_TwistedQuantumDouble_13,
  title = {Twisted {{Quantum Double Model}} of {{Topological Phases}} in {{Two--Dimension}}},
  author = {Hu, Yuting and Wan, Yidun and Wu, Yong-Shi},
  year = 2013,
  month = mar,
  eprint = {1211.3695},
  primaryclass = {cond-mat},
  doi = {10.1103/PhysRevB.87.125114},
  urldate = {2025-07-16},
  archiveprefix = {arXiv}
}

@article{HuxfordEtAl_TwistedLatticeGauge_24,
  title={Twisted lattice gauge theory: Membrane operators, three-loop braiding, and topological charge},
  author={Huxford, Joe and Nguyen, Dung Xuan and Kim, Yong Baek},
  journal={Physical Review B},
  volume={110},
  number={3},
  pages={035117},
  year={2024},
  publisher={APS}
}

@article{KetkarEtAl_NonbinaryStabilizerCodes_06,
  title = {Nonbinary {{Stabilizer Codes Over Finite Fields}}},
  author = {Ketkar, A. and Klappenecker, A. and Kumar, S. and Sarvepalli, P. K.},
  year = 2006,
  month = nov,
  journal = {IEEE Transactions on Information Theory},
  volume = {52},
  number = {11},
  pages = {4892--4914},
  issn = {1557-9654},
  doi = {10.1109/TIT.2006.883612},
  urldate = {2026-03-04}
}

@article{Kitaev_FaulttolerantQuantumComputation_03,
  title = {Fault-Tolerant Quantum Computation by Anyons},
  author = {Kitaev, {\relax A.Yu}.},
  year = 2003,
  month = jan,
  journal = {Annals of Physics},
  volume = {303},
  number = {1},
  pages = {2--30},
  issn = {00034916},
  doi = {10.1016/S0003-4916(02)00018-0},
  urldate = {2026-03-03},
  copyright = {https://www.elsevier.com/tdm/userlicense/1.0/},
  langid = {english}
}

@article{KitaevKong_ModelsGappedBoundaries_12,
  title = {Models for {{Gapped Boundaries}} and {{Domain Walls}}},
  author = {Kitaev, Alexei and Kong, Liang},
  year = 2012,
  month = jul,
  journal = {Communications in Mathematical Physics},
  volume = {313},
  number = {2},
  pages = {351--373},
  issn = {1432-0916},
  doi = {10.1007/s00220-012-1500-5},
  urldate = {2026-03-07},
  langid = {english}
}

@misc{Knill_NonbinaryUnitaryError_96,
  title = {Non-Binary {{Unitary Error Bases}} and {{Quantum Codes}}},
  author = {Knill, E.},
  year = 1996,
  month = oct,
  number = {arXiv:quant-ph/9608048},
  eprint = {quant-ph/9608048},
  publisher = {arXiv},
  doi = {10.48550/arXiv.quant-ph/9608048},
  urldate = {2026-03-04},
  archiveprefix = {arXiv}
}

@article{KobayashiEtAl_CliffordHierarchyStabilizer_25,
  title={Clifford hierarchy stabilizer codes: transversal non-Clifford gates and magic states},
  author={Kobayashi, Ryohei and Zhu, Guanyu and Hsin, Po-Shen},
  journal={Physical Review Letters},
  volume={136},
  number={25},
  pages={250802},
  year={2026},
  publisher={APS}
}

@article{KoenigEtAl_QuantumComputationTuraev_10,
  title = {Quantum Computation with {{Turaev}}--{{Viro}} Codes},
  author = {Koenig, Robert and Kuperberg, Greg and Reichardt, Ben W.},
  year = 2010,
  month = dec,
  journal = {Annals of Physics},
  volume = {325},
  number = {12},
  pages = {2707--2749},
  issn = {0003-4916},
  doi = {10.1016/j.aop.2010.08.001},
  urldate = {2026-03-07}
}

@article{KubicaBeverland_UniversalTransversalGates_15,
  title = {Universal Transversal Gates with Color Codes: {{A}} Simplified Approach},
  shorttitle = {Universal Transversal Gates with Color Codes},
  author = {Kubica, Aleksander and Beverland, Michael E.},
  year = 2015,
  month = mar,
  journal = {Physical Review A},
  volume = {91},
  number = {3},
  pages = {032330},
  publisher = {American Physical Society},
  doi = {10.1103/PhysRevA.91.032330},
  urldate = {2026-03-07}
}

@article{LaubscherEtAl_UniversalQuantumComputation_19,
  title = {Universal Quantum Computation in the Surface Code Using Non-{{Abelian}} Islands},
  author = {Laubscher, Katharina and Loss, Daniel and Wootton, James R.},
  year = 2019,
  month = jul,
  journal = {Physical Review A},
  volume = {100},
  number = {1},
  eprint = {1811.06738},
  primaryclass = {quant-ph},
  pages = {012338},
  issn = {2469-9926, 2469-9934},
  doi = {10.1103/PhysRevA.100.012338},
  urldate = {2026-06-25},
  archiveprefix = {arXiv}
}

@book{LidlNiederreiterFiniteFields,
  title = {Finite Fields},
  author = {Lidl, Rudolf and Niederreiter, Harald},
  year = 2008,
  series = {Encyclopedia of Mathematics and Its Applications},
  edition = {2},
  volume = {20},
  publisher = {Cambridge University Press},
  address = {Cambridge},
  doi = {10.1017/CBO9780511525926},
  isbn = {978-0-521-06567-2}
}

@misc{ManjunathEtAl_UniversalQuantumComputation_26,
  title = {Universal Quantum Computation with Group Surface Codes},
  author = {Manjunath, Naren and Mattei, Vieri and Tiwari, Apoorv and Ellison, Tyler D.},
  year = 2026,
  month = mar,
  number = {arXiv:2603.05502},
  eprint = {2603.05502},
  primaryclass = {quant-ph},
  publisher = {arXiv},
  doi = {10.48550/arXiv.2603.05502},
  urldate = {2026-03-06},
  archiveprefix = {arXiv}
}

@article{NaiduEtAl_FusionSubcategoriesRepresentation_09,
  title={Fusion subcategories of representation categories of twisted quantum doubles of finite groups},
  author={Naidu, Deepak and Nikshych, Dmitri and Witherspoon, Sarah},
  journal={International Mathematics Research Notices},
  volume={2009},
  number={22},
  pages={4183--4219},
  year={2009},
  publisher={OUP}
}

@article{NiEtAl_NonCommutingStabilizerFormalism_15,
  title = {A {{Non-Commuting Stabilizer Formalism}}},
  author = {Ni, Xiaotong and Buerschaper, Oliver and {Van den Nest}, Maarten},
  year = 2015,
  month = may,
  journal = {Journal of Mathematical Physics},
  volume = {56},
  number = {5},
  eprint = {1404.5327},
  primaryclass = {quant-ph},
  pages = {052201},
  issn = {0022-2488, 1089-7658},
  doi = {10.1063/1.4920923},
  urldate = {2025-09-23},
  archiveprefix = {arXiv}
}

@article{PotterVasseur_SymmetryConstraintsManybody_16,
  title = {Symmetry Constraints on Many-Body Localization},
  author = {Potter, Andrew C. and Vasseur, Romain},
  year = 2016,
  month = dec,
  journal = {Physical Review B},
  volume = {94},
  number = {22},
  pages = {224206},
  publisher = {American Physical Society},
  doi = {10.1103/PhysRevB.94.224206},
  urldate = {2026-05-28}
}

@article{RaussendorfHarrington_FaultTolerantQuantumComputation_07,
  title = {Fault-{{Tolerant Quantum Computation}} with {{High Threshold}} in {{Two Dimensions}}},
  author = {Raussendorf, Robert and Harrington, Jim},
  year = 2007,
  month = may,
  journal = {Physical Review Letters},
  volume = {98},
  number = {19},
  pages = {190504},
  issn = {0031-9007, 1079-7114},
  doi = {10.1103/PhysRevLett.98.190504},
  urldate = {2026-03-05},
  copyright = {http://link.aps.org/licenses/aps-default-license},
  langid = {english}
}

@article{ScrubyEtAl_NumericalImplementationJustInTime_22,
  title = {Numerical {{Implementation}} of {{Just-In-Time Decoding}} in {{Novel Lattice Slices Through}} the {{Three-Dimensional Surface Code}}},
  author = {Scruby, T. R. and Browne, D. E. and Webster, P. and Vasmer, M.},
  year = 2022,
  month = may,
  journal = {Quantum},
  volume = {6},
  pages = {721},
  issn = {2521-327X},
  doi = {10.22331/q-2022-05-24-721},
  urldate = {2026-06-23},
  langid = {english}
}

@book{Simon_TopologicalQuantum_23,
  title = {Topological Quantum},
  author = {Simon, Steven H.},
  year = 2023,
  publisher = {Oxford University Press},
  address = {Oxford},
  isbn = {978-0-19-888672-3}
}

@article{SongEtAl_TwistedFractonModels_19,
  title = {Twisted Fracton Models in Three Dimensions},
  author = {Song, Hao and Prem, Abhinav and Huang, Sheng-Jie and {Martin-Delgado}, M. A.},
  year = 2019,
  month = apr,
  journal = {Physical Review B},
  volume = {99},
  number = {15},
  pages = {155118},
  publisher = {American Physical Society},
  doi = {10.1103/PhysRevB.99.155118},
  urldate = {2026-06-14}
}

@article{SucharaEtAl_ConstructionsNoiseThreshold_11,
  title = {Constructions and Noise Threshold of Topological Subsystem Codes},
  author = {Suchara, Martin and Bravyi, Sergey and Terhal, Barbara},
  year = 2011,
  month = mar,
  journal = {Journal of Physics A: Mathematical and Theoretical},
  volume = {44},
  number = {15},
  pages = {155301},
  issn = {1751-8121},
  doi = {10.1088/1751-8113/44/15/155301},
  urldate = {2026-03-07},
  langid = {english}
}

@article{TantivasadakarnEtAl_HierarchyTopologicalOrder_23,
  title = {Hierarchy of {{Topological Order From Finite-Depth Unitaries}}, {{Measurement}}, and {{Feedforward}}},
  author = {Tantivasadakarn, Nathanan and Vishwanath, Ashvin and Verresen, Ruben},
  year = 2023,
  month = jun,
  journal = {PRX Quantum},
  volume = {4},
  number = {2},
  pages = {020339},
  issn = {2691-3399},
  doi = {10.1103/PRXQuantum.4.020339},
  urldate = {2025-08-01}
}

@article{VanDenNes_ClassicalSimulationQuantum_10,
  title = {Classical Simulation of Quantum Computation, the {{Gottesman-Knill}} Theorem, and Slightly Beyond},
  author = {{Van den Nest}, Maarten},
  year = 2010,
  month = mar,
  journal = {Quantum Information and Computation},
  volume = {10},
  number = {3},
  pages = {258--271},
  issn = {1533-7146}
}

@article{WanEtAl_TwistedGaugeTheory_15,
  title = {Twisted {{Gauge Theory Model}} of {{Topological Phases}} in {{Three Dimensions}}},
  author = {Wan, Yidun and Wang, Juven and He, Huan},
  year = 2015,
  month = jul,
  journal = {Physical Review B},
  volume = {92},
  number = {4},
  eprint = {1409.3216},
  primaryclass = {cond-mat},
  pages = {045101},
  issn = {1098-0121, 1550-235X},
  doi = {10.1103/PhysRevB.92.045101},
  urldate = {2025-09-25},
  archiveprefix = {arXiv}
}

@article{WarmanSchafer-Nameki_TransversalCliffordHierarchyGates_25,
  title={Constant-depth Clifford-hierarchy gates via non-Abelian surface codes},
  author={Warman, Alison and Sch{\"a}fer-Nameki, Sakura},
  journal={PRX Quantum},
  volume={7},
  number={3},
  pages={033029},
  year={2026},
  publisher={APS}
}

@article{WebsterEtAl_XPStabiliserFormalism_22,
  title = {The {{XP Stabiliser Formalism}}: A {{Generalisation}} of the {{Pauli Stabiliser Formalism}} with {{Arbitrary Phases}}},
  shorttitle = {The {{XP Stabiliser Formalism}}},
  author = {Webster, Mark A. and Brown, Benjamin J. and Bartlett, Stephen D.},
  year = 2022,
  month = sep,
  journal = {Quantum},
  volume = {6},
  eprint = {2203.00103},
  primaryclass = {quant-ph},
  pages = {815},
  issn = {2521-327X},
  doi = {10.22331/q-2022-09-22-815},
  urldate = {2025-09-24},
  archiveprefix = {arXiv}
}

@misc{CalderbankEtAl_QuantumErrorCorrection_97,
  title = {Quantum {{Error Correction}} via {{Codes}} over {{GF}}(4)},
  author = {Calderbank, A. R. and Rains, E. M. and Shor, P. W. and Sloane, N. J. A.},
  year = 1997,
  month = sep,
  number = {arXiv:quant-ph/9608006},
  eprint = {quant-ph/9608006},
  publisher = {arXiv},
  doi = {10.48550/arXiv.quant-ph/9608006},
  urldate = {2026-07-21},
  archiveprefix = {arXiv}
}

@inproceedings{NadellaKlappenecker_StabilizerCodesFrobenius_12,
  title = {Stabilizer Codes over {{Frobenius}} Rings},
  booktitle = {2012 {{IEEE International Symposium}} on {{Information Theory Proceedings}}},
  author = {Nadella, Sushma and Klappenecker, Andreas},
  year = 2012,
  month = jul,
  pages = {165--169},
  publisher = {IEEE},
  address = {Cambridge, MA, USA},
  doi = {10.1109/ISIT.2012.6283558},
  urldate = {2026-07-21},
  isbn = {978-1-4673-2579-0 978-1-4673-2580-6 978-1-4673-2578-3}
}

@misc{YangYu_ClassificationPauliStabilizer_26,
  title = {The {{Classification}} of {{Pauli Stabilizer Codes}}: {{A Lattice}} and {{Continuum Treatise}}},
  shorttitle = {The {{Classification}} of {{Pauli Stabilizer Codes}}},
  author = {Yang, Bowen and Yu, Matthew},
  year = 2026,
  month = apr,
  number = {arXiv:2604.24847},
  eprint = {2604.24847},
  primaryclass = {math-ph},
  publisher = {arXiv},
  doi = {10.48550/arXiv.2604.24847},
  urldate = {2026-07-21},
  archiveprefix = {arXiv}
}

@article{Williamson_2026Fault,
   title={Low-overhead fault-tolerant quantum computation by gauging logical operators},
   volume={22},
   ISSN={1745-2481},
   url={http://dx.doi.org/10.1038/s41567-026-03220-8},
   DOI={10.1038/s41567-026-03220-8},
   number={4},
   journal={Nature Physics},
   publisher={Springer Science and Business Media LLC},
   author={Williamson, Dominic J. and Yoder, Theodore J.},
   year={2026},
   month=Apr, pages={598--603} }

@misc{delafuente2026highthresholddecodingnonpaulicodes,
      title={High-threshold decoding of non-Pauli codes for 2D universality}, 
      author={Julio C. Magdalena de la Fuente and Noa Feldman and Jens Eisert and Andreas Bauer},
      year={2026},
      eprint={2604.02033},
      archivePrefix={arXiv},
      primaryClass={quant-ph},
      url={https://arxiv.org/abs/2604.02033}, 
}

@article{Surti_2026,
  title={Efficient simulation of logical magic state preparation protocols},
  author={Surti, Samyak and Daguerre, Lucas and Kim, Isaac H},
  journal={PRX Quantum},
  volume={7},
  number={2},
  pages={020329},
  year={2026},
  publisher={APS}
}

@misc{takada2026exactefficientsimulationnoisy,
      title={Exact efficient simulation of noisy logical magic states using Clifford stabilizers}, 
      author={Yugo Takada and Stephen D. Bartlett and Dominic J. Williamson},
      year={2026},
      eprint={2609.16929},
      archivePrefix={arXiv},
      primaryClass={quant-ph},
      url={https://arxiv.org/abs/2609.16929}, 
}

@misc{delafuente2026efficientlyestimatingfailurerates,
      title={Efficiently estimating failure rates of fault-tolerant logical non-Clifford blocks}, 
      author={Julio C. Magdalena de la Fuente and Thomas R. Scruby},
      year={2026},
      eprint={2609.19485},
      archivePrefix={arXiv},
      primaryClass={quant-ph},
      url={https://arxiv.org/abs/2609.19485}, 
}

@misc{williamson2026fastmagicstatepreparation,
      title={Fast magic state preparation by gauging higher-form transversal gates in parallel}, 
      author={Dominic J. Williamson},
      year={2026},
      eprint={2601.22939},
      archivePrefix={arXiv},
      primaryClass={quant-ph},
      url={https://arxiv.org/abs/2601.22939}, 
}

@misc{zhu2026nonabelianqldpctqftformalism,
      title={Non-Abelian qLDPC: TQFT Formalism, Addressable Gauging Measurement and Application to Magic State Fountain on 2D Product Codes}, 
      author={Guanyu Zhu and Ryohei Kobayashi and Po-Shen Hsin},
      year={2026},
      eprint={2601.06736},
      archivePrefix={arXiv},
      primaryClass={quant-ph},
      url={https://arxiv.org/abs/2601.06736}, 
}

@misc{christos2026nonabelianquantumlowdensityparity,
      title={Non-Abelian Quantum Low-Density Parity Check Codes and Non-Clifford Operations from Gauging Logical Gates via Measurements}, 
      author={Maine Christos and Chiu Fan Bowen Lo and Vedika Khemani and Rahul Sahay},
      year={2026},
      eprint={2602.12228},
      archivePrefix={arXiv},
      primaryClass={quant-ph},
      url={https://arxiv.org/abs/2602.12228}, 
}

@article{nest2011monomial,
  title={A monomial matrix formalism to describe quantum many-body states},
  author={{Van den Nest}, Maarten},
  journal={New Journal of Physics},
  volume={13},
  number={12},
  pages={123004},
  year={2011},
  publisher={IOP Publishing}
}

\onecolumngrid

\appendix 
\section{Useful results about the Clifford hierarchy}
\label{app:useful_results}

Here we prove a few useful results about the Clifford hierarchy.

\subsection{Composing operators}

\begin{lemma}
The set ${\cal C}^{(2)}$ is a group.
\end{lemma}
\begin{proof}
Suppose that $H\in {\cal C}^{(2)}$. Then, by definition, $H{\cal P}_nH^\dagger \subseteq {\cal P}_n$. So then if $H_1, H_2 \in {\cal C}^{(2)}$ we see that $(H_1H_2){\cal P}_n(H_1H_2)^\dagger \subseteq H_1{\cal P}_nH_1^\dagger \subseteq {\cal P}_n$. So then $H_1H_2\in {\cal C}^{(2)}$. We conclude that ${\cal C}^{(2)}$ is a group.
\end{proof}

We use this result mainly to note that if $H\in {\cal C}^{(2)}$ then so is $H^\dagger$.

\begin{lemma}\label{lem:conj_clifford}
Suppose that $U\in {\cal C}^{(k)}$ and $H\in {\cal C}^{(2)}$. Then $HUH^\dagger \in {\cal C}^{(k)}$.
\end{lemma}
\begin{proof}
We can prove this by induction on $k$. By the definition of ${\cal C}^{(2)}$ we know that $k=1$ is true, so assume the truth of $k$. This means that $H{\cal C}^{(k)}H^\dagger \subseteq {\cal C}^{(k)}$. Now assume $U\in {\cal C}^{(k+1)}$. Then $(HUH^\dagger){\cal P}_n(HUH^\dagger)^\dagger \subseteq HU{\cal P}_nU^\dagger H^\dagger$, again from the definition of ${\cal C}^{(2)}$. If we now use the fact that $U\in {\cal C}^{(k+1)}$ and the inductive hypothesis then we can see that $HU{\cal P}_nU^\dagger H^\dagger\subseteq H{\cal C}^{(k)}H^\dagger \subseteq {\cal C}^{(k)}$. So we conclude that $HUH^\dagger \in {\cal C}^{(k)}$ if $U\in {\cal C}^{(k)}$.
\end{proof}

Note that ${\cal C}^{(k-1)}\subseteq {\cal C}^{(k)}$. So a priori just because $HUH^\dagger \in {\cal C}^{(k)}$ does not mean it is not at a lower level, i.e. also contained in ${\cal C}^{(k-1)}$. We will make another definition to treat this case.

\begin{definition}
Let ${\cal S}^{\Delta k}$ be those operators in ${\cal C}^{(k)}$ but not in ${\cal C}^{(k-1)}$, i.e. ${\cal S}^{\Delta k} = {\cal C}^{(k)}\setminus {\cal C}^{(k-1)}$.
\end{definition}

\begin{lemma}
Suppose that $U\in {\cal S}^{\Delta k}$ and $H\in {\cal C}^{(2)}$. Then $HUH^\dagger \in {\cal S}^{\Delta k}$.
\end{lemma}
\begin{proof}
Since $U\in {\cal S}^{\Delta k}$ we know that $U\in {\cal C}^{(k)}$. Then by Lem.~\ref{lem:conj_clifford} we know that $HUH^\dagger \in {\cal C}^{(k)}$. Now suppose, by contradiction, that $HUH^\dagger \in {\cal C}^{(k-1)}$. Then since $H^\dagger \in {\cal C}^{(2)}$ we could again use Lem.~\ref{lem:conj_clifford} to see that $U\in {\cal C}^{(k-1)}$. But this is a contradiction, so we must have that $HUH^\dagger\in {\cal C}^{(k)}$ but not in ${\cal C}^{(k-1)}$. Thus $HUH^\dagger \in {\cal S}^{\Delta k}$.
\end{proof}

Two simple lemmas follow from this.

\begin{lemma}\label{lem:comp_clifford}
Suppose that $U\in {\cal S}^{\Delta k}$ and $H\in {\cal C}^{(2)}$. Then $HU$ and $UH$ are both in ${\cal S}^{\Delta k}$ (for $k\geq 3$).
\end{lemma}
\begin{proof}
By definition of ${\cal C}^{(2)}$ we know that if $H\in {\cal C}^{(2)}$ then $H{\cal P}_nH^\dagger \subseteq {\cal P}_n$. So then $(UH){\cal P}_n(UH)^\dagger \subseteq U{\cal P}_nU^\dagger \subseteq {\cal C}^{(k-1)}$, since $U\in {\cal C}^{(k)}$. But then clearly $UH\in {\cal C}^{(k)}$. (Note for this argument to work we require $k\geq 2$; clearly if $U=1\in {\cal C}^{(1)}$ then $UH\notin {\cal C}^{(1)}$). Suppose then, by contradiction, that $UH\in {\cal C}^{(k-1)}$. Then since $H^\dagger \in {\cal C}^{(2)}$ we would have that $(UH)H^\dagger = U\in {\cal C}^{(k-1)}$, a contradiction. So $UH\in {\cal S}^{\Delta k}$; by the earlier parenthetical we need $k\geq 3$.

From Lem.~\ref{lem:conj_clifford} we know that $H{\cal C}^{(k)}H^\dagger \subseteq {\cal C}^{(k)}$. So then $(HU){\cal P}_n(HU)^\dagger \subseteq H{\cal C}^{(k-1)}H^\dagger \subseteq {\cal C}^{(k-1)}$ since $U\in {\cal C}^{(k)}$. So then $HU\in {\cal C}^{(k)}$ as well. We can use nearly identical logic to above to see that $HU \in {\cal S}^{\Delta k}$.
\end{proof}

\begin{corollary}\label{cor:comp_pauli}
Suppose that $U\in {\cal S}^{\Delta k}$ and $P\in {\cal P}_n$. Then $PU$ and $UP$ are both in ${\cal S}^{\Delta k}$
\end{corollary}
\begin{proof}
This follows immediately from Lem.~\ref{lem:comp_clifford} since ${\cal P}_n = {\cal C}^{(1)}\subseteq {\cal C}^{(2)}$ by definition.
\end{proof}

\subsection{Finite difference operators}

In Def.~\ref{defn:Df} we gave a definition of a diagonal operator, $D[f]$, which took as input functions $f\colon \mathbb{Z}_p^n \rightarrow \mathbb{R}/\mathbb{Z}$. Here we will see how to determine the level of the Clifford hierarchy for this operator. To do so we introduce the finite difference operator.

\begin{definition}
Let $f\colon \mathbb{Z}_p^n \rightarrow \mathbb{R}/\mathbb{Z}$ and $\mathbf{v}\in \mathbb{Z}_p^n$. Then define $\Delta_{\mathbf{v}}f(\mathbf{x}) = f(\mathbf{x}+\mathbf{v})-f(\mathbf{x})$. Repeated $\Delta$s are defined recursively. We emphasize that addition in the argument of $f$ is performed mod $p$ because $f$ takes inputs in $\mathbb{Z}_p^n$. In particular, one should think of $\mathbf{x}+\mathbf{v}$ as $[\mathbf{x}+\mathbf{v}]_p$, where $[x]_p = x\mod p$.
\end{definition}

The finite difference operator has a few useful properties shared by derivatives. For example it is commutative, $\Delta_{\mathbf{v}}\Delta_{\mathbf{v}'}f = \Delta_{\mathbf{v}'}\Delta_{\mathbf{v}}f$. However, it does not have a natural relationship to the degree of a function. For example, consider $f(x) = x/p^2$. We might expect that if we applied $\Delta_1$ to this we would arrive at a constant. However, we will see we would actually need to apply it $p$ times.

\begin{theorem}\label{thm:Ck_finite_diff}
The group ${\cal C}^{(k),n}_d$ can be alternatively described by
\begin{equation}\label{eq:Ck_diff}
{\cal C}^{(k),n}_d = \{D[f] \mid \Delta_{\mathbf{v}_1}\cdots \Delta_{\mathbf{v}_{k+1}} f = 0 \ \forall \mathbf{v}_1,\ldots, \mathbf{v}_{k+1}\in \mathbb{Z}_p^n \}.
\end{equation}

It follows that the set ${\cal S}^{\Delta k}_d$ can be described by
\begin{equation}
{\cal S}^{\Delta k}_d = \{D[f]\mid \Delta_{\mathbf{v}_1}\cdots \Delta_{\mathbf{v}_{k+1}} f = 0 \ \forall \mathbf{v}_1,\ldots, \mathbf{v}_{k+1}\in \mathbb{Z}_p^n, \exists \mathbf{v}_1,\ldots, \mathbf{v}_{k}\in \mathbb{Z}_p^n \ \mathrm{s.t.} \ \Delta_{\mathbf{v}_1}\cdots \Delta_{\mathbf{v}_{k}} f \neq 0\}
\end{equation}
\end{theorem}
\begin{proof}
We prove this by induction on $k$ with $k=1$ as our base case. We will prove the equality of Eq. \ref{eq:Ck_diff} by showing that each set is contained within the other. 

First, we will show that $D[f]\in\mathcal{C}_d^{(1)}$ implies that $\Delta_{\mathbf{w}}\Delta_{\mathbf{v}} f = 0 \ \forall \mathbf{v},\mathbf{w}\in \mathbb{Z}_p^n$. Every element of $\mathcal{C}_d^{(1)}$ is of the form 
$e^{2\pi i d_0/p}Z_1^{d_1}\cdots Z_n^{d_n}$, where $\mathbf{d}\in\mathbb{Z}_p^{n+1}$. The corresponding phase function $f$ is given by $f(\mathbf{x})=\frac{1}{p}\left(d_0+\sum_{i=1}^n d_i x_i\right)$, which is defined mod 1. Applying the finite difference, we have
\begin{equation}
\begin{split}
\Delta_{\mathbf{w}}\Delta_{\mathbf{v}} f(\mathbf{x})&=f(\mathbf{x}+\mathbf{v}+\mathbf{w})-f(\mathbf{x}+\mathbf{v})-f(\mathbf{x}+\mathbf{w})+f(\mathbf{x})\;\text{mod}\;1,\\
&=\frac{1}{p}\left(d_0+\sum_{i=1}^n d_i [x_i+v_i+w_i]\right)-\frac{1}{p}\left(d_0+\sum_{i=1}^n d_i [x_i+v_i]\right)\\&\qquad\qquad-\frac{1}{p}\left(d_0+\sum_{i=1}^n d_i [x_i+w_i]\right)+\frac{1}{p}\left(d_0+\sum_{i=1}^n d_i x_i\right)\;\text{mod}\;1,\\
&=0\;\text{mod}\;1.
\end{split}
\end{equation}
The last equality can be most clearly understood by multiplying both sides of the second line by $p$, so that the addition on the RHS is in $\mathbb{Z}_p^n$ and the cancellations become trivial. This gives $p\Delta_{\mathbf{w}}\Delta_{\mathbf{v}} f(\mathbf{x})=0\;\text{mod}\;p$ which in turn implies that $\Delta_{\mathbf{w}}\Delta_{\mathbf{v}} f(\mathbf{x})=0\;\text{mod}\;1$. We have therefore shown that $\Delta_{\mathbf{w}}\Delta_{\mathbf{v}} f = 0 \ \forall \mathbf{v},\mathbf{w}\in \mathbb{Z}_p^n$ is satisfied whenever $D[f]\in\mathcal{C}_d^{(1)}$. 

Next, we must prove that the constraint $\Delta_{\mathbf{w}}\Delta_{\mathbf{v}} f = 0 \ \forall \mathbf{v},\mathbf{w}\in \mathbb{Z}_p^n$ is only satisfied when $D[f]\in\mathcal{C}_d^{(1)}$. Writing out the constraint, we have
\begin{equation}
\Delta_{\mathbf{w}}\Delta_{\mathbf{v}} f(\mathbf{x})=f(\mathbf{x}+\mathbf{v}+\mathbf{w})-f(\mathbf{x}+\mathbf{v})-f(\mathbf{x}+\mathbf{w})+f(\mathbf{x})=0\;\text{mod}\;1,
\end{equation}
for all $\mathbf{x},\mathbf{v},\mathbf{w}\in\mathbb{Z}_d^n$. Since this is a functional identity which applies for any $\mathbf{x}$, it must in particular apply for $\mathbf{x}=\mathbf{0}$. In this case, we have
\begin{equation}
    f(\mathbf{v}+\mathbf{w})=f(\mathbf{v})+f(\mathbf{w})-f(\mathbf{0})\;\text{mod}\;1.
\end{equation}
This equality tells us that we can define a function $g(\mathbf{x})=f(\mathbf{x})-f(\mathbf{0})$ which satisfies $g(\mathbf{v}+\mathbf{w})=g(\mathbf{v})+g(\mathbf{w})$ for all $\mathbf{v},\mathbf{w}\in\mathbb{Z}_p^n$. In other words, $g$ is a group homomorphism from $\mathbb{Z}_p^n$ to $\mathbb{R}/\mathbb{Z}$. Using this fact, we can also state that $pg(\mathbf{x})=g(p\mathbf{x})=0$, which tells us that $g(\mathbf{x})$ is valued in $\left\{\frac{k}{p}\mid k=0,\ldots,p-1\right\}$. We can then define $\tilde g(\mathbf{x})$ via $g(\mathbf{x})=\frac{1}{p}\tilde g(\mathbf{x})$. We see that $\tilde g(\mathbf{x})$ is a group homomorphism from $\mathbb{Z}_p^n$ to $\mathbb{Z}_p$. The most general such function is given by 
$\tilde g(\mathbf{x})=\sum_{i=1}^n d_i x_i$, with $\mathbf{d}\in\mathbb{Z}_p^n$. Combining everything, we have 
\begin{equation}
    f(\mathbf{x})=\frac{1}{p}\sum_{i=1}^n d_i x_i+f(\mathbf{0})\;\text{mod}\;1.
\end{equation}

Now assume the truth of $k$. Then suppose that $f$ is such that $\Delta_{\mathbf{v}_1}\cdots \Delta_{\mathbf{v}_{k+2}}f = 0$ for all $\mathbf{v}_1,\ldots, \mathbf{v}_{k+2}\in \mathbb{Z}_p^n$. Then we note that
\begin{align}
D[f]X(\mathbf{v})D^\dagger[f] &= X(\mathbf{v})D[\Delta_{\mathbf{v}}f]\\
D[f]Z(\mathbf{u})D^\dagger[f] &= Z(\mathbf{u}).
\end{align}
Now we note that by choice of $f$ for every $\mathbf{v}\in \mathbb{Z}_p^n$ we have $\Delta_{\mathbf{v}_1}\cdots \Delta_{\mathbf{v}_{k+1}}(\Delta_{\mathbf{v}}f)=0$ for every $\mathbf{v}_1,\ldots, \mathbf{v}_{k+1}\in \mathbb{Z}_p^n$. So then by the inductive hypothesis $D[\Delta_{\mathbf{v}}f]\in {\cal C}^{(k)}$ for every $\mathbf{v}\in \mathbb{Z}_p^n$. But then by Cor.~\ref{cor:comp_pauli} we conclude $X(\mathbf{v})D[\Delta_{\mathbf{v}}f]\in {\cal C}^{(k)}$. Thus $D[f]\in {\cal C}^{(k+1)}_d$.

On the other hand suppose that $D[f]\in {\cal C}^{(k+1)}_d$. Then we can use the equations above to see that $X(\mathbf{v})D[\Delta_{\mathbf{v}}f]\in {\cal C}^{(k)}$ and thus by Cor.~\ref{cor:comp_pauli} that $D[\Delta_{\mathbf{v}}f]\in {\cal C}^{(k)}$ for every $\mathbf{v}\in \mathbb{Z}_p^n$. By the inductive hypothesis this means that $\Delta_{\mathbf{v}_1}\cdots \Delta_{\mathbf{v}_{k+1}}(\Delta_{\mathbf{v}}f)=0$ for every $\mathbf{v}_1,\ldots, \mathbf{v}_{k+1},\mathbf{v}\in \mathbb{Z}_p^n$. Thus we see the truth of $k+1$ and conclude.

The explicit expression for ${\cal S}^{\Delta k}_d = {\cal C}^{(k)}_d\setminus {\cal C}^{(k-1)}_d$ follows directly.
\end{proof}

It follows immediately that applying the finite difference operator to a diagonal operator must always lower the level of the Clifford hierarchy.

\begin{corollary}\label{cor:diff_lowers}
Let $f\colon \mathbb{Z}_p^n\rightarrow \mathbb{R}/\mathbb{Z}$. Then $D[f]\in {\cal C}^{(k)}$ if and only if $D[\Delta_{\mathbf{v}}f]\in {\cal C}^{(k-1)}$ for all $\mathbf{v}$. It follows that $D[f]\in {\cal S}^{\Delta k}_d$ if and only if $D[\Delta_{\mathbf{v}}f]\in {\cal C}^{(k-1)}$ for all $\mathbf{v}$ and there exists some $\mathbf{v}$ such that $D[\Delta_{\mathbf{v}}f]\notin {\cal C}^{(k-2)}$. 
\end{corollary}
\begin{proof}
By Thm.~\ref{thm:Ck_finite_diff} we know that if $D[f]\in {\cal C}^{(k)}_d$ then $\Delta_{\mathbf{v}_1}\cdots \Delta_{\mathbf{v}_{k+1}} f = 0 \ \forall \mathbf{v}_1,\ldots, \mathbf{v}_{k+1}\in \mathbb{Z}_p^n$. Setting $\mathbf{v}_{k+1}=\mathbf{v}$ we see this means $\Delta_{\mathbf{v}_1}\cdots \Delta_{\mathbf{v}_{k}}(\Delta_{\mathbf{v}} f) = 0 \ \forall \mathbf{v}_1,\ldots, \mathbf{v}_{k}\in \mathbb{Z}_p^n$. Then by Thm.~\ref{thm:Ck_finite_diff} we know $D[\Delta_{\mathbf{v}}f]\in {\cal C}^{(k-1)}_d$. On the other hand suppose that $D[\Delta_{\mathbf{v}}f]\in {\cal C}^{(k-1)}_d$ for every $\mathbf{v}\in \mathbb{Z}_p^n$. Then by Thm.~\ref{thm:Ck_finite_diff} this means that $\Delta_{\mathbf{v}_1}\cdots \Delta_{\mathbf{v}_{k}}(\Delta_{\mathbf{v}} f) = 0. \ \forall \mathbf{v}_1,\ldots, \mathbf{v}_{k}\in \mathbb{Z}_p^n$. But then clearly $\Delta_{\mathbf{v}_1}\cdots \Delta_{\mathbf{v}_{k+1}} f = 0 \ \forall \mathbf{v}_1,\ldots, \mathbf{v}_{k+1}\in \mathbb{Z}_p^n$ and thus by Thm.~\ref{thm:Ck_finite_diff} we must have that $D[f]\in {\cal C}^{(k)}_d.$

Suppose that $D[f] \in {\cal C}^{(k)}\setminus {\cal C}^{(k-1)}$. Then by the above $D[\Delta_{\mathbf{v}}f] \in {\cal C}^{(k-1)}$ for all $\mathbf{v}$. If, however, $D[\Delta_{\mathbf{v}}f] \in {\cal C}^{(k-2)}$ for all $\mathbf{v}$ then by the above $D[f] \in {\cal C}^{(k-1)}$, a contradiction. So there must exist some $\mathbf{v}$ such that $D[\Delta_{\mathbf{v}}f]\notin {\cal C}^{(k-2)}$. On the other hand if $D[\Delta_{\mathbf{v}}f]\in {\cal C}^{(k-1)}$ for all $\mathbf{v}$ then $D[f]\in {\cal C}^{(k)}$. If, however, $D[f]\in {\cal C}^{(k-1)}$ then $D[\Delta_{\mathbf{v}}f]\in {\cal C}^{(k-2)}$ for all $\mathbf{v}$, a contradiction. So $D[f]\in {\cal S}^{\Delta k}_d$.
\end{proof}

The above corollary may seem like a simple fact, but it is nontrivial when viewed through the lens of the degree of a function. Recall again the function $f(x) = x/p^2$. This is at level $(2-1)(p-1)+1=p$ of the Clifford hierarchy. By the result above this means that $\Delta_1^{p}f$ is a nonzero constant, and thus that $\Delta_1f$ cannot be a constant. So we see that the finite difference operator does not always behave like a derivative. We demonstrate why this happens with the following example.

\begin{example}
Let $f\colon \mathbb{Z}_p\rightarrow \mathbb{Z}_{p^2}$ be defined as $f(x) = x \pmod{p^2}$. Then $f(x+1) = [x+1]_p \pmod{p^2}$, where $[x]_p = x\pmod{p}$. We know from Ref.~\cite{CuiEtAl_DiagonalGatesClifford_17} that there must exist some unique $a_k\in \mathbb{Z}_{p^2}$ such that
\begin{equation}
f(x+1) = [x+1]_p = \sum_{k=0}^{p-1}a_kx^k.
\end{equation}
We might be tempted to say that $f(x+1) = x+1 \pmod{p^2}$, so $a_0,a_1=1$ and all others are zero. However, this is clearly not correct since $f((p-1)+1)=0\pmod{p^2}$ whereas $(p-1)+1=p\pmod{p^2}$. Instead one can check the right decomposition is $f(x+1) = x + 1 - p\delta_{x,p-1}$ which, in polynomials, is
\begin{equation}
f(x+1) = x+1 - p\left(1-(x+1)^{p-1}\right)\pmod{p^2}.
\end{equation}
Since polynomial decompositions are unique it is enough to verify that this holds for all $x\in \{0,\ldots,p-1\}$. So then $\Delta_1 f = f(x+1)-f(x)$ is a degree $p-1$ polynomial over $\mathbb{Z}_{p^2}$ even though $f$ was degree $1$.
\end{example}

The above example might make us worry about something else. Usually a polynomial of a linear function has the same degree as the original function. For example if $f\colon \mathbb{R}\rightarrow \mathbb{R}$ is given by $f(x) =x^2$ then $f(ax+by)\colon \mathbb{R}\times \mathbb{R}\rightarrow \mathbb{R}$ will also have degree two as long as either $a$ or $b$ is not equal to zero. However this is clearly not the case for the functions we consider. In the example above $f(x) = x\pmod{p^2}$ had degree one, while $f(x+1)$ had degree $p-1$, over $p^2$. Nonetheless, just as with derivatives the right thing to consider is the level of the Clifford hierarchy and not the degree of the polynomial. We now show that this is unchanged upon composition with linear functions.

\begin{theorem}\label{thm:lin_preserves_level}
Let $f\colon \mathbb{Z}_p^n\rightarrow \mathbb{R}/\mathbb{Z}$ be such that $D[f]\in {\cal C}^{(k)}_d$. Further let $g\colon \mathbb{Z}_p^m \rightarrow \mathbb{Z}_p^n$ take the form $g(\mathbf{y}) = M\mathbf{y}+\mathbf{c}$, where $M \in \mathbb{Z}_p^{n\times m}$ and $\mathbf{c}\in \mathbb{Z}_p^n$. Then $D[f\circ g] \in {\cal C}^{(k)}_d$.

If $D[f]\in {\cal S}^{\Delta k}_d$ and $g$ is a surjection then $D[f\circ g]\in {\cal S}^{\Delta k}_d$.
\end{theorem}
\begin{proof}
Suppose that $D[f]\in {\cal C}^{(k)}_d$. Then by Thm.~\ref{thm:Ck_finite_diff} we must have that $(\Delta_{\mathbf{v}_1}\cdots \Delta_{\mathbf{v}_{k+1}}f)(\mathbf{x}) = 0$ for all $\mathbf{v}_1,\ldots, \mathbf{v}_{k+1},\mathbf{x}\in \mathbb{Z}_p^n$. In particular this means that for every $\mathbf{w}_1,\ldots, \mathbf{w}_{k+1},\mathbf{y}\in \mathbb{Z}_p^m$ we must have that $0 = (\Delta_{M\mathbf{w}_1}\ldots \Delta_{M\mathbf{w}_{k+1}}f)(M\mathbf{y}+\mathbf{c}) = (\Delta_{\mathbf{w}_1}\ldots \Delta_{\mathbf{w}_{k+1}}f\circ g)(\mathbf{y})$. So then we conclude that $D[f\circ g]\in {\cal C}^{(k)}_d$.

Now suppose that $D[f] \in {\cal S}^{\Delta k}_d$ and $g$ is a surjection. By the above we must have that $D[f\circ g]\in {\cal C}^{(k)}_d$. Further, by Thm.~\ref{thm:Ck_finite_diff}, we must have that there exists some $\mathbf{v}_1,\ldots, \mathbf{v}_{k},\mathbf{x}\in \mathbb{Z}_p^n$ such that $(\Delta_{\mathbf{v}_1}\cdots \Delta_{\mathbf{v}_{k}}f)(\mathbf{x}) \neq 0$. Since $g$ is a surjection then $M$ must also be a surjection. This means there must exist some $\mathbf{w}_1,\ldots, \mathbf{w}_k, \mathbf{y}$ such that $M\mathbf{w}_i = \mathbf{v}_i$ and $g(\mathbf{y}) = \mathbf{x}$. Then $0\neq (\Delta_{M\mathbf{w}_1}\cdots \Delta_{M\mathbf{w}_{k}}f)(M\mathbf{y}+\mathbf{c})=(\Delta_{\mathbf{w}_1}\ldots \Delta_{\mathbf{w}_{k}}f\circ g)(\mathbf{y})$. By Thm.~\ref{thm:Ck_finite_diff} we then know that $D[f\circ g]\in {\cal S}^{\Delta k}_d$.
\end{proof}

Note that if $g$ is not a surjection then it is easy to change the level. For example if $M=0$, so that $g(\mathbf{y})$ is a constant, then $D[f\circ g]$ will also just be a constant. There is a nice corollary of this that is useful for single qudit stabilizers.

\begin{corollary}\label{cor:lin_preserves_level}
Let $f\colon \mathbb{Z}_p\rightarrow \mathbb{R}/\mathbb{Z}$ be such that $D[f]\in {\cal S}^{\Delta k}_d$. Further let $g\colon \mathbb{Z}_p^m \rightarrow \mathbb{Z}_p$ take the form $g(\mathbf{y}) = \mathbf{m}\cdot \mathbf{y}+c$, where $\mathbf{m} \in \mathbb{Z}_p^{m}$ and $c\in \mathbb{Z}_p$. Then either $D[f\circ g]$ is a constant phase, or $D[f\circ g]\in {\cal S}^{\Delta k}_d$.
\end{corollary}
\begin{proof}
Suppose that $g(\mathbf{y}) = \mathbf{m}\cdot \mathbf{y}+c$ is not a constant phase, then at least one $m_i\in \mathbb{Z}_p$ is not zero. Then it is easy to see that $g((x-c)m_i^{-1}\mathbf{e}_i)=x$ for all $x\in \mathbb{Z}_p$, so $g$ is surjective. Thus either $g$ is a constant phase, in which case $\mathbf{m}=\mathbf{0}$, or $g$ is surjective. The result then follows from Thm.~\ref{thm:lin_preserves_level}.
\end{proof}

\section{The On-Site Clifford Hierarchy}\label{app:on_site_clifford}

In this section, we will consider a special class of \xkcd codes in which the diagonal Clifford operators admit a tensor-product decomposition into single-site operators. That is, each $D[f]$ appearing in an $n$-qudit stabilizer may be written
\begin{equation}
    D[f]=D[f_1]\otimes \cdots\otimes D[f_i]\otimes \cdots \otimes D[f_n].
\end{equation}
A diagonal operator of this form is said to be \textit{on-site}. We define the group of on-site \xkcd stabilizers on $n$ qudits as
\begin{equation}
    {\cal G}^{(k),n}=\{X(\mathbf{v})D[f]\in G^{(k),n}\mid D[f]\;\;\text{on-site}\}\subset G^{(k),n}.
\end{equation}
It is easy to see that ${\cal G}^{(k),n}$ is a subgroup of $G^{(k),n}$. Furthemore, the most general operator $g\in {\cal G}^{(k),n}$ can be written in the form
\begin{equation}
g = X(\mathbf{v})D\left[\sum_{i=1}^{n} f_i\right], \ f_i\colon \mathbb{Z}_p\rightarrow \mathbb{R}/\mathbb{Z},\; \mathbf{v} \in \mathbb{Z}_p^n,
\end{equation}
where $f_i$ is a function of $x_i$ alone (when acting on the state $\ket{\mathbf{x}}$) and the diagonal gate is at level $k$.

We now state a few basic facts about the relationship between $G^{(k),n}$ and ${\cal G}^{(k),n}$ for various $k$ and $n$. Suppose that $k=1$, then the only nontrivial elements of ${\cal C}^{(k),n}_d$ are tensor products of $Z$ operators. This means that ${\cal G}^{(1),n}=G^{(1),n}$ for any $n$. In other words, this is simply the fact that all operators in the Pauli group are tensor products of on-site operators. Likewise, by definition, we have that ${\cal G}^{(k),1}=G^{(k),1}$ for any $k$. However, suppose $n,k\geq 2$ and consider the gate $D[x_ix_j/p]=\sum_{\mathbf{x}}e^{\frac{2\pi i}{p}x_ix_j}\ketbra{\mathbf{x}}{\mathbf{x}}$, which is the qudit generalization of $CZ_{i,j}$. This operator lies in $G^{(k),n}$ but not in ${\cal G}^{(k),n}$. Thus we conclude that as long as $n,k\geq 2$ then ${\cal G}^{(k),n}$ is a proper subset of $G^{(k),n}$.

Now we would like to define stabilizer models drawn from ${\cal G}^{(k),n}$. This is analogous to defining stabilizer models drawn from $G^{(k),n}$, but we state it here for completeness.
Let ${\cal H}\subseteq {\cal G}^{(k),n}$ be a subgroup. If there exists some $n$-qudit state such that $h|\psi\rangle = |\psi\rangle$ for every $h\in {\cal H}$ then we call ${\cal H}$ an $n$-qudit \xkcd stabilizer group. The space of all states stabilized by ${\cal H}$ is denoted by ${\cal L}_{{\cal H}}$ and referred to as the $n$-qudit \xkcd stabilizer code associated with ${\cal H}$. If there is only one state in ${\cal L}_{{\cal H}}$ then we call this an $n$-qudit \xkcd stabilizer state.

\begin{definition}\label{defn:onsite_stab_sets}
We denote the set of $n$-qudit on-site \xkcd stabilizer states as $\Psi^{(k),n}_{{\cal G}}$.
\end{definition}

The set $\Psi^{(k),n}_{{\cal G}}$ and $\Psi^{(k),n}_{G}$, defined in the main text, have some straightforward inclusion relations.

\begin{lemma}
We have that $\Psi^{(k),n}_{{\cal G}}\subseteq \Psi^{(k'),n}_{{\cal G}}$ and $\Psi^{(k),n}_{G}\subseteq \Psi^{(k'),n}_{G}$ for all $k'\geq k$. Moreover we have $\Psi^{(k),n}_{{\cal G}}\subseteq \Psi^{(k),n}_G$ for any $k$.
\end{lemma}
\begin{proof}
From Ref.~\cite{CuiEtAl_DiagonalGatesClifford_17} we know that ${\cal C}^{(k),n}_d$ is a subgroup of ${\cal C}^{(k'),n}_d$ for $k'\geq k$. This means ${\cal G}^{(k),n}\subseteq {\cal G}^{(k'),n}$ and $G^{(k),n}\subseteq G^{(k'),n}$ for any $k'\geq k$. So then suppose that $|\psi\rangle\in \Psi^{(k),n}_{{\cal G}}$ ($\Psi^{(k),n}_G$). Then there exists some subgroup $H$ of ${\cal G}^{(k),n}$ ($G^{(k),n}$) that uniquely stabilizes $|\psi\rangle$. But then $H$ must be a subgroup of ${\cal G}^{(k'),n}$ ($G^{(k'),n}$) for $k'\geq k$. So it follows that $\Psi^{(k),n}_{{\cal G}}\subseteq \Psi^{(k'),n}_{{\cal G}}$ and $\Psi^{(k),n}_{G}\subseteq \Psi^{(k'),n}_{G}$ for all $k'\geq k$.

The last piece follows using identical logic from the fact that ${\cal G}^{(k),n}\subseteq G^{(k),n}\subseteq G^{(k'),n}$ for all $k'\geq k$.
\end{proof}

\begin{figure}
    \centering
    \includegraphics[width=0.38\textwidth]{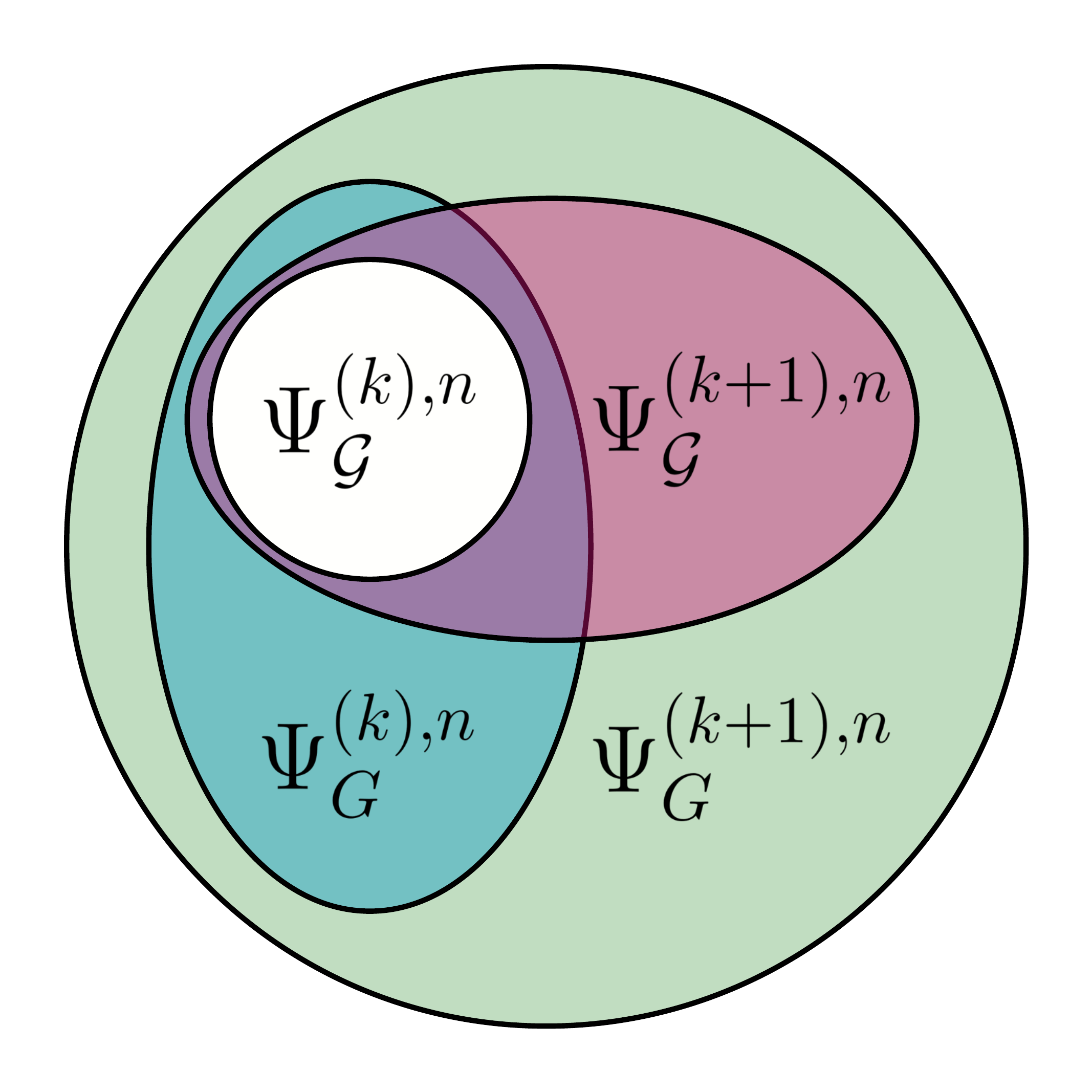}
    \caption{The structure of inclusions for the sets of stabilizer states defined in Def.~\ref{defn:onsite_stab_sets} for a generic $k$ and $n$. We do not display $k'<k$ or $k'>k+1$ and consider only a fixed number of qudits $n$.}
    \label{fig:set_inclusions}
\end{figure}

For a generic $k$ and $n$ this set of inclusions can be illustrated with Fig.~\ref{fig:set_inclusions}.

Having made our definitions of on-site Clifford hierarchy stabilizer states we can now attempt to classify these states. Using very similar logic as in Thm.~\ref{thm:unique_state} it is easy to see that if we have access to only on-site stabilizers at level $k$, i.e. if $|\psi\rangle \in \Psi_{{\cal G}}^{(k),n}$, then the phases in our wavefunction can be at most level $k+1$. However, it is much more challenging to determine whether or not it is possible to generate every possible multi-qudit function $\varphi$ at level $k+1$ using only on-site stabilizers. In general not every stabilizer state that can be stabilized with multi-qudit \xkcd stabilizers can be stabilized with single-qudit \xkcd stabilizers. We give an example of this below.

\begin{lemma} \label{lem:three_qubit_counter_example}
\begin{equation}
|\psi\rangle = \frac{1}{2^{3/2}}\sum_{x,y,z\in \mathbb{Z}_2} (-1)^{xyz}|x,y,z\rangle,
\end{equation}
is not in $\Psi^{(2),3}_{{\cal G}}$.
\end{lemma}
\begin{proof}
The states in $\Psi^{(2),3}_{{\cal G}}$ are stabilized by operators in ${\cal G}^{(2),3}$. These operators take the form $g=X(\mathbf{v})D[(ax+by+cz+d)/4]$ for $a,b,c,d\in \mathbb{Z}_4$ and $\mathbf{v}\in \mathbb{Z}_2^3$. Applying this operator to the above state we see that $g|\psi\rangle = |\psi\rangle$ if and only if
\begin{equation}
(-1)^{(x+v_0)(y+v_1)(z+v_2)} = (-1)^{xyz}i^{ax+by+cz+d},
\label{eq:psi_G_counter_example_2_3}
\end{equation}
for all $x,y,z\in \mathbb{Z}_2$. If we square this equation then it is not hard to see that $a,b,c,d=0\mod 2$ so that $a=2a',$ etc. 
\begin{equation}
(-1)^{(x+v_0)(y+v_1)(z+v_2)} = (-1)^{xyz + a'x+b'y+c'z+d'}
\end{equation}
for some $a',b',c',d'\in \mathbb{Z}_2$. Taking finite differences in two variables at a time in the phase reveals that $v_i=0$ for all $i$. But then this clearly forces $a',b',c',d'=0$ as well. Thus we conclude that the only operator in ${\cal G}^{(2),3}$ that stabilizes $|\psi\rangle$ is the identity. Thus there is no subgroup of ${\cal G}^{(2),3}$ for which $|\psi\rangle$ would be the unique stabilizer state. We conclude that this $|\psi\rangle\notin \Psi^{(2),3}_{{\cal G}}$.
\end{proof}

Despite this example, it is often the case that by adding extra qudits controlled on the first it is possible to generate an arbitrary multi-qudit phase in the wavefunction. We show how to do this for our above example below.

\begin{example} Note that this example was given in Eqs. (4) and (5) of Ref.~\cite{NiEtAl_NonCommutingStabilizerFormalism_15}.

Let $p=2$, so that we are considering qubits. Now consider ${\cal G}^{(2),6}$ and take the three elements $g_0, g_1, g_2 \in {\cal G}^{(2),6}$ given by
\begin{align}
g_0 &= X(\mathbf{e}_1+\mathbf{e}_4+\mathbf{e}_5)D[(x_2+x_3-x_6)/4]\\
g_1 &= X(\mathbf{e}_2+\mathbf{e}_4+\mathbf{e}_6)D[(x_1+x_3-x_5)/4]\\
g_2 &= X(\mathbf{e}_3+\mathbf{e}_5+\mathbf{e}_6)D[(x_1+x_2-x_4)/4].
\end{align}
One can check that these elements do not commute and all have order four. The subgroup $H = \langle g_0,g_1,g_2\rangle \subset {\cal G}^{(2),6}$ they generate is a non-abelian group.

It is particularly interesting to consider the diagonal subgroup of $H$, $D_H$. We take this to mean the subgroup of $H$ consisting of operators that are diagonal in the chosen qudit basis. It can be checked that $D_H\simeq \mathbb{Z}_2^3$ and is generated by the elements $g_0^2, g_1^2, g_2^2$ or, equivalently,
\begin{equation}
D_H = \langle Z\otimes Z \otimes 1 \otimes Z \otimes 1 \otimes 1, \ Z\otimes 1 \otimes Z \otimes 1 \otimes Z \otimes 1, \ 1\otimes Z \otimes Z \otimes 1 \otimes 1 \otimes Z\rangle.
\end{equation}
It is clear by inspection that any state stabilized by $D_H$ must be of the form
\begin{equation}
|\psi\rangle = \sum_{x,y,z\in \mathbb{Z}_2}\alpha_{x,y,z}|x,y,z,x+y,x+z,y+z\rangle,
\end{equation}
for some $\alpha_{x,y,z}$. These states form the stabilizer code ${\cal L}_{D_H}$ and since $D_H$ is a subgroup of $H$ we know the stabilizer code ${\cal L}_H$ is a subspace of ${\cal L}_{D_H}$.

We can now ask what the action of our original generators $g_0,g_1,g_2$ looks like when restricted to ${\cal L}_{D_H}$. In this subspace we see that
\begin{align}
\left.g_0\right|_{{\cal L}_{D_H}} &= X(\mathbf{e}_1+\mathbf{e}_4+\mathbf{e}_5)D[yz/2]\\
\left.g_1\right|_{{\cal L}_{D_H}} &= X(\mathbf{e}_2+\mathbf{e}_4+\mathbf{e}_6)D[xz/2]\\
\left.g_2\right|_{{\cal L}_{D_H}} &= X(\mathbf{e}_3+\mathbf{e}_5+\mathbf{e}_6)D[xy/2],
\end{align}
where we used the fact that for $x,y\in\{0,1\}$ if $[x+y]_2 = (x+y)\pmod{2}$ then
\begin{equation}\label{eqn:ancilla_qubit_eq}
(-1)^{xy} = i^{x+y-[x+y]_2}.
\end{equation}
But then it is clear that these uniquely stabilize the state 
\begin{equation}
|\psi\rangle = \sum_{x,y,z\in \mathbb{Z}_2}(-1)^{xyz}|x,y,z,x+y,x+z,y+z\rangle.
\end{equation}
\end{example}

In the example above we saw that by using only on-site stabilizers it was still possible to generate the multi-qudit function $D[\varphi] = (-1)^{xyz}$. This could be done by introducing other qubits which depended linearly on the free variables and using Eq.~\ref{eqn:ancilla_qubit_eq} to decompose products of free variables into sums of linear functions of free variables. Given this example we make the following conjecture.

\begin{conjecture}
Let $|\psi\rangle \in \Psi^{(k),n}_{{\cal G}}$. Then, up to permutation, it must take the form
\begin{equation}
|\psi\rangle = \frac{1}{p^{\ell/2}}\sum_{\mathbf{x}\in \mathbb{Z}_p^\ell} D[\phi(\mathbf{x})]|\mathbf{x},M\mathbf{x}+\mathbf{c}\rangle,
\end{equation}
where $M$ is an $(n-\ell)\times \ell$ matrix with entries in $\mathbb{Z}_p$, $\mathbf{c}\in \mathbb{Z}_p^{n-\ell}$ and $D[\phi] \in {\cal C}^{(k+1),\ell}_d$. Further, let $D[\phi]\in {\cal C}^{(k+1),\ell}_d, M\in \mathbb{Z}_p^{n_1\times \ell}, \mathbf{c}\in \mathbb{Z}_p^{n_1}$ for $n_1$ an arbitrary integer. Then there exists some $N\in \mathbb{Z}_p^{(n-n_1-\ell)\times \ell}, \mathbf{d}\in \mathbb{Z}_p^{n-n_1-\ell}$ such that
\begin{equation}
|\psi\rangle = \frac{1}{p^{\ell/2}}\sum_{\mathbf{x}\in \mathbb{Z}_p^\ell} D[\phi(\mathbf{x})]|\mathbf{x},M\mathbf{x}+\mathbf{c}, N\mathbf{x}+\mathbf{d}\rangle
\end{equation}
is contained in $\Psi^{(k),n}_{{\cal G}}$.
\end{conjecture}
Showing that all states in $\Psi^{(k),n}_{{\cal G}}$ must take this form proceeds identically to the proof in Thm.~\ref{thm:unique_state}. 

From Lem.~\ref{lem:three_qubit_counter_example} we see that if the states in the sum, $|\mathbf{x},M\mathbf{x}+\mathbf{c}\rangle$ are specified, then not every $\phi$ is allowed. However, the example above suggests that if we are allowed to supplement with additional qudits controlled on the first then we can generate an arbitrary multi-qudit function. Proving this would amount to showing that for every $D[\varphi] \in {\cal C}^{(k+1),\ell}_d$ there exists: an $(n-n_1-\ell)\times \ell$ matrix $N$ with entries in $\mathbb{Z}_p$, $\mathbf{d}\in \mathbb{Z}_p^{n-n_1-\ell}$, and $n-n_1$ functions $f_i\colon \mathbb{Z}_p\rightarrow \mathbb{R}/\mathbb{Z}$ with $D[f_i]\in {\cal C}^{(k+1),1}_d$ such that
\begin{equation}
\varphi(\mathbf{x}) = \sum_{i=1}^{\ell}f_i(x_i)+ \sum_{i=\ell+1}^{n-n_1}f_i\left(\sum_{j=1}^{\ell}N_{i-\ell,j}x_j+d_{i-\ell}\right)
\end{equation}
for all $\mathbf{x}\in \mathbb{Z}_p^\ell$. We expect that this is possible given the example above, but are not able to prove it here.

\section{Proofs related to stabilizer states}\label{app:stab_proofs}

\subsection{Proof of Lemma \ref{lem:state_req_stabilizers}}\label{ap:state_req_stabilizers}
\begin{proof}
Since this choice of $\phi(\mathbf{x})=x_0^a/p^m$ depends on only a single variable we note that all of the stabilizers $X(\mathbf{e}_i)D[\phi(\mathbf{x}+\mathbf{e}_i)-\phi(\mathbf{x})]$ lie in ${\cal G}^{(k+1),n}\subseteq G^{(k+1),n}$. So then $|\psi\rangle \in \Psi^{(k+1),n}_{{\cal G}}\subseteq \Psi^{(k+1),n}_G$. If we can show that $|\psi\rangle \notin \Psi^{(k),n}_G$ then because $\Psi^{(k),n}_{{\cal G}}\subseteq \Psi^{(k),n}_G$ we will have $|\psi\rangle \notin \Psi^{(k),n}_{{\cal G}}$ and we will be done.

First we consider all possible stabilizers of $|\psi\rangle$ in $G^{(k),n}$. Recall that a generic stabilizer takes the form $X(\mathbf{v})D_{m'}[f]$ where $(p-1)(m'-1) + \mathrm{deg}(f) \leq k$. Note that this means $a\geq(p-1)(m'-m)+\mathrm{deg}(f) + 2$. But since $1\leq a\leq p-1$ this means that $m'\leq m$. Now one can check that for $X(\mathbf{v})D_{m'}[f]$ to stabilize the above state we need that
\begin{equation}
D_{m'}[f(\mathbf{x})] = D[\varphi(\mathbf{x}+\mathbf{v}) - \varphi(\mathbf{x})] = D_m[[x_0+v_0]_p^a-x_0^a], \ \forall \mathbf{x}\in \mathbb{Z}_p^n,
\end{equation}
where $[x]_p = x\pmod{p}$. Now raise this expression to the power of $p^{m-1}$. If $m'< m$ then it reduces to $1=D_1[(x_0+v_0)^a-x_0^a]$ for all $x_0$. But this is clearly true if and only if $v_0=0$ (otherwise consider $x_0=0$ for the contradiction). On the other hand suppose $m'=m$. Then from above we have $a\geq  \mathrm{deg}(f)+2$. Now again raise the expression above to the power of $p^{m-1}$, then $D_1[f(\mathbf{x})] = D_1[(x_0+v_0)^a-x_0^a]$ for all $\mathbf{x}$. However, it is not hard to see that if $v_0\neq 0$ then the polynomial on the right hand side has degree $a-1$, so $a = \mathrm{deg}(f)+1$ and we have a contradiction with the derived lower bound on $a$. Thus in every case we must have $v_0=0$. Returning to the original equation we see this means $D_{m'}[f(\mathbf{x})] = 1$ for all $\mathbf{x}$. So we conclude that the stabilizers of this state in $G^{(k),n}$ must have the form $X(\mathbf{v})$ for $\mathbf{v}$ such that $v_0=0$. But these stabilizers cannot uniquely stabilize the above state. Thus we conclude there is no way to uniquely stabilize the above state with any subgroup of stabilizers in $G^{(k),n}$ and hence $|\psi\rangle \notin \Psi^{(k),n}_G$.
\end{proof}

\subsection{Proof of Lemma \ref{lem:wavefct_construction}}\label{ap:wavefct_construction}

\begin{proof}
We know from Lem.~\ref{lem:group_comp_wavefct} that 
\begin{equation}
|\psi\rangle = \frac{1}{\sqrt{|H|}}\sum_{g\in H}g|\mathbf{x}\rangle,
\end{equation}
for every $\mathbf{x}\in \mathbb{Z}_p^n$ such that $\sum_{g\in H}g|\mathbf{x}\rangle\neq 0$. Choose one of these $\mathbf{x}$. Now recall from Rem.~\ref{rem:orbit} that for every $g\in H$ there is a unique $g'\in H/D_H$ and $h\in D_H$ such that $g=g'h$ (where we are suppressing $p$). Now let $D_m[f]\in D_H$ for $f$ a map into $\mathbb{Z}_{p^m}$, then this means that 
\begin{align}
|\psi\rangle &= \frac{1}{\sqrt{|H|}}\sum_{g'\in H/D_H}\sum_{h\in D_H}g'h|\mathbf{x}\rangle \nonumber\\
&= \frac{1}{\sqrt{|H|}}\sum_{g'\in H/D_H}\sum_{h\in D_H/\langle D_m[f]\rangle}g'h \nonumber\\
&\times \left(\sum_{k=0}^{p^m-1}D_m[kf(\mathbf{x})]\right)|\mathbf{x}\rangle.
\end{align}
The term in parentheses is equal to zero if $f(\mathbf{x})\neq 0\pmod{p^m}$, otherwise it is equal to $p^m$. Since we have specified that $\mathbf{x}$ is such that $\sum_{g\in H}g|\mathbf{x}\rangle \neq 0$ then it must be the case that $f(\mathbf{x}) = 0\pmod{p^m}$. Further since $D_m[f]\in D_H$ was arbitrary we conclude that it must be the case that if $\sum_{g\in H}g|\mathbf{x}\rangle \neq 0$ then $f(\mathbf{x}) = 0\pmod{1}$ for every $D[f]\in D_H$. This tells us that $h|\mathbf{x}\rangle = |\mathbf{x}\rangle$ for all $h\in D_H$. So then we see that if $\sum_{g\in H}g|\mathbf{x}\rangle \neq 0$ then 
\begin{equation}
|\psi\rangle = \sqrt{\frac{|D_H|}{|H|}}\sum_{g'\in H/D_H}g'|\mathbf{x}\rangle.
\end{equation}

We have thus proven the simplified presentation of the stabilizer state, and established that if $\sum_{g\in H}g|\mathbf{x}\rangle \neq 0$ then $f(\mathbf{x})=0\pmod{1}$ for all $D[f]\in D_H$. Now we prove the reverse direction. Suppose that $f(\mathbf{x})=0\pmod{1}$ for all $D[f]\in D_H$. Then, as above, $h|\mathbf{x}\rangle = |\mathbf{x}\rangle$ for all $h\in D_H$. In particular this means that if we define
\begin{equation}
|\psi'\rangle = \frac{1}{\sqrt{|H|}}\sum_{g\in H}g|\mathbf{x}\rangle,
\end{equation}
then by the logic above
\begin{equation}
|\psi'\rangle =\sqrt{\frac{|D_H|}{|H|}}\sum_{g'\in H/D_H}g'|\mathbf{x}\rangle.
\end{equation}
Now consider $\langle \mathbf{x}|\psi'\rangle$. From Rem.~\ref{rem:orbit} we know that there is a unique correspondence between $g'$ and $\mathbf{v}\in \mathbb{Z}_p^n$ such that $g' = X(\mathbf{v})D[h]$ for some $D[h]$. In particular there is only one $g'$ such that $\langle\mathbf{x}|g'|\mathbf{x}\rangle\neq 0$, and this is $g'=1$. So then
\begin{align}
\langle \mathbf{x}|\psi'\rangle &= \sqrt{\frac{|D_H|}{|H|}}\sum_{g'\in H/D_H}\langle \mathbf{x}|g'|\mathbf{x}\rangle\\
&= \sqrt{\frac{|D_H|}{|H|}}\langle \mathbf{x}|1|\mathbf{x}\rangle\\
&= \sqrt{\frac{|D_H|}{|H|}}.
\end{align}
So we conclude that $|\psi'\rangle \neq 0$ and thus that $\sum_{g\in H}g|\mathbf{x}\rangle \neq 0$.
\end{proof}

\subsection{Proof of Lemma \ref{lem:Vpsi_descrip}}\label{ap:Vpsi_descrip}

\begin{proof}
Suppose that $\mathbf{x}\in \mathbb{Z}_p^n$ is such that $\sum_{g\in H}g|\mathbf{x}\rangle \neq 0$. Then by Lem.~\ref{lem:wavefct_construction} we have that 
\begin{equation}
|\psi\rangle = \sqrt{\frac{|D_H|}{|H|}}\sum_{g'\in H/D_H}g'|\mathbf{x}\rangle.
\end{equation}
Then $\langle \mathbf{x}|\psi\rangle = \sqrt{|H/D_H|}\langle \mathbf{x}|1|\mathbf{x}\rangle = \sqrt{|H/D_H|}$, by the logic of Lem.~\ref{lem:wavefct_construction}. So we see that $\mathbf{x}\in V_\psi$.

On the other hand suppose that $\mathbf{x}\in V_\psi$. We will show that this must mean that $f(\mathbf{x}) = 0\pmod{1}$ for all $D[f]\in D_H$. We prove this result by contradiction. Assume that there is some $D_m[f]\in D_H$ such that $f(\mathbf{x})\neq 0 \pmod{p^m}$. Further take $\mathbf{y}$ to be such that $\sum_{g\in H}g|\mathbf{y}\rangle \neq 0$. Then 
\begin{align}
\langle \mathbf{x}|\psi\rangle &= \frac{1}{\sqrt{|H|}}\sum_{g\in H}\langle \mathbf{x}|g|\mathbf{y}\rangle\nonumber\\
&= \frac{1}{\sqrt{|H|}}\sum_{g\in H/\langle D_m[f]\rangle} \langle \mathbf{x}|\left(\sum_{k=0}^{p^m-1}D_m[kf(\mathbf{x})]\right)g|\mathbf{y}\rangle\nonumber\\
&= 0.
\end{align}
Thus $\mathbf{x}\notin V_\psi$. We conclude, by contradiction, that if $\mathbf{x}\in V_\psi$, then $f(\mathbf{x})=0\pmod{1}$ for all $D[f]\in D_H$. But then by Lem.~\ref{lem:wavefct_construction} this means that $\sum_{g\in H}g|\mathbf{x}\rangle \neq 0$.
\end{proof}

\section{Restrictions on $G$ such that ${\cal D}(G) = {\cal D}^\omega({\cal A})$ with ${\cal A}$ abelian}\label{app:quant_double_rest}

In this appendix we establish necessary and sufficient conditions for a nonabelian quantum double ${\cal D} (G)$ to be equivalent to a twisted abelian quantum double ${\cal D}^{\omega} ({\cal A})$. We find that $G$ must be a nil-2 group, and that if it is a nil-2 group there always exists some ${\cal A}$ and some $\omega \in H^3({\cal A},\mathrm{U}(1))$ such that ${\cal D}(G) = {\cal D}^\omega({\cal A})$. If we further require that ${\cal A}=\mathbb{Z}_p^n$, for $p$ an odd prime, then $G$ must be a $p$-group with exponent $p$ or $p^2$. If we even further require that $\omega$ be a pure type-III cocycle then $G$ must have exponent $p$. Conversely, if $p$ is an odd prime then for every $G$ which is a nil-2 $p$ group of exponent $p$ there exists some $n$ and some pure type-III cocycle $\omega \in H^3(\mathbb{Z}_p^n, \mathrm{U}(1))$ such that ${\cal D}(G)={\cal D}^\omega(\mathbb{Z}_p^n)$.
This generalizes the well known case of ${\cal D} (D_4) = {\cal D} ^{\omega_\text{III}} (\mathbb{Z}_2^3)$, where $p=2$ is not an odd prime. Interestingly, quantum doubles of larger dihedral groups ${\cal D}(D_{> 4})$ cannot be equivalent to abelian twisted quantum doubles as $D_{> 4}$ are not nil-2.

We begin by proving a theorem.

\begin{theorem}\label{thm:if_and_only_if}
Suppose that $G$ is a group with nilpotency class two. Then let ${\cal A} \simeq Z(G)\times G/Z(G)$. Then there exists some $\omega \in H^3({\cal A},\U)$ such that ${\cal D}(G) = {\cal D}^\omega({\cal A})$.

Further, let $G$ be a generic group and suppose that there exists some abelian group ${\cal A}$ and some $\omega \in H^3({\cal A},\U)$ such that ${\cal D}(G)={\cal D}^\omega({\cal A})$. Then $G$ has nilpotency class two.
\end{theorem}
\begin{proof}
If $G$ is a nil-2 group then the extension $1\rightarrow Z(G)\rightarrow G \rightarrow G/Z(G)\rightarrow 1$ is central with $Z(G)$ and $G/Z(G)$ abelian. Thus ${\cal A}=Z(G)\times G/Z(G)$ is abelian. Then from the discussion in Appendix C.2 of \cite{TantivasadakarnEtAl_HierarchyTopologicalOrder_23} it follows that there exists some $\omega \in H^3({\cal A},\U)$ such that ${\cal D}(G)={\cal D}^\omega({\cal A})$.

The other direction is demonstrated in the rest of this section.
\end{proof}

\begin{remark}
Note that this theorem does {\bf not} say that for every ${\cal A}$ and every $\omega \in H^3({\cal A},\U)$ there exists some nil-2 $G$ such that ${\cal D}^\omega({\cal A})={\cal D}(G)$. A counterexample to this statement is ${\cal D}^\omega(\mathbb{Z}_2)$ with nontrivial $\omega$. This is equivalent to the doubled semion model which is not equal to any untwisted quantum double.
\end{remark}

In order to prove that if ${\cal D}(G) = {\cal D}^\omega({\cal A})$ for some abelian ${\cal A}$ and some $\omega \in H^3({\cal A},\U)$, then $G$ is nil-2, we begin with a lemma about abelian Lagrangian subalgebras.

\begin{lemma}\label{lem:abelianTQD}
Every twisted quantum double (TQD) of an abelian group ${\cal A}$, ${\cal D}^\omega({\cal A})$, has an abelian Lagrangian subalgebra.
\end{lemma}
\begin{proof}
This follows from the fact that every abelian TQD can be generated by gauging an $\omega$-SPT protected by an ${\cal A}$ zero-form symmetry, where $\omega \in H^3({\cal A},\U)$. Doing so will leave dual one-form symmetries in ${\cal D}^\omega({\cal A})$ which are generated by ${\cal A}$. If the anyons of ${\cal A}$ are condensed it will trivialize the theory, so ${\cal A}\subseteq {\cal D}^\omega({\cal A})$ is an abelian Lagrangian subalgebra. For a good reference on this see Section 4.4 of Ref.~\cite{DrinfeldEtAl_BraidedFusionCategories_10}. Another good reference is Ref.~\cite{BarkeshliEtAl_SymmetryFractionalizationDefects_19}.
\end{proof}

Now we give a number of useful and short definitions and lemmas which we will need to prove our main theorem.

\begin{definition}
A group $G$ is nil-2 if its commutator subgroup, $[G, G]$, lies within its center, $Z(G)$.
\end{definition}

\begin{definition}
Let $G^{ab} = G/[G,G]$ be the abelianization of $G$, where $[G,G]$ is again the commutator subgroup.
\end{definition}

\begin{remark}
As its name suggests, the abelianization $G^{ab}$ is abelian since we have quotiented out by all elements which do not commute.
\end{remark}

\begin{definition}
Let $G$ be a finite group and define $\widehat{G} = {\rm Hom}(G, \U)$. Note that this is the set of 1-dimensional representations of $G$.
\end{definition}

\begin{lemma}\label{lem:hatG}
We have that $\widehat{G} \cong \widehat{G^{ab}}$.
\end{lemma}
\begin{proof}
Let $\phi:G\rightarrow G^{ab}\cong G/[G,G]$ denote the quotient map from $G$ to $G^{ab}$. Then for any $\chi^{ab}\in\widehat{G^{ab}}$, $\chi = \chi^{ab} \circ \phi$ is a one-dimensional representation on $G$. Conversely given $\chi \in \widehat{G}$, $\chi(aba^{-1}b^{-1})=1$ for all $a,b\in G$. Thus $\chi(g)=1$ for all $g\in[G,G]$ and hence $\chi$ is well defined on cosets of $[G,G]$, which precisely form $G^{ab}$. Then each $\chi \in \widehat{G}$ maps to a $\chi^{ab}\in \widehat{G^{ab}}$, establishing the bijection.
\end{proof}

\begin{definition}
Let $G$ be a group and $S\subseteq G$ be a subgroup. Then define the annihilator of $S$ in $\widehat{G}$ as 
\begin{equation}
{\rm Ann}_G(S) = \left\{\chi \in \widehat{G}: \ \chi(s)=1 \quad \forall s\in S\right\}.
\end{equation}
\end{definition}

\begin{lemma}\label{lem:Hsub}
Let $H$ be a subgroup of a finite abelian group $G$. Then ${\rm Ann}_G(H)$ is isomorphic to $G/H$.
\end{lemma}
\begin{proof}
Follows from Ref.~\cite{LidlNiederreiterFiniteFields}.
\end{proof}

\begin{lemma}\label{lem:Gab_size}
Let $\varphi$ be the quotient map from $G$ to $G^{ab}$ and $S\subseteq G$ be a subgroup. Then
\begin{equation}
|G^{ab}| = |{\rm Ann}_G(S)|\cdot |\varphi(S)|.
\end{equation}
\end{lemma}
\begin{proof}
We note that $\varphi(S)\subseteq G^{ab}$ is a subgroup of the finite abelian group $G^{ab}$. From Lem.~\ref{lem:Hsub} we then have that ${\rm Ann}_{G^{ab}}(\varphi(S))$ is isomorphic to $G^{ab}/\varphi(S)$. Then we see that
\begin{equation}
|G^{ab}| = |{\rm  Ann}_{G^{ab}}(\varphi(S))|\cdot |\varphi(S)|,
\end{equation}
and it remains only to prove that $|{\rm  Ann}_{G^{ab}}(\varphi(S))| = |{\rm Ann}_G(S)|$.

From Lem.~\ref{lem:hatG} we have that for every $\chi \in \widehat{G}$ there exists some $\chi^{ab}\in \widehat{G^{ab}}$ such that $\chi = \chi^{ab}\circ \varphi$ and vice versa. If $\chi \in {\rm Ann}_G(S)$ then $1=\chi(S)=\chi^{ab}(\varphi(S))$ so $\chi^{ab} \in {\rm Ann}_{G^{ab}}(\varphi(S))$ and vice versa. So we see that $|{\rm  Ann}_{G^{ab}}(\varphi(S))| = |{\rm Ann}_G(S)|$ and we are done.
\end{proof}

Having provided these definitions and lemmas, we now use them to prove our main result.

\begin{theorem}\label{thm:abelian_implies_nil2}
For a finite group $G$, the quantum double ${\cal D}(G)$ contains an abelian Lagrangian subalgebra if and only if $G$ is nil-2.
\end{theorem}
\begin{proof}
We note that one direction is already proven. Ref.~\cite{TantivasadakarnEtAl_HierarchyTopologicalOrder_23} showed that if $G$ is a nil-2 group then ${\cal D}(G)$ contains an abelian Lagrangian subalgebra. Thus it remains only to prove that if ${\cal D}(G)$ has an abelian Lagrangian subalgebra then $G$ is necessarily nil-2.

First recall that the anyons in ${\cal D}(G)$ are labeled by $(C, R)$, where $C$ is a conjugacy class of $G$ and $R$ is an irreducible representation of $C_G(r_C)$, where $r_C\in C$ is an arbitrary representative element of $C$ and $C_G(r_C)$ is the centralizer of $r_C$ in $G$ \cite{Simon_TopologicalQuantum_23}. The quantum dimension of this anyon is given by $d_{(C,R)} = |C|d_R$, where $d_R$ is the dimension of the representation. Thus an abelian anyon $(C,R)\in {\cal A}$ must be such that $|C|=1$ and $d_R=1$. Having $|C|=1$ implies that the conjugacy class is a single $g$ that commutes with every other element in $G$, i.e. is in $Z(G)$. Then $C_G(g)=G$ and $d_R=1$ forces the representation on $G$ to be 1-dimensional. Thus ${\cal A} = Z(G)\times \widehat{G}$. 

Let us now determine the spins and braidings of the anyons in ${\cal A}$. The topological spin is given by \cite{Simon_TopologicalQuantum_23}
\begin{equation}
    e^{i\theta}=\dfrac{\chi^R(r_c)^\ast}{d_R}
\end{equation}
where $\chi^R$ is the character of the representation $R$. Letting $(g, \chi)$ denote elements of ${\cal A}$, with $g\in Z(G)$ and $\chi \in \widehat{G}$, the spin of such an anyon is simply $\chi(g)^\ast$. The $S$-matrix elements of a quantum double are given by \cite{Simon_TopologicalQuantum_23}:
\begin{equation}
    S_{[C, R],[C', R']}=\dfrac{1}{| C_G(r_{C'})| | C_G(r_C)|} \sum_{p: \; p^{-1}r_{C'}p \in C_G(r_C)}
    \chi^{R'}(p \ r_{C} \ p^{-1})\chi^R(p^{-1} \ r_{C'} p)
\end{equation}
For elements of ${\cal A}$, this simplifies to
\begin{equation}
    S_{(g, \chi), (g', \chi')} = \frac{1}{| G |}\chi(g')\chi'(g)
\end{equation}
The braiding phase can then be calculated from the ``monodromy scalar component'', Eq.~(38) of \cite{BarkeshliEtAl_SymmetryFractionalizationDefects_19},
\begin{equation}
M_{ab} = \frac{S^*_{ab}S_{00}}{S_{0a}S_{0b}} = e^{i\theta_{a,b}} = \left(\chi(g'
)\chi'(g)\right)^\ast
\end{equation}

Now suppose that ${\cal D}(G)$ possesses an abelian Lagrangian subalgebra, ${\cal L}\subseteq {\cal A}$. Then the anyons in ${\cal L}$ must be self-bosons and braid trivially with one another \cite{Simon_TopologicalQuantum_23}. Thus at least two conditions for ${\cal L}$ to be Lagrangian must be satisfied,
\begin{enumerate}
    \item $\chi(g)=1 \quad \forall (g, \chi)\in {\cal L}$
    \item $\chi_1(g_2)\chi_2(g_1)=1 \quad \forall (g_1, \chi_1), (g_2, \chi_2)\in {\cal L}$.
\end{enumerate}
Additionally, for ${\cal L}$ to be Lagrangian $| {\cal L} |  = | G|$. This follows from the fact that ${\cal D}_{{\cal D}(G)} = |G|$ and condensing ${\cal L}$ reduces the quantum dimension by a factor of $|{\cal L}|$. Since condensing a Lagrangian subalgebra must trivialize the quantum double it follows $|{\cal L}| = |G|$.

Now since ${\cal A} = Z(G)\times \widehat{G}$, there exists projectors $\pi_1:{\cal A}\rightarrow Z(G)$ and $\pi_2:{\cal A}\rightarrow \widehat{G}$. With the inclusion map $i: {\cal L} \rightarrow {\cal A}$, $\pi_1$ and $\pi_2$ extend to a projectors on ${\cal L}$; $\pi_1':{\cal L}\rightarrow Z(G)$ and $\pi_2':{\cal L}\rightarrow \widehat{G}$. Let $K = {\rm ker}(\pi_1')$ and $S={\rm im}(\pi_1')$. Clearly $K$ is composed of all elements of the form $(0,\chi)$ in ${\cal L}$.

Now we note $\pi_2'(K) \subseteq {\rm Ann}_G(S)$. To see this, let $(0, \chi) \in K$ and $g\in S \subseteq Z(G)$. Since $S = {\rm im}(\pi_1')$, there exists an anyon $(g,\eta)\in {\cal L}$ for which $\pi_1(g, \eta) = g$. By the Lagrangian braiding condition applied to $(0, \chi)$ and $(g, \eta)$ we require that $1=\chi(g)\eta(0) = \chi(g)$. This is true for all $g \in S$, so by the definition of the annihilator $\pi_2'(K) \subseteq {\rm Ann}_G(S)$. We further note that because $K$ is composed of all elements of the form $(0,\chi)\in {\cal L}$ that $|\pi_2'(K)|=|K|$.

Now we calculate group sizes.
By the definition of $G^{ab}$ we have that $| [G, G] | \cdot | G^{ab} | = | G|$.
By the Lagrangian assumption $| G| = | {\cal L}|$. Further $|{\cal L}|=|K|\cdot |S|$ by the first isomorphism theorem. Then $| [G, G] | \cdot | G^{ab} | = | K| \cdot | S|$.
From the above we also have that 
\begin{equation}
| K| =| \pi_2'(K)|\leq | {\rm Ann}_G(S) | 
\end{equation}
Thus 
\begin{equation}
|[G,G]|\cdot |G^{ab}| = |K|\cdot |S| \leq |{\rm Ann}_G(S)|\cdot |S|.
\end{equation}
Next we can use the result of Lem.~\ref{lem:Gab_size} to see that $|{\rm Ann}_G(S)| = |G^{ab}|/|\varphi(S)|$. Inserting this into the above equation, dividing by $|G^{ab}|$ and multiplying by $|\varphi(S)|$ allows us to see that
\begin{equation}
| [G, G] | \cdot | \phi(S) | \leq | S |
\end{equation}

Now consider the first isomorphism theorem applied to the restriction of $\varphi$ to $S$, $\varphi_S\colon S\rightarrow G^{ab}$. Clearly ${\rm im}(\varphi_S) = \varphi(S)$ and ${\rm ker}(\varphi_S) = S\cap [G,G]$. Then we see that 
\begin{equation}
    | S |  = | S \cap [G, G]| \cdot | \phi (S)|
\end{equation}
Therefore $| [G, G]| \leq | S \cap [G, G]|$. This can only be true if $[G, G] \subseteq S$. But $S\subseteq Z(G)$, so $[G,G] \subseteq Z(G)$ and $G$ must be nil-2.
\end{proof}

\subsection{Restrictions on $G$ such that ${\cal D}(G) = {\cal D}^\omega(\mathbb{Z}_p^n)$}
Specializing to when the abelian group ${\cal A}$ is equal to $\mathbb{Z}_p^n$ we will be able to obtain even more restrictions on $G$. The most basic restriction is on the size of $G$ and follows from the quantum dimension of the theories.

\begin{lemma}\label{lem:qd_size}
Let ${\cal D}^\omega(G)$ be a twisted quantum double. Then its total quantum dimension ${\cal D}_{{\cal D}^\omega(G)}=|G|$.
\end{lemma}
\begin{proof}
This follows from Lemma 5.9 of Ref.~\cite{NaiduEtAl_FusionSubcategoriesRepresentation_09}.
\end{proof}

\begin{corollary}
Let $G$ be a group and suppose there exists ${\cal A}=\mathbb{Z}_p^n$ and $\omega \in H^3(\mathbb{Z}_p^n, \U)$ such that ${\cal D}(G)={\cal D}^\omega(\mathbb{Z}_p^n)$, where $p$ is a prime. Then $G$ is a $p$-group of size $p^n$.
\end{corollary}
\begin{proof}
This follows directly from Lem.~\ref{lem:qd_size}.
\end{proof}

Next we can get a useful restriction on the exponent of $G$ using the $T$-matrix of these theories.

\begin{lemma}\label{lem:ord_T}
Let $T$ be the $T$-matrix of the untwisted quantum double ${\cal D}(G)$. Define the order of $T$, $\mathrm{ord}(T)$, to be the smallest integer, $n$, such that $T^n=1$. Then $\mathrm{ord}(T)=\mathrm{exp}(G)$. (Here $\mathrm{exp}(G)$ is the smallest integer $n$ such that $g^n=e$ for all $g \in G$.)
\end{lemma}
\begin{proof}
This is Proposition 1(b) of \cite{CosteEtAl_FiniteGroupModular_00}.
\end{proof}

\begin{lemma}\label{lem:types}
Let $T$ be the $T$-matrix of the twisted quantum double ${\cal D}^\omega({\cal A})$ for ${\cal A}=\mathbb{Z}_p^n$ for $p$ an odd prime. Then $\mathrm{ord}(T)=p$ if and only if $\omega$ has only type-III components. Otherwise $\mathrm{ord}(T)=p^2$.
\end{lemma}
\begin{proof}
It is shown in \cite[\S5.3]{CosteEtAl_FiniteGroupModular_00} that $\mathrm{ord}(T)$ divides $n\mathrm{exp}({\cal A})$, where $n$ is the cohomological order of $\omega$. Since ${\cal A}=\mathbb{Z}_p^n$ we know that $\mathrm{exp}({\cal A}) = p$. Further
\[
H^3({\cal A},\U)\cong \mathbb{Z}_p^{\;n+\binom{n}{2}+\binom{n}{3}}
\qquad\text{(Eq.~(102) of \cite{HuEtAl_TwistedQuantumDouble_13}),}
\]
so every cocycle has order dividing $p$. Thus $\mathrm{ord}(T)$ must divide $p^2$, so it can be $1, p,$ or $p^2$.

Now from Ref.~\cite{HuEtAl_TwistedQuantumDouble_13} the anyons of ${\cal D}^\omega({\cal A})$ are labeled by $(g,\mu)$ with $g\in {\cal A}$ and $\mu$ labeling an irreducible projective representation. The topological spin of an anyon $(g,\mu)$ is then given by $\theta^\mu_g$, where these appear as the diagonal elements of the $T$ matrix. Further Ref.~\cite{HuEtAl_TwistedQuantumDouble_13} showed that 
\begin{equation}\label{eq:theta-p}
(\theta_g^\mu)^p=\Omega_g,\qquad
\Omega_g=\prod_{m=0}^{p-1}\omega(g,mg,g),
\end{equation}
where $\Omega_g$ is Eq.~(71) of \cite{HuEtAl_TwistedQuantumDouble_13} rewritten in additive notation.

We know that the most general 3-cocycle in $H^3(\mathbb{Z}_p^n,U(1))$ is given up to coboundary by
\begin{equation}
    \omega(a,b,c)=\prod_{i=1}^{N} \omega^{(i)}_\mathrm{I}(a,b,c)^{n_i}\prod_{i<j}^{N} \omega^{(ij)}_\mathrm{II}(a,b,c)^{n_{ij}}\prod_{i<j<k}^{N} \omega^{(ijk)}_\mathrm{III}(a,b,c)^{n_{ijk}},
\end{equation}
where 
\begin{equation}\label{eq:general_cocycle}
    \omega_\mathrm{I}^{(i)}(a,b,c)=e^{\frac{2\pi i}{p^2}a_i f(b_i,c_i)},\quad \omega_\mathrm{II}^{(ij)}(a,b,c)=e^{\frac{2\pi i}{p^2}a_i f(b_j,c_j)},\quad \omega_\mathrm{III}^{(ijk)}(a,b,c)=e^{\frac{2\pi i}{p}a_ib_jc_k}.
\end{equation}
This tells us that
\begin{equation}
    \omega(g,mg,g)=\prod_{i=1}^{N} e^{\frac{2\pi i}{p^2}n_i g_i f(m g_i,g_i)}\prod_{i<j}^{N} e^{\frac{2\pi i}{p^2}n_{ij} g_i f(m g_j,g_j)}\prod_{i<j<k}^{N} e^{\frac{2\pi i}{p} n_{ijk} m g_ig_jg_k}.
\end{equation}

First, consider the topological spin $\theta_{\hat e_\ell}^\mu$ of the anyons labeled by $(\hat e_\ell,\mu)$, where $\hat e_\ell \in\mathbb{Z}_p^n$ is the vector with all entries 0 except the $\ell$th entry, which is 1. Eq.~\ref{eq:theta-p} tells us that 
\begin{equation}
    (\theta_{\hat e_\ell}^\mu)^p=\Omega_{\hat e_\ell}=\prod_{m=0}^{p-1} e^{\frac{2\pi i}{p^2}n_\ell f(m,1)}=e^{\frac{2\pi i}{p}n_\ell},
\end{equation}
where in the last equality we have used the fact that $f(m,1)=m+1-[m+1]_p$ equals $0$ for $m=0,\dots,p-2$ and equals $p$ for $m=p-1$. Since $n_\ell\in\{0,\ldots,p-1\}$, $e^{\frac{2\pi i}{p}n_\ell}$ is a $p$th root of unity. Whenever $\omega$ contains a type-I cocycle ($n_\ell\neq 0 \;(\text{mod}\;p)$ for any $\ell$), we can thereby conclude that ${\cal D}^\omega(A)$ will have anyons with topological spins of order $p^2$. In these cases, $\mathrm{ord}(T)=p^2$.

Now suppose that $n_\ell$ vanishes for all $\ell$. In this case, we may consider the topological spin of the anyons labeled by $(\hat e_\ell+\hat e_r,\mu)$. Setting $n_\ell=0$ for all $\ell$ in Eq.~\ref{eq:general_cocycle} and applying Eq.~\ref{eq:theta-p}, we find
\begin{equation}
    (\theta_{\hat e_\ell+\hat e_r}^\mu)^p=\Omega_{\hat e_\ell+\hat e_r}=\prod_{m=0}^{p-1} e^{\frac{2\pi i}{p^2}n_{\ell r} f(m,1)}=e^{\frac{2\pi i}{p}n_{\ell r}},
\end{equation}
where the last equality is obtained through the same considerations as above. This tells us that any whenever $\omega$ contains a type-II cocycle ($n_{\ell r}\neq 0 \;(\text{mod}\;p)$ for any pair $(\ell,r)$), we can conclude that ${\cal D}^\omega(A)$ will again have topological spins of order $p^2$ so that $\mathrm{ord}(T)=p^2$.

Now suppose $\omega$ is generated by type--III cocycles only.
It suffices to check a single generator $\alpha_{III}^{(ijk)}$. For fixed $g\in {\cal A}$,
\[
\alpha_{III}^{(ijk)}(g,mg,g)=\exp\!\left(\frac{2\pi i}{p}\,g_i\,(m g_j)\,g_k\right)
=\exp\!\left(\frac{2\pi i}{p}\,m\,g_i g_j g_k\right).
\]
Therefore
\[
\prod_{m=0}^{p-1}\alpha_{III}^{(ijk)}(g,mg,g)
=\exp\!\left(\frac{2\pi i}{p}\,\Big(\sum_{m=0}^{p-1}m\Big)\,g_i g_j g_k\right)
=\exp\!\left(\frac{2\pi i}{p}\,\frac{p(p-1)}{2}\,g_i g_j g_k\right)=1,
\]
because $p\mid \frac{p(p-1)}{2}$ for odd $p$. Hence $\Omega_g=1$ for all $g$. In turn we know that $T^p=I$, so $\mathrm{ord}(T)\leq p$. We noted above that $\mathrm{ord}(T)\in \{1,p,p^2\}$ and it cannot be 1 unless the theory is trivial \cite{BarkeshliEtAl_SymmetryFractionalizationDefects_19}, so we may conclude that $\mathrm{ord}(T)=p$ in this case.
\end{proof}

\begin{corollary}\label{cor:G_p_group}
Let $G$ be a group and suppose there exists ${\cal A} = \mathbb{Z}_p^n$ and $\omega \in H^3(\mathbb{Z}_p^n, \U)$ such that ${\cal D}(G) = {\cal D}^\omega(\mathbb{Z}_p^n)$, with $p$ an odd prime. Then $G$ is a $p$-group with exponent $p$ or $p^2$. Further, if $G$ has exponent $p^2$ then $\omega$ must possess a cocycle of type-I or II.
\end{corollary}
\begin{proof}
Follows directly from Lem.~\ref{lem:ord_T} and Lem.~\ref{lem:types}.
\end{proof}

So far we have been obtaining restrictions on what types of $G$ can satisfy ${\cal D}(G) = {\cal D}^\omega(\mathbb{Z}_p^n)$. Now we see that, at least in certain cases, we can turn this into an if and only if.

\begin{theorem}\label{thm:nil2_exp}
Suppose that $G$ is a nil-2 $p$ group with exponent $p$, where $p$ is an odd prime. Then there exists some $n$ and some $\omega \in H^3(\mathbb{Z}_p^n,\U)$ which is only a product of type-III cocycles such that ${\cal D}(G)={\cal D}^\omega(\mathbb{Z}_p^n)$.

On the other hand suppose that ${\cal D}(G)={\cal D}^\omega(\mathbb{Z}_p^n)$ where $\omega$ is a product of only type-III cocycles and $p$ is an odd prime. Then $G$ is a nil-2 $p$-group with exponent $p$.
\end{theorem}
\begin{proof}
Suppose that $G$ is a nil-2 $p$ group with exponent $p$. Then both $Z(G)$ and $G/Z(G)$ must be abelian $p$ groups of exponent $p$. This means that they are both products of $\mathbb{Z}_p$. In particular this means that there exists some $n$ such that ${\cal A} = Z(G)\times G/Z(G)\simeq \mathbb{Z}_p^n$. Then we can use Thm.~\ref{thm:if_and_only_if} to see that there must exist some $\omega \in H^3(\mathbb{Z}_p^n,\U)$ such that ${\cal D}^\omega({\cal A})={\cal D}(G)$. Now using Lem.~\ref{lem:types} and the fact that $G$ has exponent $p$ we learn that $\omega$ must be made up of only type-III cocycles.

For the other direction we can use Thm.~\ref{thm:if_and_only_if} to see that $G$ is nil-$2$. That it is a $p$-group with exponent $p$ then follows from Cor.~\ref{cor:G_p_group}.
\end{proof}

Now from our earlier work we have seen that we can prepare all twisted quantum doubles of the form ${\cal D}^\omega(\mathbb{Z}_p^n)$ with $\omega$ a type-III cocycle using only Clifford stabilizers. Further, we conjectured that we \textit{cannot} prepare ${\cal D}^\omega(\mathbb{Z}_p^n)$ with any type-I or II cocycles using only Clifford stabilizers. This lends itself to a very speculative conjecture. Suppose that $G$ is a $p$-group, with $p$ an odd prime. Then we can prepare ${\cal D}(G)$ using only Clifford stabilizers if and only if $G$ is also nil-$2$ with exponent $p$.

\section{Additional magic state preparation details}\label{ap:magic-state}

In this section, we present figures which visualize conclusions discussed in the main text. In Fig.~\ref{fig:color-auto}, we show that $\color{red}\mathcal{A}_{v,r}$ commutes with $\color{Green}\hat C_g$ up to a product of $\color{blue}B_{f,b}$, which enables us to compute the logical action by acting on a single representative state. Fig.~\ref{fig:full-color-logs} depicts this computation.

\begin{figure}[h]
  \centering
  $\vcenter{\hbox{\begin{tikzpicture}[x=2.5cm,y=2.5cm]
  \path[use as bounding box, clip] (-0.65,-1.55) rectangle (1.65,0.65);

  \lattice{\cellB}{(0,-0.5773503)}
  \lattice{\cellR}{(0.5,-0.2886751)}
  \lattice{\cellG}{(0,0)}

  \foreach \y in {-2,...,1}{%
    \foreach \x in {-1,...,2}{%
      \node[stamp] at (\x+0.5*\y,\hh*\y) {\usebox\czbox};}}

  \begin{scope}[shift={(0.5,-\hh/3)},red,line width=1.5pt,line cap=round]
    \draw[dirA] ( 0  , 0  ) -- ( 1  , 0  );
    \draw[dirA] (-1  , 0  ) -- ( 0  , 0  );
    \draw[dirB] ( 0  , 0  ) -- ( 0.5, \hh);
    \draw[dirB] (-0.5,-\hh) -- ( 0  , 0  );
    \draw[dirC] ( 0  , 0  ) -- (-0.5, \hh);
    \draw[dirC] ( 0.5,-\hh) -- ( 0  , 0  );
    \foreach \a in {120,180,240}{\node[op,fill=red] at (\a:0.5) {\fixw{$X^\dagger$}};}
    \foreach \a in {300,0,60}  {\node[op,fill=red] at (\a:0.5) {$X$};}
  \end{scope}

\end{tikzpicture}}}\;\;\mylongmapsto{$\quad{\color{Green}\hat C_g}\quad$}\;\;\vcenter{\hbox{\begin{tikzpicture}[x=2.5cm,y=2.5cm]
  \path[use as bounding box, clip] (-0.65,-1.55) rectangle (1.65,0.65);

  \lattice{\cellB}{(0,-0.5773503)}
  \lattice{\cellR}{(0.5,-0.2886751)}
  \lattice{\cellG}{(0,0)}

  \begin{scope}[blue,line width=1.5pt,line cap=round]
    \draw[dirA] (-0.5, \hh/3   ) -- ( 0.5, \hh/3   );
    \draw[dirA] ( 0.5, \hh/3   ) -- ( 1.5, \hh/3   );
    \draw[dirA] ( 0  ,-2*\hh/3 ) -- ( 1  ,-2*\hh/3 );
    \draw[dirB] ( 1  ,-2*\hh/3 ) -- ( 1.5, \hh/3   );
    \draw[dirB] ( 0.5,-5*\hh/3 ) -- ( 1  ,-2*\hh/3 );
    \draw[dirC] ( 0  ,-2*\hh/3 ) -- (-0.5, \hh/3   );
    \draw[dirC] ( 0.5,-5*\hh/3 ) -- ( 0  ,-2*\hh/3 );
  \end{scope}

  \begin{scope}[shift={(0.5,-\hh/3)},red,line width=1.5pt,line cap=round]
    \draw[dirA] ( 0  , 0  ) -- ( 1  , 0  );
    \draw[dirA] (-1  , 0  ) -- ( 0  , 0  );
    \draw[dirB] ( 0  , 0  ) -- ( 0.5, \hh);
    \draw[dirB] (-0.5,-\hh) -- ( 0  , 0  );
    \draw[dirC] ( 0  , 0  ) -- (-0.5, \hh);
    \draw[dirC] ( 0.5,-\hh) -- ( 0  , 0  );
    \foreach \a in {120,180,240}{\node[op,fill=red] at (\a:0.5) {\fixw{$X^\dagger$}};}
    \foreach \a in {300,0,60}  {\node[op,fill=red] at (\a:0.5) {$X$};}
  \end{scope}

  \node[op ,fill=blue] at ( 0   , \hh/3  ) {$Z$};
  \node[op ,fill=blue] at ( 1   , \hh/3  ) {$Z$};
  \node[op ,fill=blue] at (-0.25,-\hh/6  ) {\fixw{$Z^\dagger$}};
  \node[op ,fill=blue] at ( 1.25,-\hh/6  ) {\fixw{$Z^\dagger$}};
  \node[op ,fill=blue] at ( 0.25,-7*\hh/6) {$Z$};
  \node[op ,fill=blue] at ( 0.75,-7*\hh/6) {$Z$};
  \node[opl,fill=blue] at ( 0.5 ,-2*\hh/3) {$Z^{\dagger2}$};

\end{tikzpicture}}}$
  \caption{We demonstrate $\color{Green}\hat C_g$ preserves the codespace by deriving its action on the red vertex stabilizers. Notice that $\color{Green}\hat C_g$ attaches to each red vertex stabilizer a product of four blue plaquette stabilizers. The action of $\color{Green}\hat C_g$ on the blue vertex stabilizers is analogous.}
  \label{fig:color-auto}
\end{figure}

\begin{figure}
  \centering
  $\vcenter{\hbox{\begin{tikzpicture}[x=2.5cm,y=2.5cm]
  \path[use as bounding box,clip] (-2.5,-3) rectangle (4,1);

  \lattice{\cellB}{(0,-0.5773503)}
  \lattice{\cellR}{(0.5,-0.2886751)}
  \lattice{\cellG}{(0,0)}

  \draw[dash pattern={on 8pt off 5pt}, blue, line width=1pt] (-3,-\hh/6) -- (5,-\hh/6);
  \draw[dash pattern={on 8pt off 5pt}, blue, line width=1pt] (0.5,-14*\hh/3) -- (3.5,4*\hh/3);
  \draw[dash pattern={on 8pt off 5pt}, red , line width=1pt] (-2,-13*\hh/3) -- (1,5*\hh/3);
  \draw[dash pattern={on 8pt off 5pt}, red , line width=1pt] (-3,-11*\hh/6) -- (5,-11*\hh/6);

  \foreach \y in {-4,...,2}{%
    \pgfmathtruncatemacro{\xmin}{-4-\y/2}%
    \pgfmathtruncatemacro{\xmax}{ 5-\y/2}%
    \foreach \x in {\xmin,...,\xmax}{%
      \node[stamp] at (\x+0.5*\y,\hh*\y) {\usebox\czbox};}}

  \draw[line width=1.5pt,line cap=round]       (-30:0.28) to[bend left=20] ( 30:0.28);
  \draw[line width=1.5pt,line cap=round]       (210:0.28) to[bend left=20] (270:0.28);
  \draw[line width=1.5pt,line cap=round,brown] (270:0.28) to[bend left=20] (330:0.28);
  \begin{scope}[shift={(2,-2*\hh)}]
    \draw[line width=1.5pt,line cap=round]       ( 90:0.28) to[bend left=20] (150:0.28);
    \draw[line width=1.5pt,line cap=round,brown] ( 30:0.28) to[bend left=20] ( 90:0.28);
    \draw[line width=1.5pt,line cap=round,brown] (150:0.28) to[bend left=20] (210:0.28);
  \end{scope}

  \begin{scope}[blue,line width=1.5pt,line cap=round]
    \foreach \xs in {-2,...,6}{
      \draw[dirA] (-1.5+\xs,\hh/3) -- (-1.0+\xs,-2*\hh/3);
      \node[opl,fill=blue] at (-1.25+\xs,-\hh/6) {$-n$};
      \draw[dirA] (-1.0+\xs,-2*\hh/3) -- (-0.5+\xs,\hh/3);
      \node[op,fill=blue] at (-0.75+\xs,-\hh/6) {\fixw{${n}$}};}
  \end{scope}

  \begin{scope}[blue,line width=1.5pt,line cap=round]
    \foreach \xs in {-3,...,3}{
      \draw[dirA] (1.5+\xs/2,-5*\hh/3+\hh*\xs) -- (2.5+\xs/2,-5*\hh/3+\hh*\xs);
      \node[op,fill=blue] at (2+\xs/2,-5*\hh/3+\hh*\xs) {\fixw{$v$}};
      \draw[dirA] (2+\xs/2,-2*\hh/3+\hh*\xs) -- (2.5+\xs/2,-5*\hh/3+\hh*\xs);
      \node[op,fill=blue] at (2.25+\xs/2,-5*\hh/3+\hh*\xs+\hh/2) {\fixw{$v$}};}
  \end{scope}

  \begin{scope}[red,line width=1.5pt,line cap=round]
    \foreach \xs in {-2,...,6}{
      \draw[dirA] (-1+\xs,-4*\hh/3) -- (-0.5+\xs,-7*\hh/3);
      \node[opl,fill=red] at (-0.75+\xs,-11*\hh/6) {$-u$};
      \draw[dirA] (-0.5+\xs,-7*\hh/3) -- (\xs,-4*\hh/3);
      \node[op,fill=red] at (-0.25+\xs,-11*\hh/6) {\fixw{${u}$}};}
  \end{scope}

  \begin{scope}[red,line width=1.5pt,line cap=round]
    \foreach \xs in {-3,...,3}{
      \draw[dirA] (-1+\xs/2,-4*\hh/3+\hh*\xs) -- (\xs/2,-4*\hh/3+\hh*\xs);
      \node[op,fill=red] at (-0.5+\xs/2,-4*\hh/3+\hh*\xs) {\fixw{$m$}};
      \draw[dirA] (-0.5+\xs/2,-\hh/3+\hh*\xs) -- (\xs/2,-4*\hh/3+\hh*\xs);
      \node[op,fill=red] at (-0.25+\xs/2,-4*\hh/3+\hh*\xs+\hh/2) {\fixw{$m$}};}
  \end{scope}

  \node[opl,fill=blue] at ( 2.75,-\hh/6   ) {$v-n$};
  \node[opl,fill=red ] at (-0.75,-11*\hh/6) {$m-u$};
\end{tikzpicture}}}$
  \caption{We take a representative state in the ${\color{red}\ket{\overline{m}}_{rV}}{\color{blue}\ket{\overline{n}}_{bH}}{\color{red}\ket{\overline{u}}_{rH}}{\color{blue}\ket{\overline{v}}_{bV}}$ homology class, where all qudits not explicitly labeled are taken to be $\ket{0}$. We draw the six local operators which contribute a nontrivial phase to $\color{Green}\hat{C}_g$ in full saturation. These phases multiply to $\zeta^{-mn+uv}$, producing an effective logical action $\overline{CZ}_{{\color{red}rV},{\color{blue}bH}}^\dagger\overline{CZ}_{{\color{red}rH},{\color{blue}bV}}$.}
  \label{fig:full-color-logs}
\end{figure}

\end{document}